\documentclass[11pt]{article}
\usepackage[a4paper, left=2.5cm, right=2.5cm, top=2.5cm, bottom=2.5cm]{geometry}
\usepackage{graphicx} 
\usepackage{caption}
\usepackage{subcaption} 
\usepackage{hyperref} 
\usepackage{enumitem}
\usepackage[round, authoryear]{natbib}
\usepackage{amsmath}
\usepackage{amsfonts}
\usepackage{comment}
\usepackage{mathtools} 
\mathtoolsset{showonlyrefs} 
\usepackage{booktabs} 

\usepackage{authblk}

\title{Reinforcement Learning for Trade Execution \\
with Market and Limit Orders\thanks{We thank Robert Almgren, Shankar Narayanan, Isaac Scheinfeld and two anonymous referees for helpful comments. Support from Swiss NSF Grant 10003723 is gratefully acknowledged.}
}
\date{} 

\author{Patrick Cheridito  and 
	Moritz Weiss\thanks{Corresponding author: moritz.weiss@math.ethz.}  \\
	Department of Mathematics, ETH Zurich, Switzerland}

\usepackage{upgreek}
\usepackage{algorithm}
\usepackage{algorithmic}

\DeclareMathOperator{\N}{\mathbb{N}}

\DeclareMathOperator{\E}{\mathbb{E}}
\DeclareMathOperator{\Cov}{\mathrm{Cov}}

\DeclareMathOperator{\Var}{\mathrm{Var}}
\DeclareMathOperator{\SI}{\mathbb{S}}
\DeclareMathOperator{\R}{\mathbb{R}}

\newcommand{\inventory}{M}
\newcommand{\activeorders}{m}
\newcommand{\traderstd}{\delta}

\newcommand{\gaegamma}{\gamma^\text{GAE}}
\newcommand{\gaelambda}{\lambda^\text{GAE}}

\newcommand{\expsub}{\substack{a \sim \pi_\theta \\  s_0 \sim \rho}}

\usepackage{amsthm}
\theoremstyle{remark}
\newtheorem{remark}{Remark}[section]

\begin{document}

\maketitle

\begin{abstract}
	In this paper, we introduce a novel reinforcement learning framework for optimal trade execution in a limit order book. We formulate the trade execution problem as a dynamic allocation task whose objective is the optimal placement of market and limit orders to maximize expected revenue. By modeling market and limit order allocations with multivariate logistic-normal distributions, the framework enables efficient training of the reinforcement learning algorithm. Numerical experiments show that the proposed method outperforms traditional benchmark strategies in simulated limit order book environments featuring noise traders submitting random orders, tactical traders responding to order book imbalances, and a strategic trader seeking to acquire or liquidate an asset position.	

\end{abstract}

\section{Introduction} 
The trade execution problem is a key concern for traders attempting to minimize trading costs and maximize execution speed. Its mathematical formulation goes back to \citet{BERTSIMAS19981} and \citet{almgren2001optimal}. Subsequent works such as 
\cite{almgren2003optimal, almgren2012optimal, gatheral2011optimal,	gatheral2013dynamical, obizhaeva2013optimal, 
	cheridito2014optimal, 
	gueant2015general,
	cartea2016incorporating,
	 curato2017optimal, 
	 di2024optimal}
 have generalized the price dynamics and form of the trade impact. These works all focus on finding an optimal execution schedule. Related works such as \cite{guilbaud2013optimal, guilbaud2015optimal, cartea2013modelling, cartea2015optimal, huitema2014optimal, cartea2015algorithmic} investigate the optimal placement of market and limit orders, but they model the mid-price together with a reduced-form market impact to keep the optimization problem tractable. 
 
 In contrast to these approaches, our work studies the optimal use of market and limit orders, cancellations and their interactions with the full limit order book. Since this results in high-dimensional state and action spaces, we train a reinforcement learning algorithm to find an optimal execution strategy.
Various authors have applied reinforcement learning to trade execution using market simulations based on historical data. But typically, they impose certain restrictions on how orders can be placed. For instance, in \citet{nevmyvaka2006reinforcement}, the entire inventory is replaced in every step, ignoring the queue positions of limit orders. \citet{moallemi2022reinforcement} frame the trade execution problem as an optimal stopping problem, determining the optimal time to execute a market order. Other approaches, such as \citet{hendricks2014reinforcement, ning2021double, leal2022learning} develop optimal execution schedules, but are not concerned with order placement.

Modeling market impact in simulations based on historical data faces the limitation that the data is given and does not react to simulated trades. While direct market impact due to the consumption of liquidity on the other side of the book can be modelled using real limit order book data, indirect impact, which results from market participants adjusting to an algorithm’s behavior, cannot.
As a result, many reinforcement learning approaches model only direct impact or ignore market impact altogether. A more realistic alternative uses stochastic simulations with interacting trading agents, where both direct and indirect impacts emerge naturally. 
Typically, these agents trade according to Poisson processes as in \citet{cont2010stochastic}. Other works use hand-crafted trading agents such as \citet{byrd2020abides}. Algorithms trained in such simulations \citep{hafsi2024optimal, byun2023practical} still face limitations due to restricted state and action spaces.

We introduce a new approach. Our main contributions are:
\begin{itemize}
	\item     
	The development of a novel reinforcement learning algorithm that uses the logistic-normal distribution to model random allocations as distributions over the simplex.  We follow an actor-critic approach and derive a new expression for the associated policy gradient. It is not limited to trade execution and can be used for any dynamic allocation task. 
	\item 
	A formulation of the trade execution problem with general state and action spaces. While the states provide detailed information on current and past market conditions and the positions of all limit orders posted by the algorithm, the actions allow for precise and adaptable order placement and cancellations. This extends existing works, which typically have restricted state and action spaces. 
	\item 
	A detailed empirical analysis of the algorithm's performance 
	in simulated markets. The simulations contain trading agents reacting to the algorithm's order placement, causing direct and indirect market impact. The algorithm’s performance is compared with common trade execution benchmarks used by practitioners.  
\end{itemize}

The remainder of the paper is structured as follows. In Section \ref{LOBs}, we review limit order books.
In Section \ref{sec:trade_execution_problem}, we describe the trade execution problem in detail. In Section \ref{RL}, we derive the policy gradient under the logistic-normal distribution and discuss the resulting algorithm. In Section \ref{SIM}, we introduce the financial market. In Section \ref{sec:experiments}, we discuss numerical results in various market simulations.

\section{Limit order books} 
\label{LOBs}

A limit order book is a mechanism used by most trading exchanges to process orders submitted by market participants. Market participants may submit market orders, limit orders, or cancellations. Arriving market orders are immediately matched with limit orders on the opposite side of the book. 
Limit orders allow agents to post buy or sell offers away from the best market offers, but they are not guaranteed to be filled. Cancellations allow an agent to withdraw active limit orders from the book. Prices are given in discrete units called {\sl ticks}\footnote{A tick is the minimum price increment. It is typically 0.01 or 0.05 USD for common equity markets.}. 
The best bid price $p^b(t)\in\R_{+}$ is the highest price at which buy orders are available in the limit order book. Conversely, the best ask price $p^a(t)\in\R_{+}$ is the lowest price at which sell orders are present in the book.
One always has  $p^b(t) < p^a(t).$ Usually, the mid-price $p(t) = (p^b(t) + p^a(t))/2$ is used as a reference price. The spread is given by $s(t) = p^a(t) - p^b(t) \in \R_+$ and is a common indicator of market liquidity. An order book can be described by a vector $v(t) = (v^{b}(t), v^{a}(t))$ that describes the volume on each price level, where 
\begin{equation}
	v^{b}(t) = (v^{b,1}(t), \dots, v^{b,D}(t)) \in \R^D_{+}
	\quad \mbox{and} \quad 
	v^{a}(t) =  (v^{a,1}(t), \dots, v^{a,D}(t)) \in \R^D_{+} 
\label{volume} 
\end{equation} 
for $ D \in \N.$ For $k\in\{1,\dots,D\} $, the quantity $v^{b,k}(t)$ is the volume $k-1$ ticks below the best bid price. In particular, $v^{b,1}(t)$ is the volume at the best bid price. The volumes on the ask side $v^{a}(t)$ are defined analogously.   
The volume vector $v(t)$ reveals information about demand and supply in the market.  
For example, if there is more volume on the bid than the ask side, it is more likely that the mid-price will go up next. 
The volumes on each price level consist of individual limit orders by market participants and form a queue. 
If an incoming market order arrives at a price level, the limit orders that arrived first in the queue will be matched first. Any optimal execution strategy should keep track of price levels and queue positions of active limit orders. For instance, the lower the price level and the queue position of a limit order, the more likely it will be filled. For an extensive overview of limit order books we refer to the review paper \citet{gould2013limit} or the textbook \citet{bouchaud2018trades}. 

\section{The trade execution problem}
\label{sec:trade_execution_problem} 

We consider a trade execution algorithm selling $\inventory\in\N$ lots\footnote{A {\sl lot} refers to a single unit of a financial asset.} over a given time interval $[0,T]$ at discrete time steps $t_n=n
\Delta t$ for $n\in\{0,1,\dots,N-1\},$ where $N\in\N,$ and $\Delta t = T/N.$ We denote the algorithm's inventory at time $t\in\{t_0,t_1,\dots, t_N\}$ by $M(t)\in \{0,1,\dots,M\}.$ The algorithm starts with an initial inventory $M(0)=M$ and must sell everything by the final time $ T$, so that $M(T)= 0$. At each time $t_n,$ the algorithm observes a market state $s_n,$ takes an action $a_n,$ and receives a random reward $r_n (s_n, a_n).$
We assume at each time $t_n,$ the algorithm takes into account what happened during the preceding subinterval $(t_{n-1}, t_n],$ and takes an action $a_n$ immediately after $t_n.$ 
The states reflect the condition of the order book, including the current time, while the actions describe the algorithm’s order placement or cancellation. 
The algorithm learns an optimal policy, which we model as a conditional density $\pi$ on the action space, such that 
\begin{equation} 
	\label{objective} 
	J(\pi) = \underset{\expsub}{\E} \left[ \sum_{n={0}}^{N-1} r(s_n, a_n)  \right]
\end{equation}
is maximized. The subscript $a\sim\pi$ means that each action $a_n$ in the sequence of actions $a=(a_0, a_1, \dots, a_{N-1})$ is sampled from a conditional distribution with density $\pi(\cdot\mid s_n)$ in each step $n\in\{0,1,\dots,N-1\}.$ The transition from one state $s_n$ to the next $s_{n+1}$ results from the interaction of our algorithm with the other market participants. The subscript $s_0\sim \rho$ means that the initial state $s_0$ is sampled from an initial distribution $\rho.$

\subsection{State space}
\label{subsec:observation_space} 
We divide states into market states, which are visible to all market participants, and private states, only visible to the algorithm.

\paragraph{Market states}
The market states consist of the following quantities:
\begin{itemize}
	\item The best bid and ask prices $p^b(t)$ and $p^a(t).$ 	
	\item The first $K-1$ entries of both volume vectors defined in \eqref{volume}, that is $(v^{b,1}(t), \dots, v^{b,K-1}(t))$ and $(v^{a,1}(t), \dots, v^{a,K-1}(t))$ for $K\in\N$ with $K-1\leq D.$  
	\item 
	The market order flow $\Delta^M(t)\in\R,$ 	which is the difference between the total volume of market buy and sell orders in the time interval $(t-\Delta t, t].$ 
	\item The limit order flow $\Delta^L(t)\in\R,$ which is the difference between the total volume of limit buy and sell orders in the time interval
	 $(t-\Delta t, t].$ 
     \item The cancellation order flow $\Delta^C(t)\in\R,$ which is the difference between the total volume of sell and buy limit order cancellations in the interval $(t-\Delta t, t]$. 
	\item The mid-price drift from time $t-\Delta t$ to $t,$ given by $\Delta p(t) = p(t) - p(t-\Delta t)\in\R.$ 				
\end{itemize}

\paragraph{Private states} The private states consist of the following quantities:

\begin{itemize}
	\item The time $t\in\{t_0,t_1,\dots,t_{N-1}\}$ within the execution period. 
	\item The inventory $\inventory(t)\in\{1,2,\dots,M\},$ which is the number of lots the algorithm still has to sell.  
	\item The number of limit orders $\activeorders(t) \in \{0,1,\dots, \inventory(t)\},$ currently resting in the limit order book.
	\item 
	For $i\in\{1,2,\dots, \activeorders(t)\},$ the levels and queue positions $\left(l^i(t), q^i(t)\right) \in \N \times \N$ of the active limit orders. E.g.\  $\left(l^i(t), q^i(t)\right)=(l,q)$ means that the $i$-th active limit order is located $l^i(t)-1$ ticks above the best ask price, and at position $q^i(t)$ in the order queue.   
\end{itemize}
We normalize the features above to speed up the learning process. Furthermore, we add an additional feature that represents the allocation of the algorithm's orders per price level. Both are explained in more detail in Appendix \ref{sec:feature_normalization} below. 

\subsection{Action space}
\label{subsec:action_space}
The algorithm observes states at the discrete time steps $t\in\{t_0, t_1, \dots, t_{N-1}\}$ right before taking an action $a(t).$ For $K\in \N,$ the action space is the simplex
\[
\SI^K=\left\{a=(a^0, a^1, \dots, a^K): \,  \sum_{k=0}^K a^k = 1 , a^k\geq0 \right\}. 
\] 
An action $a = (a^0, a^1, \dots, a^K),$ represents the allocation of orders to market and limit orders. More precisely, $a^0$ is the fraction of the remaining inventory $M(t)$ to be sent as a market order, $a^k$ for $k\in\{1,2,\dots K-1\}$ is the fraction of the remaining inventory $M(t)$ to be placed $k$ ticks above the best bid price, and $a^K$ is the fraction of the remaining inventory $M(t)$ to be held outside of the order book. We require that $K-1\leq D,$ such that the algorithm always posts orders within the visible order book levels defined in \eqref{volume}. If the vector $a(t)\inventory(t)$ includes non-integer values, we round the entries  $a^0(t)\inventory(t), a^1(t)\inventory(t), \dots, a^{K-1}(t)\inventory(t)$ to the closest integer sequentially, always assuring that total allocated orders to not exceed $\inventory(t).$ To preserve queue positions, the algorithm only cancels orders if necessary to reach the new allocation when reallocating orders, prioritizing canceling orders with higher queue positions first. 

\begin{figure}[htbp]	
	\begin{center}
		\centerline{\includegraphics[width=\linewidth]{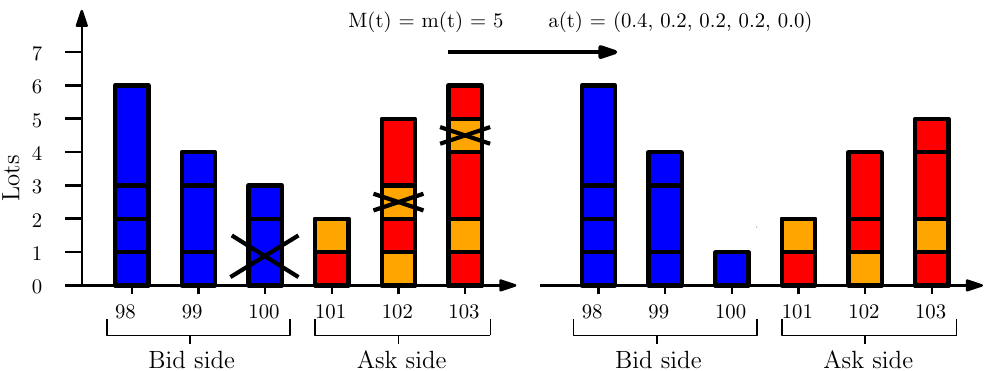}}
		\centering 
		\caption{Limit order book with the algorithm's orders in orange. Limit buy and sell orders are colored in blue and red, respectively. The algorithm takes action $a(t).$ The left panel shows the order book before the action. The right panel shows the order book after the action. The crosses indicate order cancellations (for the orange orders) or market order fills (for the blue orders).} 
		\label{fig:order_book_with_action}
	\end{center}
\end{figure}

Figure \ref{fig:order_book_with_action} illustrates the states and actions. 
At time $t,$ the entire inventory $M(t) = m(t) = 5$ is sitting as limit orders in the limit order book. The algorithm then decides to reallocate the orders according to the action $a(t)=(0.4,0.2,0.2,0.2, 0.0).$ The left panel of the figure shows the order book before the action. The right panel
shows the order book at time $t+$, immediately after the action $a(t).$ 
The best bid and ask prices are $p^b(t)=100$ and $p^a(t)=101.$ The bid and ask volumes before taking the action  are  $(v^{b,1}(t),v^{b,2}(t),v^{b,3}(t))=(3,4,6)$ and $(v^{a,1}(t),v^{a,2}(t),v^{a,3}(t))=(2,5,6).$  The volumes right after the action are given by $(v^{b,1}(t+),v^{b,2}(t+),v^{b,3}(t+))=(1,4,6)$ and $(v^{a,1}(t+),v^{a,2}(t+),v^{a,3}(t+),v^{a,4}(t+))=(2,4,5).$ The algorithm's orders are colored in orange. The buy and sell limit orders of other market participants are shown in blue and red, respectively. To reach the new allocation, the algorithm must cancel one order on the second and third ask levels. Each of those levels contains two orders. On each price level, the algorithm cancels the order with the worst queue position, as indicated by the two crossed orange boxes. Here, we assume that it is always better to cancel orders with a higher queue position first. After canceling the two limit orders, it sends a market sell order of size two, as indicated by the crossed blue box. The levels and queue position of the algorithm at time $t$ are given by $(l^1(t), q^1(t))=(1,2), (l^2(t), q^2(t))=(2,1), (l^3(t), q^3(t))=(2,3), (l^4(t),q^4(t))=(3,2), (l^5(t),q^5(t))=(3,5).$ The levels and queue positions at time $t+$ are given by $(l^1(t+), q^1(t+))=(1,2), (l^2(t+), q^2(t+))=(2,1), (l^3(t+), q^3(t+))=(3,2).$ The inventory and the number of limit orders resting in the book are $M(t+)=m(t+)=3.$

\subsection{Rewards}
\label{sec:rewards}
At each time $t_n,$ the algorithm places market and limit orders according to an action $a_n.$ The market orders are filled immediately after $t_n,$ leading to an instant cash flow, while the limit orders remain in the book waiting to be filled. The total cash flow from market and limit order fills in the interval $(t_n, t_{n+1}]$ is denoted by $\bar{r}(s_n, a_n).$ If any inventory is left at time $t_N,$ the algorithm is forced to send a market order of size $\inventory(t_N).$ In that case, the cash flow $\bar{r}(s_{N-1}, a_{N-1})$ also contains the proceeds generated by the market order sent at time $t_N.$ Suppose that the algorithm sells $\gamma_n$ lots in the interval $(t_n, t_{n+1}].$ Then, we define a normalized reward function, which compares the reward $\bar{r}(s_n, a_n)$ to the cash flow that would have been generated by trading at the initial bid price. More precisely, we define
\begin{equation*}
	\label{normalized_rewards} 
	r(s_n, a_n) = \frac{\bar{r}(s_n, a_n) - \gamma_n p^b(0)}{\inventory}. 
\end{equation*}
Then objective \eqref{objective} becomes
\begin{align} 
	\label{total_normalized_rewards}	
	J(\pi)
	=
	\frac{1}{\inventory}
	\underset{\expsub}{\E}
	\left[ 
	\sum_{n=0}^{N-1} \bar{r}(s_n, a_n)  
	\right]- p^b(0), 
\end{align} 
which describes the average revenue per lot relative to the initial bid price. The metric in \eqref{total_normalized_rewards} is a normalized version of the industry-standard measure used to evaluate execution algorithms, commonly known as {\sl implementation shortfall}. We observed in our numerical experiments that normalizing the cash flows increases the training speed of the reinforcement learning algorithm. 

\begin{remark} Rather than forcing the algorithm to sell any leftover inventory, one could also introduce a soft constraint by using a penalty for the leftover inventory. However, forcing liquidation at terminal time is common in academic papers \citep{almgren2001optimal, nevmyvaka2006reinforcement} and is also a reasonable assumption for real-world applications. It is fairly common for traders to set a time limit for a trade execution algorithm to complete. 
\end{remark} 

\subsection{Benchmark algorithms}
\label{ss:bm} 
\label{sec:benchmark_algorithms}
We compare our trained reinforcement learning algorithm with two heuristic benchmarks and an alternative reinforcement learning algorithm using the Dirichlet distribution. The Dirichlet algorithm is explained in Appendix \ref{sec:dirichlet_rl_implementation_details}. 
The heuristic strategies trade according to the following rules:
\begin{itemize}
	\item Submit and leave algorithm (SL algorithm): Submits the entire position $M$ at $t=0$ as a limit order at the best ask price. 
	\item Time-weighed average price\footnote{This is abbreviated with TWAP and represents a standard execution algorithm banks and hedge funds use.} algorithm (TWAP algorithm): 
     Divides the position into blocks of size $M/N$ and posts a limit sell order of size $M/N$ at each time $t_n, n =0,\dots, N-1.$ 
\end{itemize}
Both algorithms sell any leftover inventory $\inventory(t_N)$ with a market order.

\section{Actor-critic policy gradient with logistic-normal distribution} 
\label{RL} 

\subsection{Actor-critic policy gradient} 
The trade execution problem described in Section \ref{sec:trade_execution_problem} is a stochastic control problem. The transition probabilities from state $s_n$ to  $s_{n+1}$ and the reward function $r(\cdot)$ are not available in closed form, but we can simulate them. Therefore, the problem can be solved with reinforcement learning. Actor-critic policy gradient algorithms are a class of reinforcement learning algorithms that aim to find an optimal stochastic policy $\pi_\theta(\cdot \mid s_n),$ referred to as {\sl actor}, by estimating the gradient of the objective function $J(
\pi_\theta).$ A separate parameterized value or advantage function, called {\sl critic}, is used to reduce the variance of the gradient estimate. 
We use a neural network with weights $\theta$ to map states $s_n$ to the parameters of the policy's density. The gradient of the objective function can be computed as 
\begin{equation}
	\label{policy_gradient} 
	\nabla_\theta J(\pi_\theta)
	= 
	\underset{  
    \expsub
    }{\E}
	\left[\sum_{n=0}^{N-1} A^{\pi_\theta}(s_n, a_n) \nabla_\theta \log \pi_\theta(a_n \mid s_n ) 
	\right],
\end{equation}
by the policy gradient theorem; see e.g.~\cite{schulman2015high}. 
We note that the actions $a_0, \dots a_{N-1}$ are sampled from a conditional distribution with density $\pi_\theta(\cdot \mid s_n)$ and the states $(s_0,\dots,s_{N-1})$ are sampled from the market environment.
Here, $A^{\pi_\theta}$ is the advantage function, which is defined together with the value function $V^{\pi_{\theta}}$ by
\begin{align}
        \label{eq:value_function}
        V^{\pi_{\theta}}(s) &= \underset{a\sim\pi_\theta}{\E} \left[ \sum_{l=n}^{N-1} r(s_l, a_l) \mid s_n =s \right],        \\
		A^{\pi_\theta}(s, a) &=  
		\underset{a\sim\pi_\theta}{\E} \left[ \sum_{l=n}^{N-1} r(s_l, a_l) \mid s_n =s, a_n=a \right]- 
        V^{\pi_{\theta}}(s),  
        \label{advantage} 		
\end{align}
where $s$ is a state and $a$ an action at time step $n.$

\subsection{Policy gradient with logistic-normal distribution}
\label{sec:policy_gradient_with_logistic_normal}

In most applications of policy gradients with continuous action spaces, the multivariate normal distribution is used as an action distribution, see e.g. \cite{schulman2017proximal}. However, in our application, the actions must be contained in the simplex. 
Therefore, a natural choice for the action distribution is the {\sl logistic-normal distribution}, which has support in $\SI^K.$

The logistic transformation $h$ is a function that maps vectors $x=(x^0, x^1, \dots, x^{K-1})$ in $\R^K$ to actions $a=(a^0, a^1, \dots, a^K)$ in $\SI^K,$ where   
\begin{align*}
	a^k &= \frac{e^{x^k}}{1 + \sum_{l=0}^{K-1} e^{x^l} }, \quad k = 0,1,\dots,K-1, 
	\\ 
	a^K &= \frac{1}{1+\sum_{l=0}^{K-1} e^{x^l}} . 
\end{align*} 
The inverse mapping $h^{-1}$ is given by 
\begin{align*}
	x^k = \log\left(\frac{a^{k}}{a^{K}} \right), \quad k = 0,1,\dots K-1. 
\end{align*} 
Let $X$ be a random variable with multivariate normal distribution on $\R^K$ with mean $\mu=(\mu^0, \dots, \mu^{K-1}) \in \R^{K}$ and covariance matrix $\Sigma = (\Sigma^{ij})_{i,j = 0, \dots ,K-1} \in \R^{K\times K}.$  Then, the transformed random variable $h(X)$ has a logistic-normal distribution. The distribution has support $\SI^K,$ and its density function is given by 
\begin{equation}
	\pi(a)= 
	\frac{1}{ (2\uppi)^{K} |\Sigma|^{\frac{1}{2} }\prod_{k=0}^K a^k }
	e^{-\frac{1}{2}  
	\left(h^{-1}(a) - \mu\right)^T  
	\Sigma^{-1}
	\left(h^{-1}(a) - \mu \right)
}
,
\quad 
\text{for } a \in \SI^K;
\label{density}
\end{equation}
see e.g.~\citet{atchison1980logistic} for a detailed overview of the logistic-normal distribution. 

We choose a policy that has logistic-normal distribution with density $\pi_\theta(\cdot\mid s)$ for each state $s.$ The parameters $\theta$ are the weights of a neural network (or any other parametric family of functions) that maps states $s$ to the mean of the underlying normal distribution $\mu.$ 
We model the covariance matrix $\Sigma$ as a hyperparameter that is iteratively reduced during training, as explained further below. Let $\varphi_\theta$ be the density of the underlying normal distribution on $\R^K$ with mean $\mu_\theta$ and covariance matrix $\Sigma.$ 
Since the normalizing constant in \eqref{density} does not depend on $\theta,$ we obtain for each state $s$ and action $a\in\SI^K$ that 
\begin{equation}
	\label{eq:gradient_of_density}
	\nabla_\theta \log \left(\pi_\theta(a \mid s)\right) = 
	\nabla_\theta \log \left(\varphi_\theta \left( h^{-1}(a) \mid s \right)\right). 
\end{equation}
Combining equations \eqref{policy_gradient} and \eqref{eq:gradient_of_density} yields 
\begin{align}  
		\nabla_\theta J(\pi_\theta) &= 
		\underset{\expsub}{\E}
		\left[\sum_{n=0}^{N-1} 
        A^{\pi_\theta}(s_n, a_n)
		\nabla_\theta \log 
		\left(\varphi_\theta\left( h^{-1}(a_n) | s_n \right)\right) 
		\right] \\ 
		& =
		\underset{\substack{x \sim \varphi_\theta \\  s_0 \sim \rho}}{\E}
		\left[\sum_{n=0}^{N-1} A^{\pi_\theta}(s_n, h(x_n)) \nabla_\theta 
		\log \left(\varphi_\theta(x_n \mid s_n ) \right) 
		\right]. 
\label{policy_gradient_logistic_normal}
\end{align}
Here $x\sim\varphi_\theta$ means that each element $x_n$ in the sequence $x=(x_0,x_1,\dots,x_{N-1})$ is sampled from a normal distribution with conditional density $\varphi_\theta(\cdot \mid s_n)$ for each state $s_n$ in the sequence $(s_0, s_1,\dots,s_{N-1}).$ 
The main difference between our algorithm and other actor-critic reinforcement learning algorithms is the use of the logistic-normal distribution, which ensures actions lie in the simplex. The representation \eqref{policy_gradient_logistic_normal} still allows us to compute gradients efficiently.

\begin{remark}
Instead of using the multivariate logistic-normal distribution, one could combine the multivariate normal distribution with a softmax activation function. More precisely, if $X$ is a multivariate normal distribution on $\R^{K+1}$ we could parametrize the policy as $\text{softmax}(X)$, where $\text{softmax}(\cdot)$ maps $x\in\R^{K+1}$ to $a\in\SI^K$ with 
\[
a^k = 
\frac{e^{x^k}}{\sum_{l=0}^{K} e^{x^l}}, \quad \text{for } k = 0,1,\dots,K.  
\] 
However, $\text{softmax}(X)$ is not bijective. Therefore, it is not possible to derive the policy gradient in closed form as in \eqref{policy_gradient_logistic_normal}.
Prior works such as \cite{tian2022prescriptive} and \cite{winkel2023simplex} have employed the Dirichlet distribution to model actions on the simplex. We use an actor-critic algorithm based on the Dirichlet distribution as a benchmark in our experiments, but we find that it performs sub-optimally, even with extensive hyperparameter tuning. 
\end{remark}

\subsection{Initialization and scaling of policy parameters}
\label{sec:initialization_scaling_parameters}
The initialization of the mean and reduction of the variances of the stochastic policy $\pi_\theta$ is an important aspect of training the policy.
The mean and covariance of the logistic-normal distribution are not known in closed form. However, for the log ratios, we have 
\begin{align}
	\label{mean}
	\E \left[ \log \left( \frac{a^j}{a^k} \right) \right] &= \mu^j - \mu^k,  \\ 
	\Cov \left( \log\left(\frac{a^j}{a^k} \right), \log \left( \frac{a^l}{a^m} \right) \right) &= \Sigma^{jl} + \Sigma^{km} - \Sigma^{jm} - \Sigma^{kl}, 
	\label{cov}  
\end{align}
for $j,k,l,m \in \{0,1,2, \dots K\},$ with the convention $\mu^{K}=0$ and $\Sigma^{j, K}=0.$ 
At initialization, we set the parameters of the neural network’s final policy layer close to zero, except for the bias term $b\in\R^K.$ Then, the bias $b$  controls the initial allocations of the policy.
We model $\Sigma$ as $\sigma I_K,$ where $I_K$ is the $K$-dimensional identity matrix and $\sigma$ is reduced after each gradient step. This allows for exploration at the beginning of training but forces the algorithm to gradually reduce the variance of the stochastic policy until it becomes deterministic. With a diagonal covariance matrix, for different $j,k,l$ and $m,$ equation \eqref{cov} implies 
\[ 
\Cov \left( \log\left(\frac{a^j}{a^k} \right), \log \left( \frac{a^l}{a^m} \right) \right) = 0. 
\]
For indices $j$ and $k$ with $j\neq k,$ we obtain for the variances of the log ratios that 
\begin{equation*}
	\label{variances}
	\Var \left( \log\left(\frac{a^j}{a^k} \right) \right)
	= 
	\begin{cases}
		\sigma:   \text{ if } j=K \text{ or } k=K, \\ 
		2\sigma: \text{ otherwise}. 
	\end{cases}
\end{equation*} 
The initial parameters of the covariance matrix $\sigma I_K$ determine the policy's initial exploration rate. As training progresses, we gradually lower the entries in the covariance matrix to discourage exploration and encourage exploitation. For $H\in\N$ gradient steps and for an initial variance $\sigma_{\text{init}}>0$ and final variance $\sigma_{\text{final}}>0,$ we set the covariance matrix to $\sigma_i I_K,$ where     
\begin{equation} 
\sigma_i = (\sigma_{\text{final}}-\sigma_{\text{init}})\left(\frac{i-1}{H-1}\right) + \sigma_{\text{init}}, \quad \text{for } i = 1,2,\dots, H.
\label{covariance_scaling} 
\end{equation}
The choice for the initial and final variance and the bias term for the trade execution problem is discussed further below. For other applications, they should be chosen so that exploration is encouraged at the beginning and discouraged at the end of the training cycle. For example, in optimal portfolio allocation problems, one should set the bias such that the initial allocation is balanced to all assets in the portfolio, ensuring that the algorithm observes a complete set of market scenarios.

\begin{remark}
Rather than manually scaling the algorithm's action variances as in \eqref{covariance_scaling}, they can also be encoded and learned during training. We explore this option in Appendix \ref{section:std_learning}, where we model the covariance matrix as a learnable diagonal matrix. We find that manual scaling performs significantly better in one of our test cases, but similarly in all other test cases. Although learning a full (non-diagonal) covariance matrix is also possible, we expect this to complicate the learning process and do not pursue it here.
\end{remark}

\subsection{Empirical policy gradient and advantage function estimation} 

In practice, the expectation \eqref{policy_gradient_logistic_normal} is not available, but can be replaced with a sample-based estimate. This also requires an estimate of the advantage function \eqref{advantage}. We estimate it using an estimate of the value function $V_\vartheta^{\pi_\theta},$ defined in \eqref{eq:value_function},  where $\vartheta$ are the weights of a neural network. 
We make $H\in\N$ gradient steps with respect to the policy parameters $\theta$ and the parameters $\vartheta$ corresponding to the value function. At the beginning of the training cycle, the parameters are initialized as $\theta_1$ and $\vartheta_1.$ In each training epoch $i\in\{1,\dots,H\},$ we collect $\tau\in\N$ trajectories  
\begin{equation}
	\mathcal{T}=\{(s_{n,k}, a_{n,k}, r(s_{n,k}, a_{n,k})), \;  n\in\{0, \dots, N-1 \}, k\in\{1,\dots,\tau\} \},
	\label{batch_of_trajectories}
\end{equation}
where the actions $a_{n,k}$ are generated by sampling $x_{n,k}$ from the underlying normal distribution with density $\varphi_{\theta_i}(\cdot\mid s_{n,k}),$ which are then transformed with $h.$ For each state-action pair $(s_{n,k},a_{n,k}),$ the advantage function is estimated as
\begin{equation}
	\label{advantage_estimate} 	
	A_{\vartheta_i}^{\pi_{\theta_i}}(s_{n,k}, a_{n,k}) 
	= \sum_{l=n}^{N-1} r(s_{l,k}, a_{l,k}) - V^{\pi_{\theta_i}}_{\vartheta_i}(s_{n,k}).
\end{equation}
The parameters $\theta$ of the policy network are updated, from $\theta_i$ to $\theta_{i+1},$ by making a gradient step with learning rate $\eta>0$ using the loss function 
\begin{equation}
	\theta \mapsto 
- 
\frac{1}{\tau N}
\sum_{k=1}^\tau 	 
\sum_{n=0}^{N-1}
A_{\vartheta_i}^{\pi_{\theta_i}}(s_{n,k}, a_{n,k})
\log \varphi_{\theta}(x_{n,k} \mid s_{n,k} ). 		
\label{logistic_normal_loss} 
\end{equation}
The parameters $\vartheta$ of the value function neural network  are updated, from $\vartheta_i$ to $\vartheta_{i+1},$ by making a gradient step with learning rate $\eta>0$ using the loss function 
\begin{equation}
\vartheta \mapsto 
\frac{1}{\tau N}
\sum_{k=1}^\tau
\sum_{n=0}^{N-1} \left\lvert V^{\pi_{\theta_i}}_{\vartheta}(s_{n,k}) - \sum_{l=n}^{N-1} r(s_{l,k}, a_{l,k}) \right\rvert^2.
\label{value_loss} 
\end{equation}
Algorithm \ref{alg:actor_critic} summarizes the whole training process. 
\begin{algorithm}[tb]
	\caption{Actor-critic policy gradient method with logistic-normal}
	\begin{algorithmic}
		\STATE {\textbf{Initialize} policy $\pi_\theta,$ value function $V_\vartheta,$ number of iterations $H,$ number of trajectories $\tau,$ learning rate $\eta,$ bias $b$, initial variance $\sigma_{\text{init}},$ final variance $\sigma_{\text{final}}$:}
		\FOR{$i = 1, \dots, H$}
		\STATE \textbf{Collect} trajectories $\mathcal{T} = \left\{(s_{n,k}, a_{n,k}, r(s_{n,k}, a_{n,k})), n\in \{0,\dots,N \}, k \in \{1,\dots,\tau\} \right\},$ by sampling $x_{n,k}\sim\varphi_\theta,$  and transforming  $a_{n,k}=h(x_{n,k}).$  
		\STATE{\textbf{Update} policy $\pi_\theta$ with learning rate $\eta$ and loss function \eqref{logistic_normal_loss}}. 
		\STATE{\textbf{Update} value function $V^{\pi_\theta}_\vartheta$ with learning rate $\eta$ and loss function \eqref{value_loss}.}
		\STATE{\textbf{Scale} down values in covariance matrix $\Sigma$ according to \eqref{covariance_scaling}.}
		\ENDFOR
	\end{algorithmic}
	\label{alg:actor_critic}
\end{algorithm}

\begin{remark}
\citet{schulman2015high} describe a general method, which they call {\sl generalized advantage function estimation (GAE)}, for estimating advantage functions that interpolates between high-variance/low-bias estimates and low-variance/high-bias estimates. The interpolation is controlled with a parameter\footnote{We use the symbol $\gaelambda$ rather than $\lambda,$ as in their paper, to avoid confusion with the symbol for trading intensities below.} $\gaelambda\in[0,1].$ In \eqref{advantage_estimate}, we choose a high-variance/low-bias estimate, by setting $\gaelambda=1$ and reduce the variance of the estimate by using a large number of trajectories $\tau$ at each gradient step. The hyperparameters of the algorithm are discussed in \ref{sec:hyper_parameters}. For comparison, we conduct an additional experiment with a smaller number of trajectories per gradient step and different choices of $\gaelambda,$ to reduce variance, in Appendix \ref{sec:gae_experiments}. We find that choosing a smaller value for $\gaelambda$ does not lead to better results.
\end{remark}

\section{Market environment}
\label{SIM} 

We test our RL algorithm in a market environment populated by three different types of traders: noise traders who place and cancel orders randomly, tactical traders who react to order book imbalance, and a strategic trader who buys or liquidates a large position with a TWAP strategy, and therefore, trades in one direction at constant speed. We simulate the order book during the time interval $[-\Delta t, T].$ 

\subsection{Noise traders} 
Our first category of traders are noise traders who submit market, limit, and cancellation orders according to independent Poisson processes \citep{cont2010stochastic, abergel2013mathematical}. The intensities of market orders, limit orders, and cancellations are modeled in the following way.
\begin{itemize}
	\item Market buy and sell orders arrive with intensity $\lambda^M.$ 
	\item 
	For $k \in \{1, \dots, D\},$ limit buy and sell orders arrive $k$ ticks below the best ask price or $k$ ticks above the best bid price with intensity $\lambda^{L,k}.$ 
	\item If the spread is $j$ ticks, then for $ k \in \{j, \dots, D\},$ cancellations of limit buy orders arrive $k$ ticks below the best ask price with intensity $\lambda^{C,k} v^{b,{k-j+1}}$, and cancellations of limit sell orders arrive $k$ ticks above the best bid price with intensity with intensity $\lambda^{C,k} v^{a,{k-j+1}}.$   	
\end{itemize}

\subsection{Tactical traders} 
In contrast to the noise traders, the intensities of the tactical traders depend on volume imbalance. We define weighted volumes on the bid and ask sides by
\begin{equation*}
	V^b = \sum_{k=1}^{D} v^{b,k} e^{-c(k-1)} \quad \text{and} \quad V^a = \sum_{k=1}^{D} v^{a,k} e^{-c(k-1)}, 
\end{equation*}
where $c \in \R_+$ is a damping factor. The weighted volume imbalance is then defined by 
\begin{equation*}  
	I = \frac{V^b-V^a}{V^b+V^a}.
\end{equation*}
The damping factor $c$ effectively controls how sensitive the imbalance is to volumes posted at deeper levels in the book. Let $I^+$ and $I^-$ be the positive and negative parts of the imbalance. The intensities for different order types are defined as follows. 
\begin{itemize}
	\item For $d^M \in \R_+,$  market buy orders arrive with intensity $d^MI^+$ and market sell orders arrive with intensity $d^MI^-.$ 
	\item For $k\in\{1,2, \dots, D\},$ limit buy orders arrive $k$ ticks below the best ask price with intensity $d^{L,k}I^+,$  and limit sell orders arrive $k$ ticks above the best bid price with intensity $d^{L,k}I^-.$   
	\item If the spread is $j$ ticks, then for $k\in\{j,\dots,D\},$ cancellation of limit buy orders arrive $k$ ticks below the best ask price with intensity $d^{C,k}I^- v^{b,{k-j+1}},$  and cancellations  of limit sell orders arrive
    $k$ ticks above the best bid price with intensity $d^{C,k}I^+ v^{a,{k-j+1}}.$ 
\end{itemize}
For $k\in\{1,\dots, D\},$ market orders, limit orders and cancellations of the noise traders arrive with sizes $1+\traderstd^{M}_\text{noise}|Z|, 1+\traderstd^{L,k}_\text{noise}|Z|$ and $1+\traderstd^{C,k}_\text{noise}|Z|$ respectively, where $Z$ is a random variable with standard normal distribution for both noise and tactical traders. Similarly, orders of the tactical traders arrive  with sizes $1+\traderstd^{M}_\text{tactical}|Z|, 1+\traderstd^{L,k}_\text{tactical}|Z|$ and $1+\traderstd^{C,k}_\text{tactical}|Z|.$ Order sizes are always rounded to the closest integer. Both types of traders always cancel orders with the highest queue positions first. No more orders can be canceled than exist at any given price level.

\subsection{Strategic trader} The strategic trader buys or sells a position in the market using market and limit orders.
\newcommand{\ordersizestrategic}{\nu}
\begin{itemize}
	\item The direction, either buy or sell, is drawn randomly at the start time $-\Delta t$ with probability $1/2$ each. 
	\item 
	The trader sends a market order of size  $\ordersizestrategic^M \in \N$ at time steps $-\Delta t, -\Delta t +\Delta t^M, -\Delta t + 2\Delta t^M,  ...$  and stops trading at terminal time $T.$  
	\item 
The trader sends a limit order of size $\ordersizestrategic^L \in \N$ at the best ask price at time steps $-\Delta t, -\Delta t +\Delta t^L, -\Delta t + 2\Delta t^L,  ...$  and stops trading at terminal time $T.$ 
\end{itemize}

\section{Numerical results}
\label{sec:experiments} 
In this section, we test our RL algorithm based on the logistic-normal distribution (LN) in different simulated market environments and compare it with the two heuristic benchmarks from Section \ref{sec:benchmark_algorithms} and another RL algorithm based on the Dirichlet distribution (DR). Our source code is available at \url{https://github.com/moritzweiss/rlte}.
 Although our simulations are stylized, they capture key characteristics of real-world market behavior. Importantly, our simulations differ significantly from historical market data replay because in our simulations, trades generate market impact. 
First, large market orders have direct impact since they consume the best levels on the opposite side of the order book. Second, limit orders have indirect impact as the tactical traders react to order book imbalance. Those effects cannot be incorporated into a simulation that relies on historical data replay, because historical data cannot respond to the algorithm's trading decisions.

\subsection{Simulation setup}
\label{sec:simulation_setup}
We simulate three increasingly complex market environments. The first market contains only noise traders who behave purely randomly and do not react to other market participants. The second market includes both noise traders and tactical traders. In this setting, order flows react to order book imbalances, a behavior commonly observed in empirical studies of limit order book dynamics; see e.g.\ \cite{stoikov2018micro}. The third market adds a strategic trader to the previous participants. It simulates a market that exhibits upward or downward price drift due to the presence of a strategic trader. We will see that our RL algorithm learns to adapt its execution strategy by front-or back-loading orders.

As mentioned in Section \ref{sec:trade_execution_problem}, the execution algorithms trade at discrete time steps $t_n= n\Delta t $ for $n\in\{0,1,\dots,N-1\}.$ For our simulations, we choose $\Delta t = 15s$ for a total of $N=10$ steps, such that the total execution duration is $150$ seconds. This is a typical time horizon in high-frequency trading. There is substantial market activity within 150 seconds. On average, roughly 1000 order book updates (limit orders, market orders, or cancellations) occur in that time horizon, and a volume of 100 lots is traded; see Appendix \ref{sec:market_statistics}. 
In Appendix \ref{sec:long_time_horizon_experiment}, we also evaluate the execution algorithms on a longer time horizon of 300 seconds, achieving good results. Training over a much longer time horizon is also possible, but would require more computational resources. Our simulations are relatively expensive, as documented in Appendix \ref{sec:hyper_parameters}. Our reinforcement learning algorithm can be
viewed as a specialized order-placing algorithm that can be integrated into a long-term execution schedule, such as \citet{almgren2001optimal}.

In each market, the execution algorithms trade a small (20) and a large (60) number of lots. What can be considered a small or large number of lots depends on the average traded volume in the interval $[0,T].$ In our examples, the average traded volume over the time interval is roughly 100 lots in the first two markets, and slightly more in the last one. Precise statistics on traded volumes are documented in Appendix \ref{sec:market_statistics}. 
Hence, 20 lots correspond to roughly 20\% and 60 lots correspond to 60\% of the total traded volume.

In all three markets, we start the simulation at time $-\Delta t,$ where at initialization, we start the order book from its long-term average shape. Therefore, when the execution algorithms start trading at $t=0,$ the order book has moved from its initial state to a random state. The average shape of the order book can be estimated with a Monte Carlo simulation, which is described in more detail in Appendix \ref{subsec:stationary_shapes}. The specific parameter choices for each trading agent are reported in Appendix \ref{sec:parameters_market_sim}. 
We use the two heuristic algorithms of Section \ref{sec:benchmark_algorithms} as benchmarks and introduce an additional benchmark based on the Dirichlet distribution, described in Section \ref{sec:dirichlet_rl_implementation_details}. One could also compare against other trade execution benchmark algorithms, such as the volume-weighted-average price algorithm. However, our algorithm is not directly comparable with other reinforcement learning algorithms from the literature since they do not use the same state and action spaces.  

\begin{figure}[htbp]
	\centering
	\includegraphics[width=0.75\columnwidth]{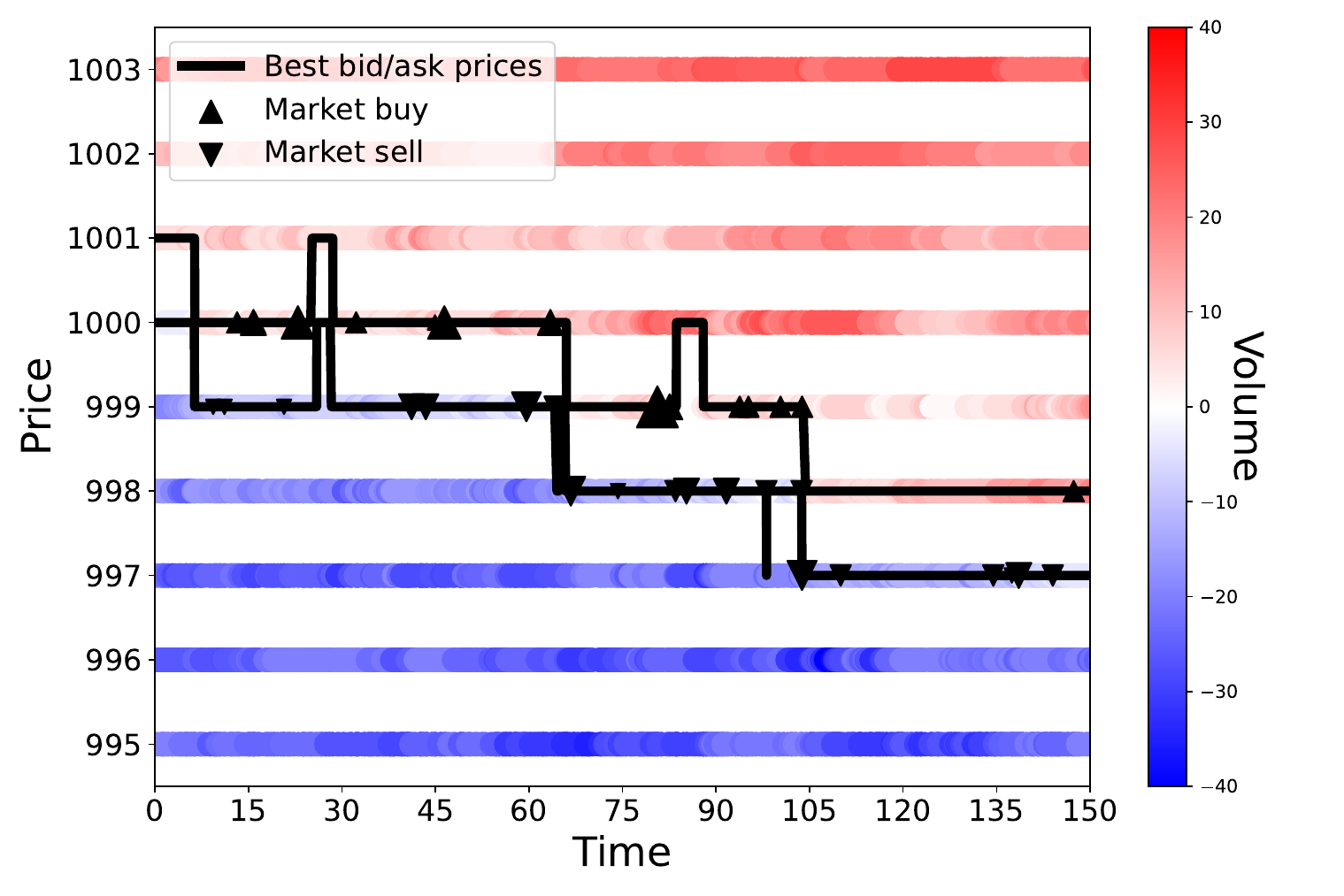}
	\caption{The black solid lines are the best bid and ask prices. The triangles indicate buy and sell market orders. The size of the triangles corresponds to the size of the market orders. The colors indicate volumes. Darker colors correspond to larger volumes. Buy limit orders have a blue color, and sell limit orders have a red color.} 
	\label{fig:price_path} 
\end{figure}

Figure \ref{fig:price_path} shows an evolution of the order book in the interval $[0,T]$ for the market with only noise traders. The bid and ask prices move from $p^b(0)=1000$ and $p^a(0)=1001$  three ticks down to $p^b(T)=997$ and  $p^a(T)=998.$ The composition of the order book changes due to the arrival of market orders, limit orders, and cancellations.

\subsection{Market impact study}
\begin{figure}[htbp]
	\centering
	\begin{subfigure}[t]{0.3\textwidth}
		\includegraphics[width=\textwidth]{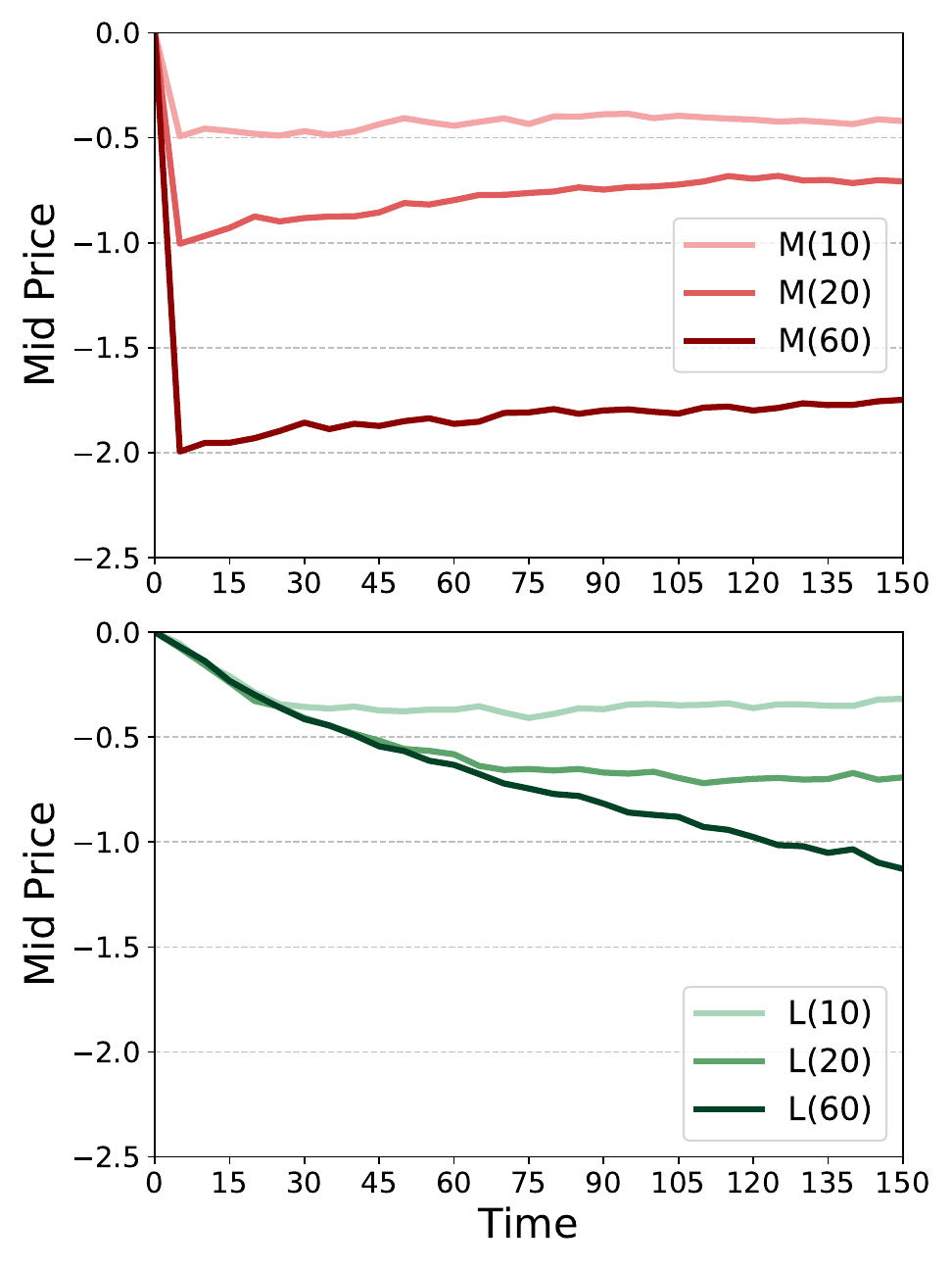}
		\caption{Noise Traders}
		\label{fig:impact_noise}
	\end{subfigure}\textbf{}
	\hfill
	\begin{subfigure}[t]{0.3\textwidth}
		\includegraphics[width=\textwidth]{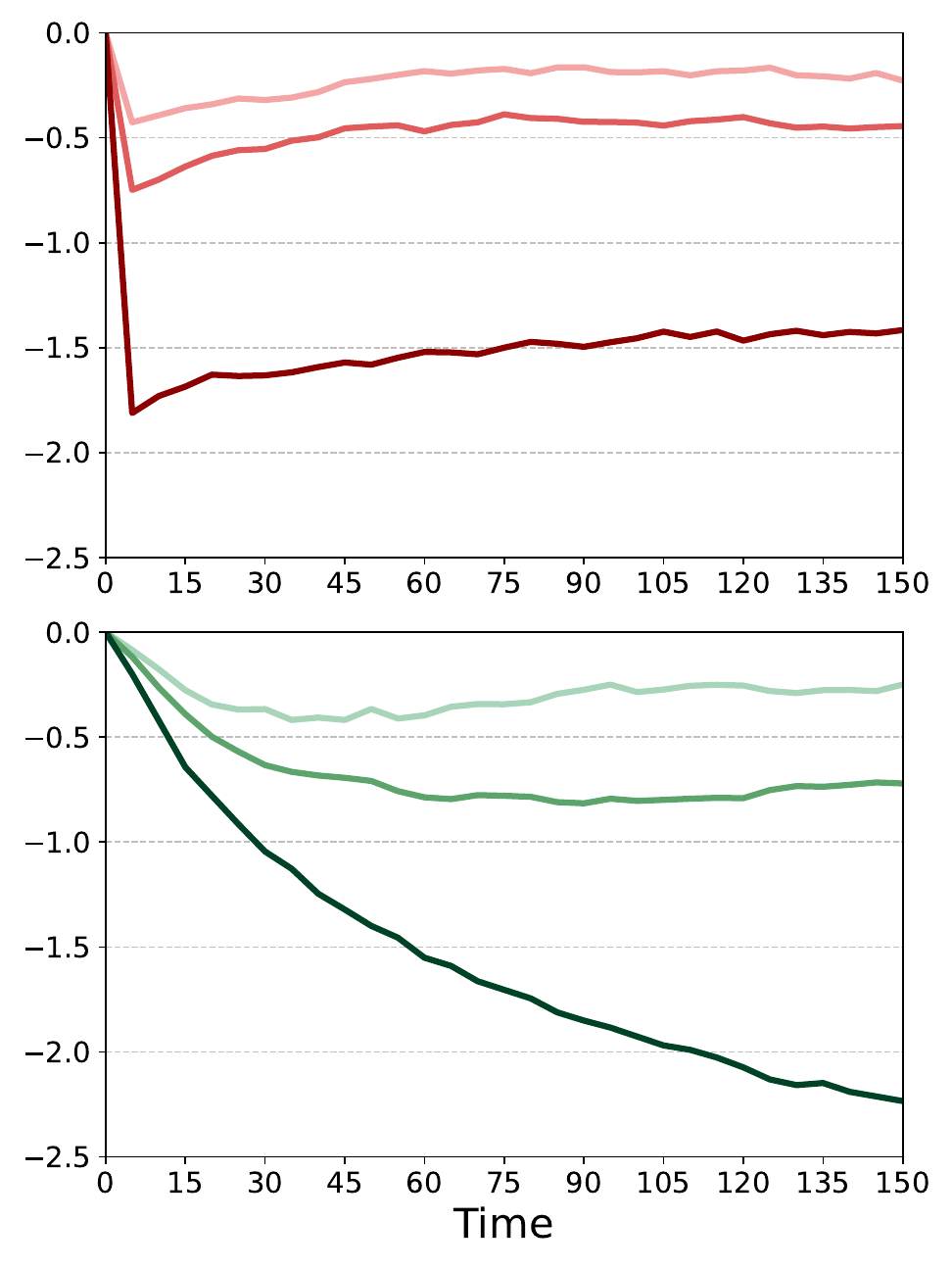}
		\caption{Noise \& Tactical Traders}
		\label{fig:impact_tactical}
	\end{subfigure}
	\hfill
	\begin{subfigure}[t]{0.3\textwidth}
		\includegraphics[width=\textwidth]{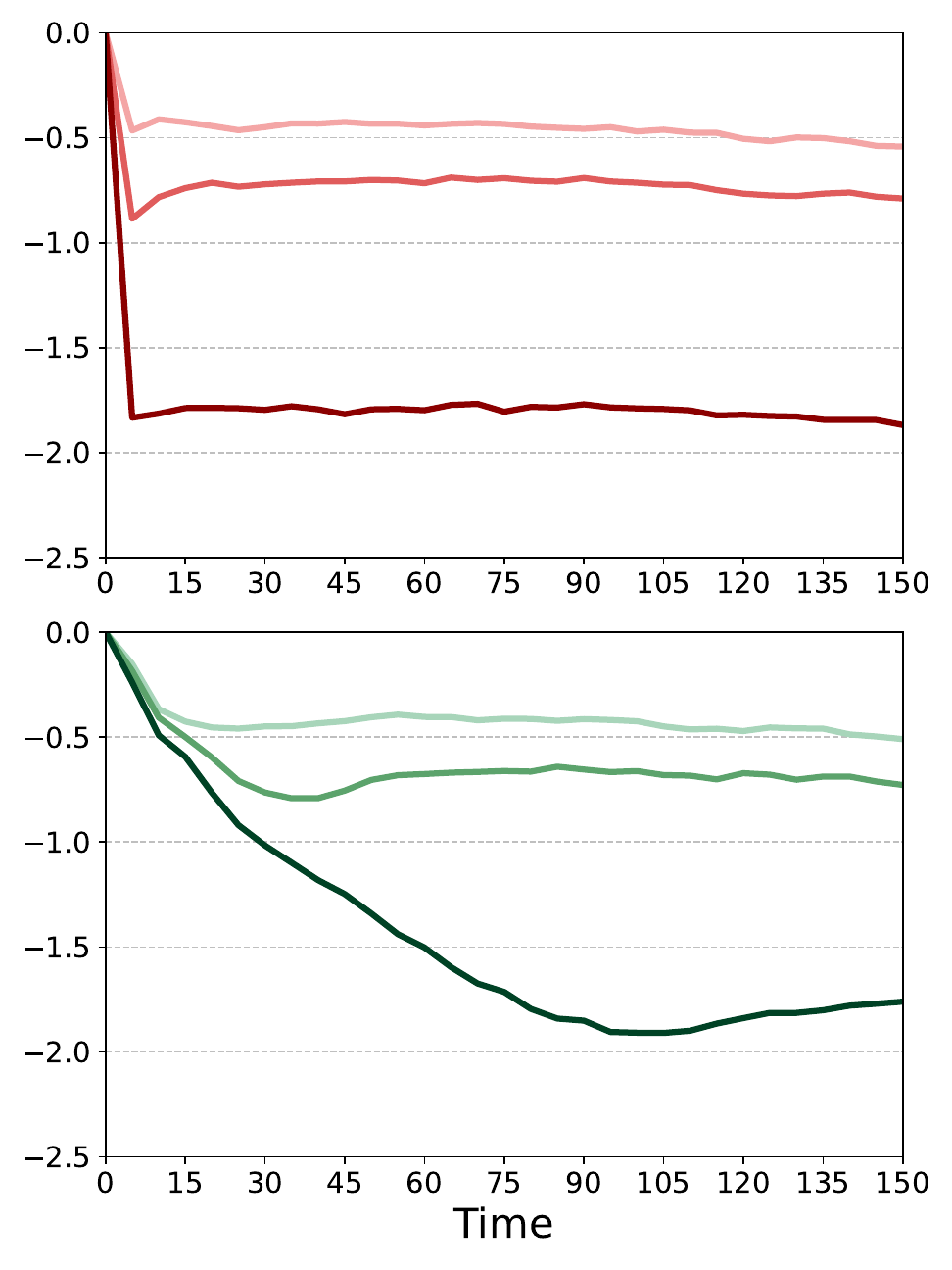}
	        \caption{Noise, Tactical, \& Strategic Traders}	
		\label{fig:impact_strategic}
	\end{subfigure}
	
	\caption{Mid price evolutions after a market or limit order of size 10, 20, or 60 lots is placed at time $t=0.$ The upper panel shows mid price evolutions after a market order was placed. The lower panel shows mid price evolutions after a limit order was placed. The legend indicates the order sizes in lots. Deeper color tones correspond to larger order sizes. 
	}
	\label{fig:impact_study}
\end{figure}

Before evaluating the performance of the reinforcement learning trading algorithms, we first conduct a market-impact study to quantify how market and limit order placement influences price dynamics. In each experiment, we initialize the market simulation at time $t = -\Delta t$. At time $t = 0$, we submit either a market sell order or a limit sell order, placed at the best ask price, with sizes of 10, 20, or 60 lots. Figure~\ref{fig:impact_study} reports the average evolution of the mid-price relative to its value at time zero, $p(t) - p(0),$ up to the terminal time $T = 150,$ for each market and for different position sizes. Averages are computed based on 10,000 market simulations.  

For market orders, we observe a sharp price drop immediately after the order is placed, followed by a slight recovery across all market environments. Naturally, larger market orders lead to a larger price impact in all cases. 
The mid-price recovery after the price drop is stronger in the market with noise and tactical traders than in the market with only noise traders. This is because the imbalance reaction from tactical traders has a mean-reverting effect, which pushes the price back up after a price level in the order book is removed. For the market consisting of noise, tactical, and strategic traders, the prices do not recover as much as in the other cases because when the strategic traders are selling, they amplify the downward move by continuing to push prices lower. For the limit order study, it is noticeable that the 60 lot limit order has much more market impact in the market with tactical traders. This is because the large limit order on the first level makes the order book very imbalanced. Since tactical traders react to order book imbalance, they send more market sell and limit sell orders, which drives the price down. As for market orders, we observe that larger limit orders have a larger impact in all cases. We conclude that both market and limit orders have substantial market impact in our simulations. This differentiates our work from works that rely on historical data simulations and usually ignore market impact.

\label{sec:market_impact}

\subsection{Reinforcement learning setup and evaluation}
\label{sec:rl_setup}
We train the LN algorithm with Algorithm \ref{alg:actor_critic}. As explained in Sections \ref{subsec:observation_space}--\ref{subsec:action_space}, the parameter $K$ controls how many order book levels the algorithm observes and how deep it may post orders into the order book. 
In our simulations, the price rarely moves more than five ticks during the execution period. Therefore, we choose $K=6$ for the dimension of the simplex $\SI^K.$ This means that the RL algorithm may place limit orders up to $K-1=5$ ticks above the best bid price 
and observes the volumes on the first five levels. To encode the policy $\pi_\theta$ and the value function $V_\vartheta,$ we use two separate feed-forward neural networks with two hidden layers, $\tanh$-activation functions in the hidden layers, and no activation functions in the last layers. The output layer of the policy network, specifying the mean $\mu_\theta,$ has $K$ nodes, and the last layer of the value function network has one node. We initialize the covariance matrix $\Sigma$ to the identity matrix and then scale it down following the schedule \eqref{covariance_scaling}. We initialize the bias of the output layer of the policy network to $b=(-1,-1,\dots,-1)\in\R^{K}$ and set the other weights of the output layer of the policy network close to zero. Then, we obtain from \eqref{mean} that
\begin{equation}
	\label{bias}
	\E \left[ \log\left( \frac{a^K}{a^k} \right)  \right] \approx 1 
	\quad 
	\text{and}
	\quad 
	\E \left[ \log\left( \frac{a^k}{a^K} \right)  \right] \approx -1, 
	\quad 
	k = 0,1,\dots,K-1, 
\end{equation}
showing that action $a_K$ is more likely than actions $a_0, \dots, a_{K-1}.$ Therefore, the algorithm is less likely to place orders at the beginning of the training cycle. This increases the likelihood of observing complete trajectories up to the terminal time $T$ early in the training process, thereby helping to prevent convergence to local minima. When we evaluate the algorithm after training is completed, we use a deterministic version of it. More precisely, we use the action $h(\mu_\theta),$ where $\mu_\theta$ is the mean of the underlying normal distribution, parametrized by the trained neural network weights, and $h$ is the logistic-normal transform; see Section~\ref{sec:policy_gradient_with_logistic_normal}. Further implementation details concerning the training of the LN algorithm are discussed in Appendix~\ref{sec:hyper_parameters}.

We compare the LN algorithm with heuristic benchmarks and an algorithm based on the Dirichlet distribution (DR). For evaluation, we use the deterministic version of the DR algorithm, selecting the mean of the algorithm’s action distribution as the action. More implementation details of this algorithm are discussed in Appendix \ref{sec:dirichlet_rl_implementation_details}.

\begin{figure}[htbp]
	\centering
	\begin{subfigure}[t]{0.3\textwidth}
		\includegraphics[width=\textwidth]{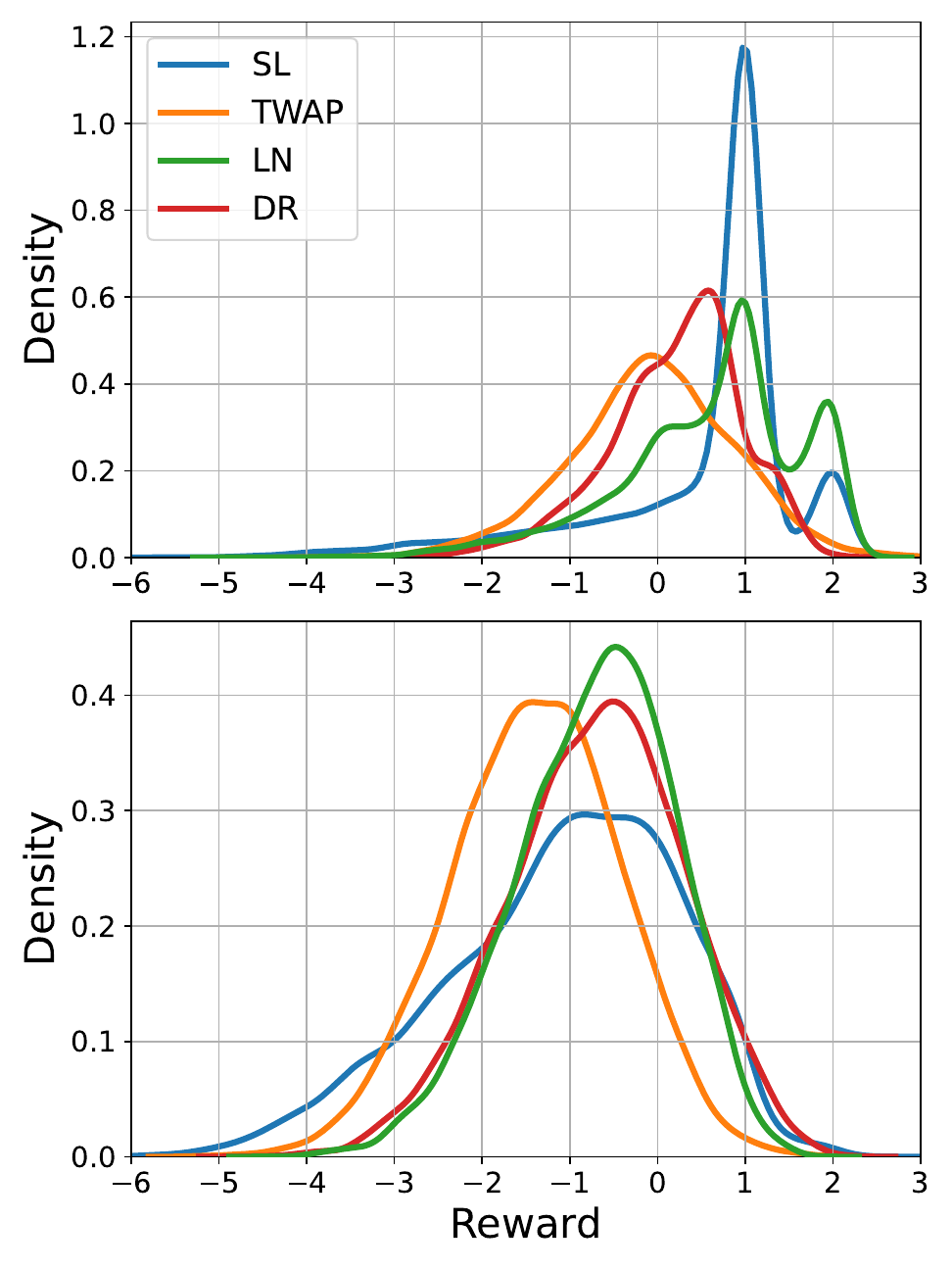}
		\caption{Noise Traders}
		\label{fig:noise_results}
	\end{subfigure}
	\hfill
	\begin{subfigure}[t]{0.3\textwidth}
		\includegraphics[width=\textwidth]{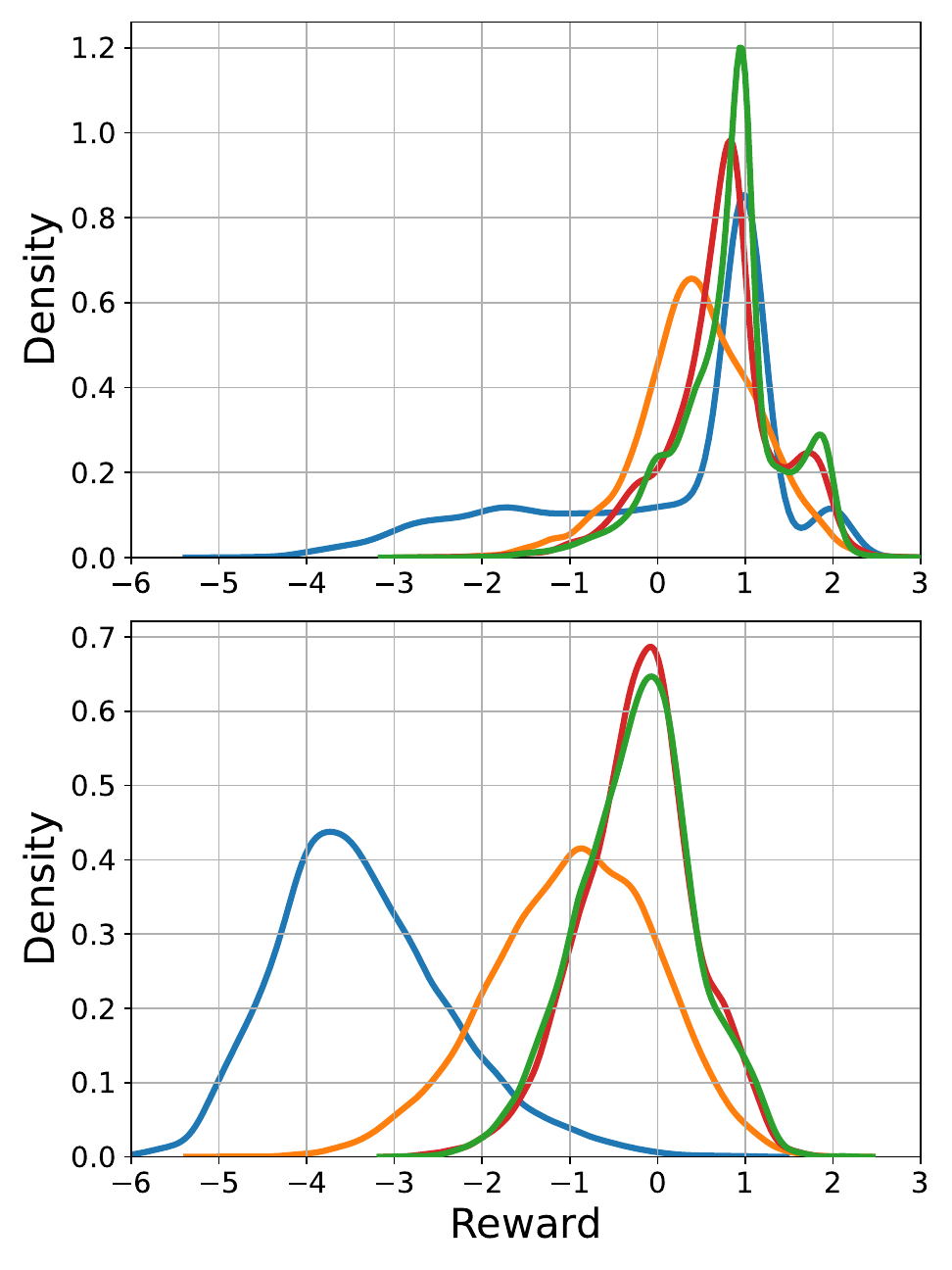}
		\caption{Noise \& Tactical Traders}
		\label{fig:tactical_results}
	\end{subfigure}
	\hfill 
	\begin{subfigure}[t]{0.3\textwidth}
	\includegraphics[width=\textwidth]{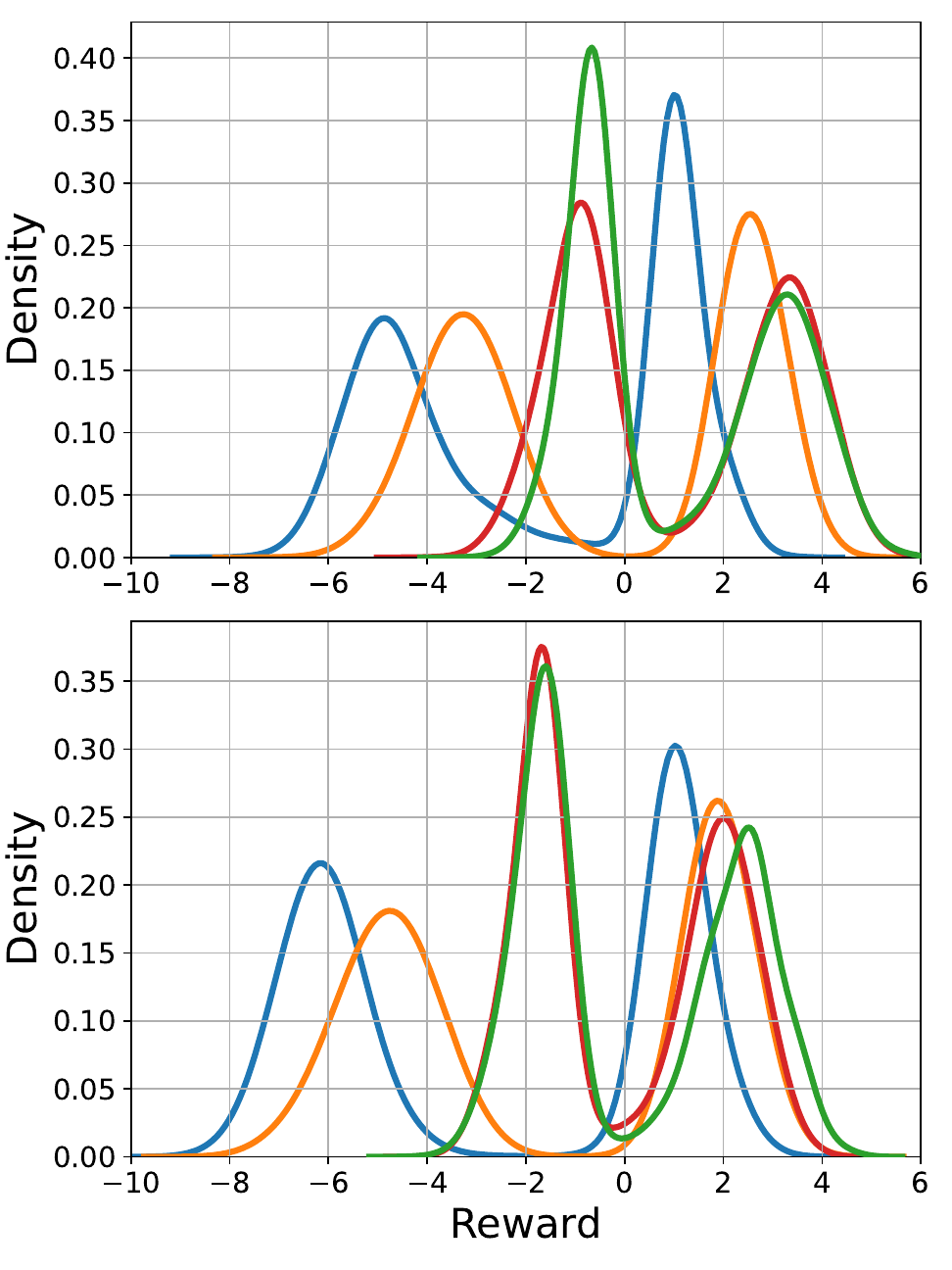}
	\caption{Noise, Tactical \& Strategic Traders}
	\label{fig:strategic_results}			
	\end{subfigure}	
	\caption{The different panels show reward distributions of the submit and leave algorithm (SL), the time-weighted average price algorithm (TWAP), the logistic-normal algorithm (LN) and the Dirichlet algorithm (DR) for three simulated markets. The first row corresponds to 20 lots, and the second to 60 lots.}
	\label{fig:densities_for_all_markets}
\end{figure}

In the following sections, we turn to a detailed discussion of the results. The test results are summarized in Table \ref{table:results}, which presents the expected value and standard deviation of the sum of normalized rewards defined in \eqref{total_normalized_rewards} across all market environments, position sizes, and execution algorithms.  For example, the reward of $-1.09$ for the SL algorithm at 60 lots in the market with noise traders means that on average, the algorithm sold each of the 60 lots at $-1.09$ ticks less than the initial bid price. Figure \ref{fig:densities_for_all_markets} shows histograms of the sum of normalized rewards. Expected values, standard deviations, and reward histograms are calculated from 10,000 market simulations. The random seed for the market environments used to evaluate the algorithms differs from that used during training, thereby avoiding overfitting to the training simulations. 

\subsection{Logistic-normal vs Dirichlet distribution}
\label{sec:logistic_vs_dirichlet_discussion}

\begin{figure}[htbp]
	\centering
	\begin{subfigure}[t]{0.3\textwidth}
		\includegraphics[width=\textwidth]{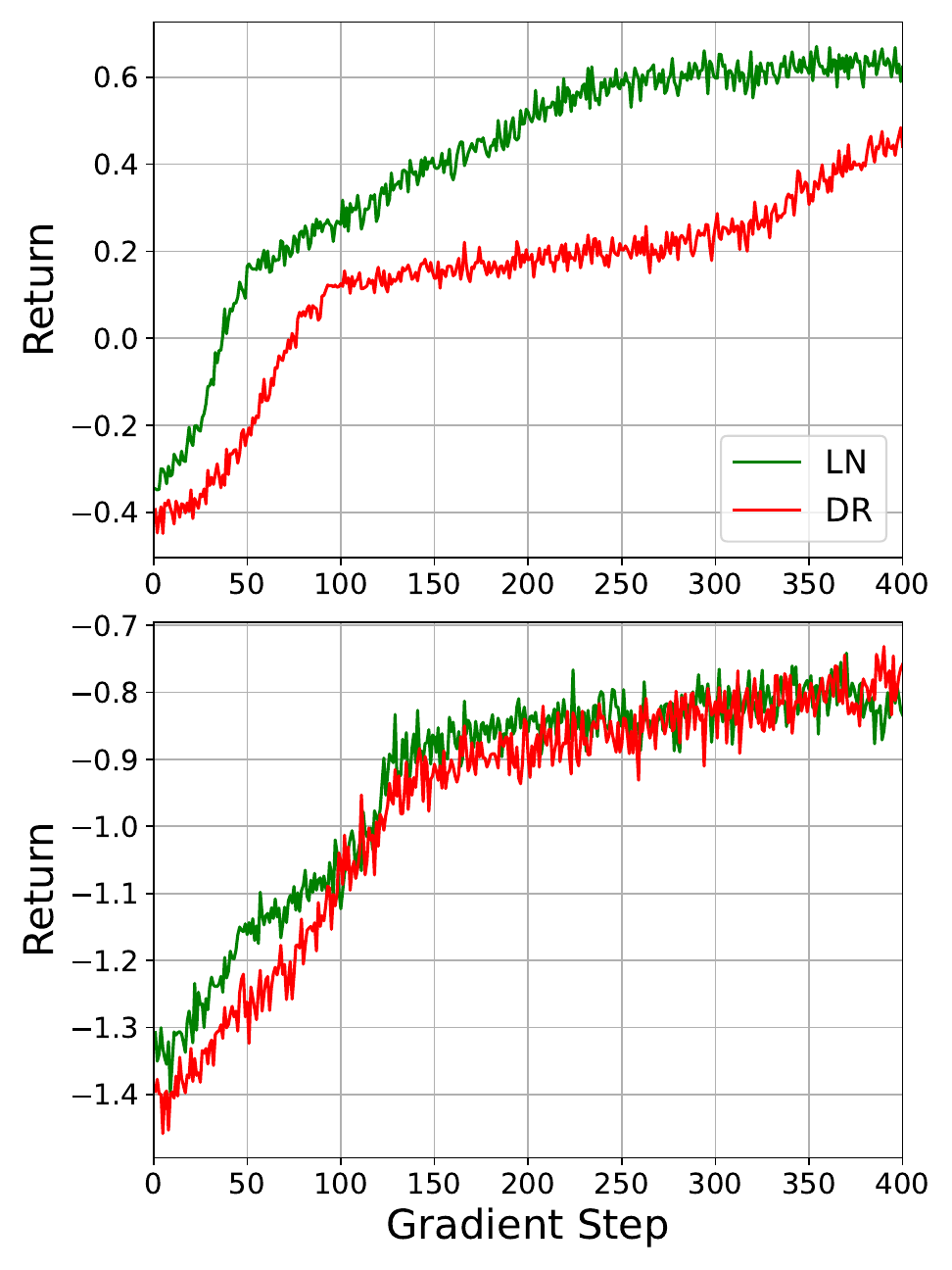}
		\caption{Noise Traders}
		\label{fig:noise_convergence}
	\end{subfigure}
	\hfill
	\begin{subfigure}[t]{0.3\textwidth}
		\includegraphics[width=\textwidth]{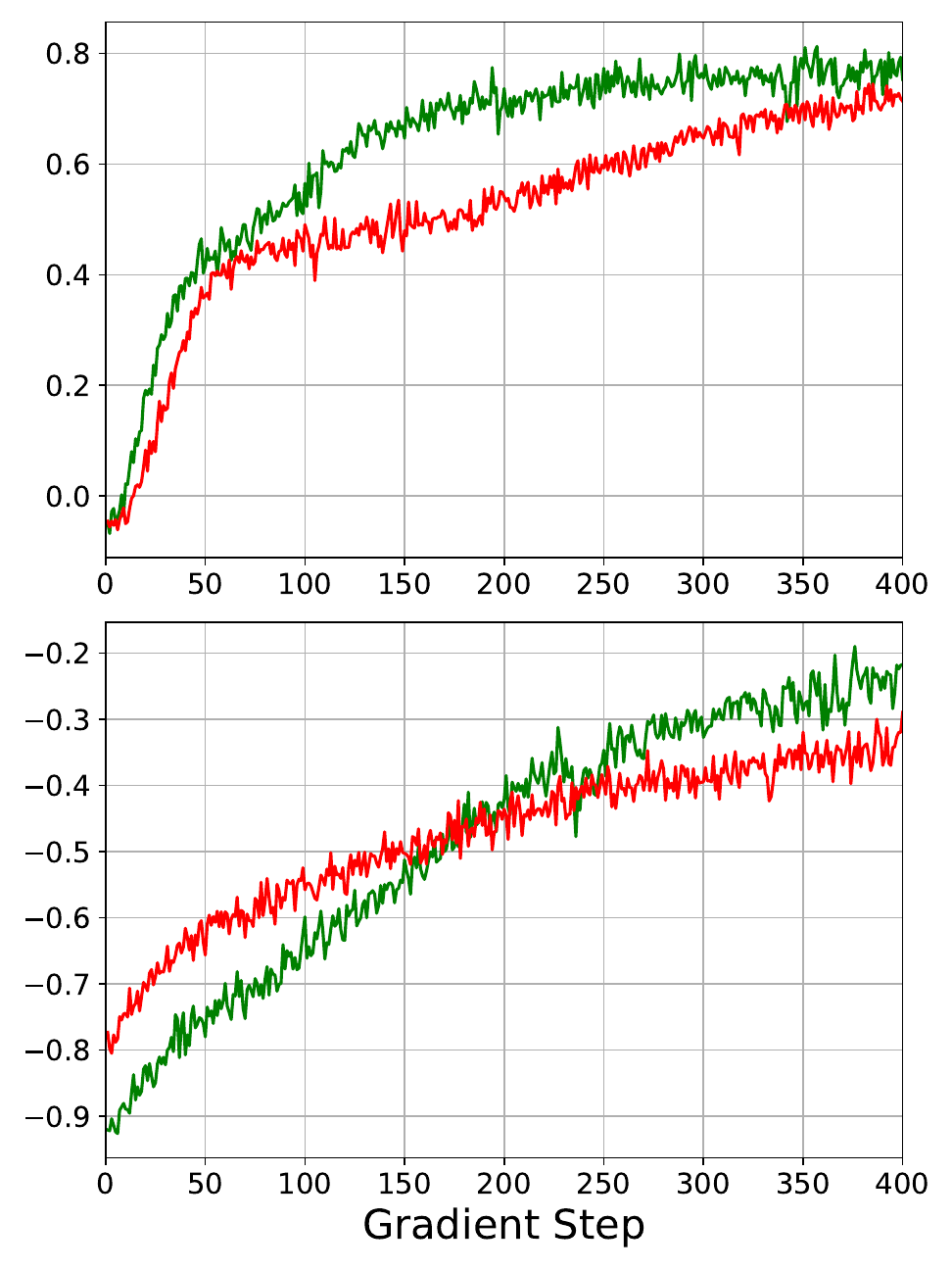}
		\caption{Noise \& Tactical Traders}
		\label{fig:tactical_convergence}
	\end{subfigure}
	\hfill
	\begin{subfigure}[t]{0.3\textwidth}		\includegraphics[width=\textwidth]{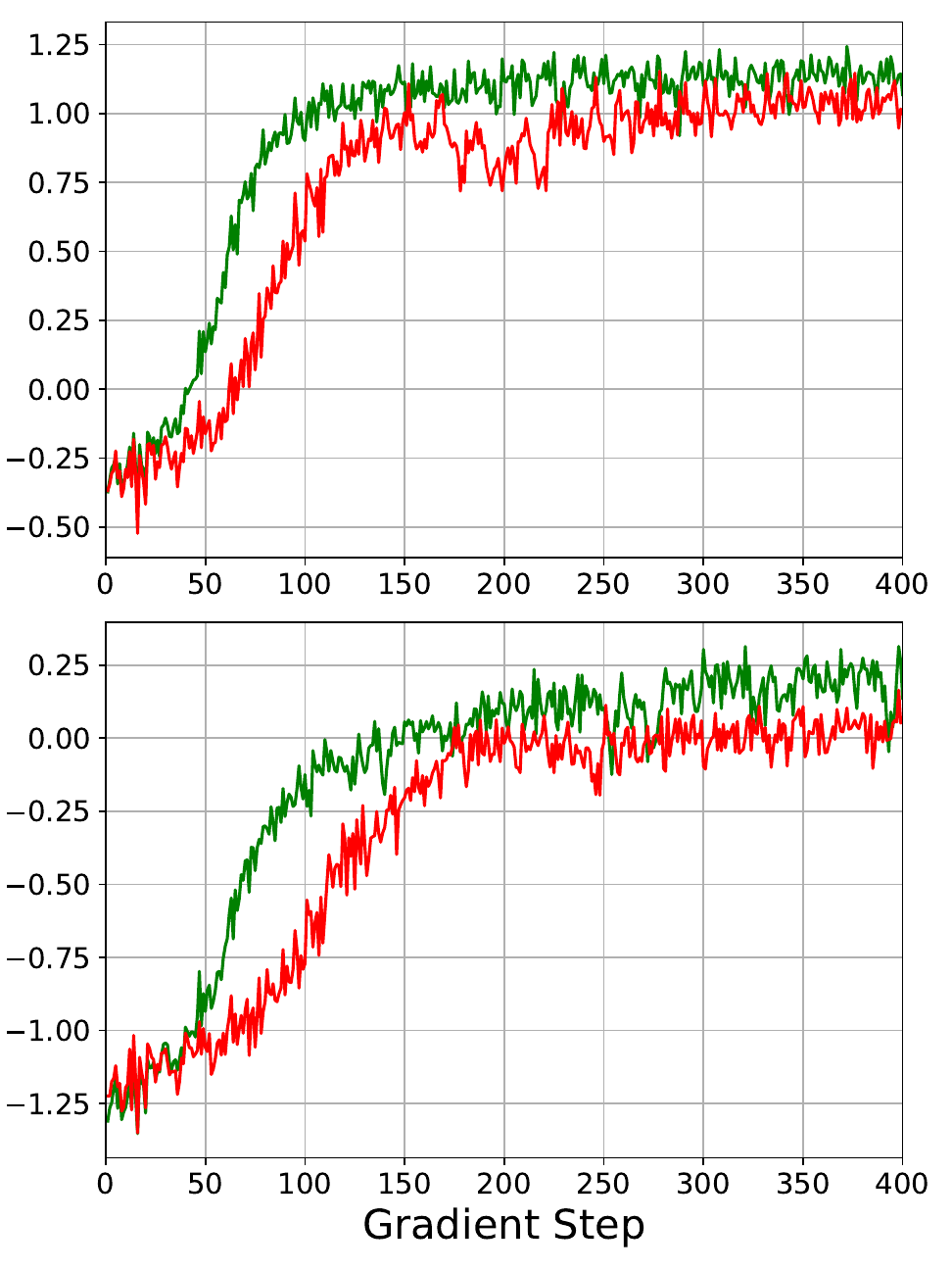}
	        \caption{Noise, Tactical, \& Strategic Traders}	
		\label{fig:strategic_convergence}
	\end{subfigure}
	
	\caption{
		Average episode return per batch during training for different environments, initial position sizes, and algorithms. The x-axis shows the number of gradient steps.
		The top row shows the average returns for 20 lots. The bottom row shows the average returns for 60 lots. The green line corresponds to the LN algorithm, and the red line corresponds to the DR algorithm. 
	}
	\label{fig:return_convergence}
\end{figure}

We start by comparing the LN and DR algorithms. We observe that across all three market simulations, the LN algorithm outperforms the DR algorithm, except in the market with noise traders and in the market with noise and tactical traders when 60 lots are traded, as shown in Table \ref{table:results}. However, when the LN algorithm performs worse than the DR algorithm, the difference is only marginal. Figure \ref{fig:return_convergence} shows the average rewards per batch over the full training cycle. The upper panel shows the rewards for 20 lots, and the lower panel shows the rewards for 60 lots. We observe that the rewards for the LN algorithm converge better than for the DR algorithm, except when it sells 60 lots in the market that only contains noise traders. 
Overall, this suggests that the LN algorithm is a superior alternative to the DR algorithm for the trade execution problem. 

\newcommand{\SL}{\text{SL}}
\newcommand{\TWAP}{\text{TWAP}}
\newcommand{\RL}{\text{RL}}

\begin{table}[htbp]
	\caption{The table shows the expected rewards and their standard deviations for the submit and leave algorithm (SL), the time-weighted average price algorithm (TWAP), the logistic-normal algorithm (LN), and the Dirichlet algorithm (DR) in three simulated markets and for initial positions of 20 and 60 lots. The best expected value per row is highlighted in bold. 
	}
\label{table:results}\begin{center}    \begin{scriptsize}        \begin{sc}\begin{tabular}{lccccccccc}
\toprule
Market & \#Lots & $\mathbb{E}[\text{SL}]$ & $\sigma[\text{SL}]$ & $\mathbb{E}[\text{TWAP}]$ & $\sigma[\text{TWAP}]$ & $\mathbb{E}[\text{DR}]$ & $\sigma[\text{DR}]$ & $\mathbb{E}[\text{LN}]$ & $\sigma[\text{LN}]$ \\
\midrule
Noise & 20 & 0.52 & 1.19 & -0.06 & 0.94 & 0.21 & 0.76 & \textbf{0.61} & 1.03 \\
 & 60 & -1.09 & 1.34 & -1.40 & 0.98 & \textbf{-0.71} & 1.0 & -0.72 & 0.90 \\
Noise \& Tactical & 20 & 0.10 & 1.43 & 0.48 & 0.68 & 0.73 & 0.65 & \textbf{0.81} & 0.64 \\
 & 60 & -3.36 & 0.99 & -0.96 & 0.95 & \textbf{-0.23} & 0.65 & -0.25 & 0.67 \\
Noise \& Tactical & 20 & -1.64 & 2.95 & -0.36 & 3.03 & 1.06 & 2.24 & \textbf{1.13} & 2.08 \\
\& Strategic & 60 & -2.51 & 3.67 & -1.45 & 3.46 & 0.03 & 1.97 & \textbf{0.23} & 2.15 \\
\bottomrule
\end{tabular}
        \end{sc}    
        \end{scriptsize}\end{center}\end{table}

\subsection{Market with noise traders} 
\label{sec:result_market_with_noise}
We note that all algorithms across all markets exhibit high standard deviation. This is due to the highly stochastic environment, combined with the fact that minimizing the standard deviation is not part of the training criterion. Sending a market order at the beginning of the execution period would result in a strategy with almost zero standard deviation; however, the execution price would be worse than the initial best bid and, therefore, worse than most of our execution strategies. The first two rows of Table \ref{table:results} and the reward histograms in Figure~\ref{fig:noise_results} demonstrate that the LN algorithm outperforms the heuristic benchmark strategies for 20 and 60 lots in the market with noise traders. 
It is better than the DR algorithm for 20 lots and slightly worse for 60 lots. The rewards per batch in Figure~\ref{fig:noise_convergence} show much better convergence for the LN algorithm compared to the DR algorithm when it trades 20 lots, and similar convergence when it trades 60 lots. Interestingly, the reward histograms of the LN and DR algorithms differ when the initial position is 20 lots, with the LN algorithm's histogram exhibiting more pronounced peaks at one and two ticks above the best bid price. This indicates that the algorithms employ different order placement strategies.  For 60 lots, the LN algorithm's histogram exhibits less skew to the left compared with the heuristic benchmark strategies, suggesting that it crosses the spread when the price moves down. The TWAP algorithm performs the worst for both position sizes. Since the market does not react to volume imbalance, placing the entire position passively at the start works relatively well. On the other hand, spreading the order over the entire execution interval leads to a worse performance for two reasons. First, due to the shape of the order book (Figure \ref{fig:average_shape} below), limit orders at the best ask are less likely to be filled when the market is moving up, and more likely to be filled when the market is moving down. Second, it is more likely that the algorithm does not sell its inventory by the end of the execution period and is forced to cross the spread at terminal time.

\subsection{Market with noise \& tactical traders}
The third and fourth rows of Table \ref{table:results} and Figure \ref{fig:tactical_results} show the results for the market with noise and tactical traders. Both demonstrate the LN algorithm's superior performance relative to the heuristic benchmark algorithms for small and large position sizes.
The LN algorithm performs better than the DR algorithm for an initial position of 20 lots and slightly worse for 60 lots. As before, for 20 lots, we observe bumps in the histogram one and two ticks above the best bid price for the LN algorithm. The presence of tactical traders reacting to order book imbalances introduces indirect market impact. Consequently, posting the entire position passively at the beginning, as the SL algorithm does, triggers tactical traders to place sell orders, leading to the worst performance for both initial positions. Spreading the order out over the entire execution interval, as done by the TWAP algorithm, works better. Still, the LN algorithm performs better, suggesting intelligent order placement strategies that don't provoke a reaction from tactical traders. 

\subsection{Market with noise \& tactical \& strategic traders}
The fifth and sixth rows of Table \ref{table:results} and Figure \ref{fig:strategic_results} show results for the market consisting of noise, tactical, and strategic traders and demonstrate a strong performance of the LN algorithm. The presence of the strategic trader introduces an upward or downward drift in the market. The LN algorithm detects this drift using the historical market, limit, and cancellation order flow features and the mid-price drift feature defined in Section \ref{subsec:observation_space}. The LN algorithm outperforms the DR algorithm, but the DR algorithm's returns have a lower standard deviation when the initial position is 60 lots. The LN algorithm has less skew to the left than the DR algorithm for an initial position of 20 lots, and more skew to the right for an initial position of 60 lots. The relative shift of the LN algorithm's histogram compared to those of the heuristic benchmark algorithms suggests that it trades more aggressively when the strategic trader is selling and more patiently when the strategic trader is buying.

\subsection{Analyzing the algorithms' trading behavior}

In this section, we run two experiments that illustrate the trading behavior of the LN and DR algorithms. Averages and standard deviations are computed over 10,000 simulations, after the algorithms have been trained. In addition to these experiments, we also conduct an ablation study in Appendix~\ref{sec:feature_importance} to assess feature importance. 

\begin{figure}[htbp]
	\centering
	\begin{subfigure}[t]{0.3\textwidth}
		\includegraphics[width=\textwidth]{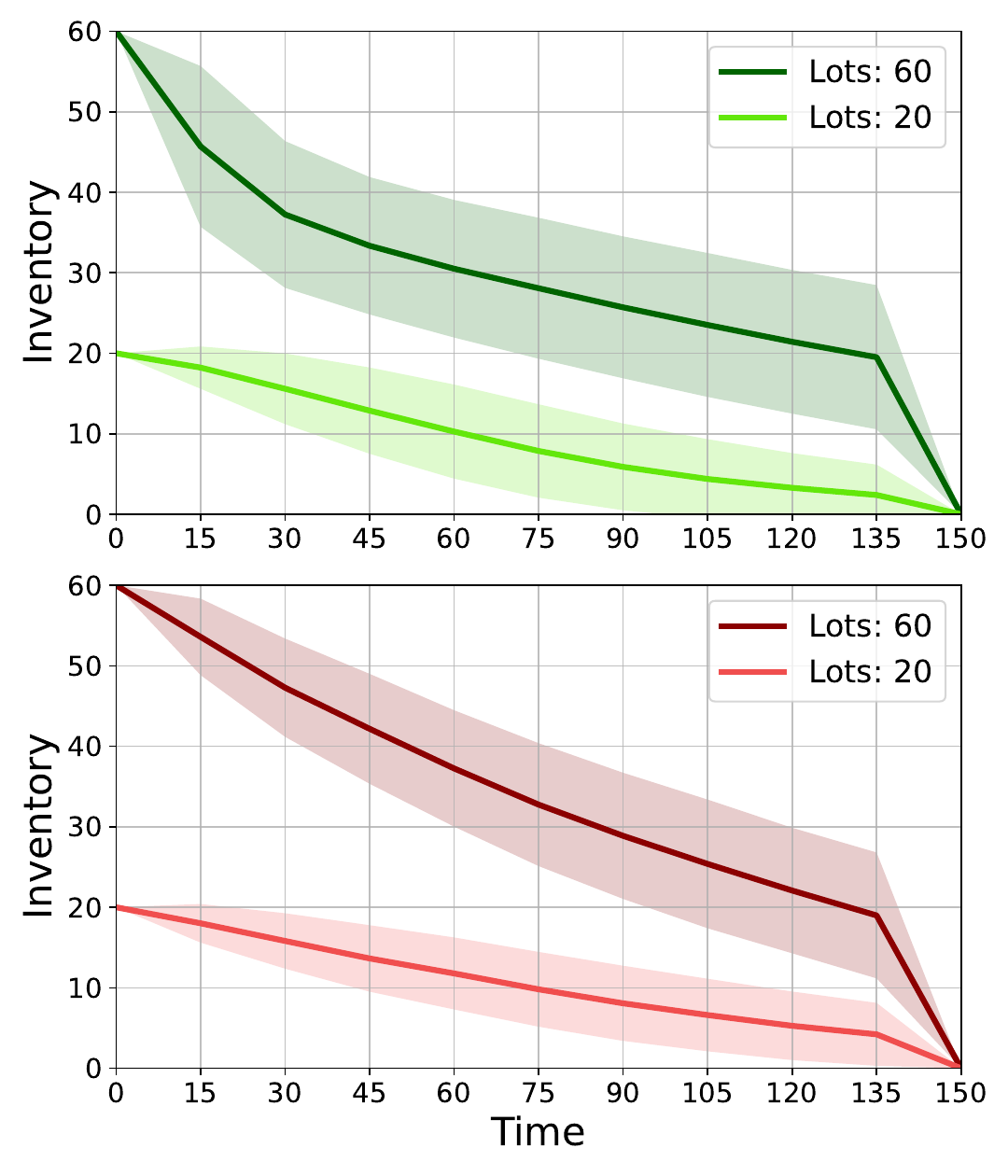}
		\caption{Noise Traders}
		\label{fig:vcurces_noise}
	\end{subfigure}\textbf{}
	\hfill
	\begin{subfigure}[t]{0.3\textwidth}
		\includegraphics[width=\textwidth]{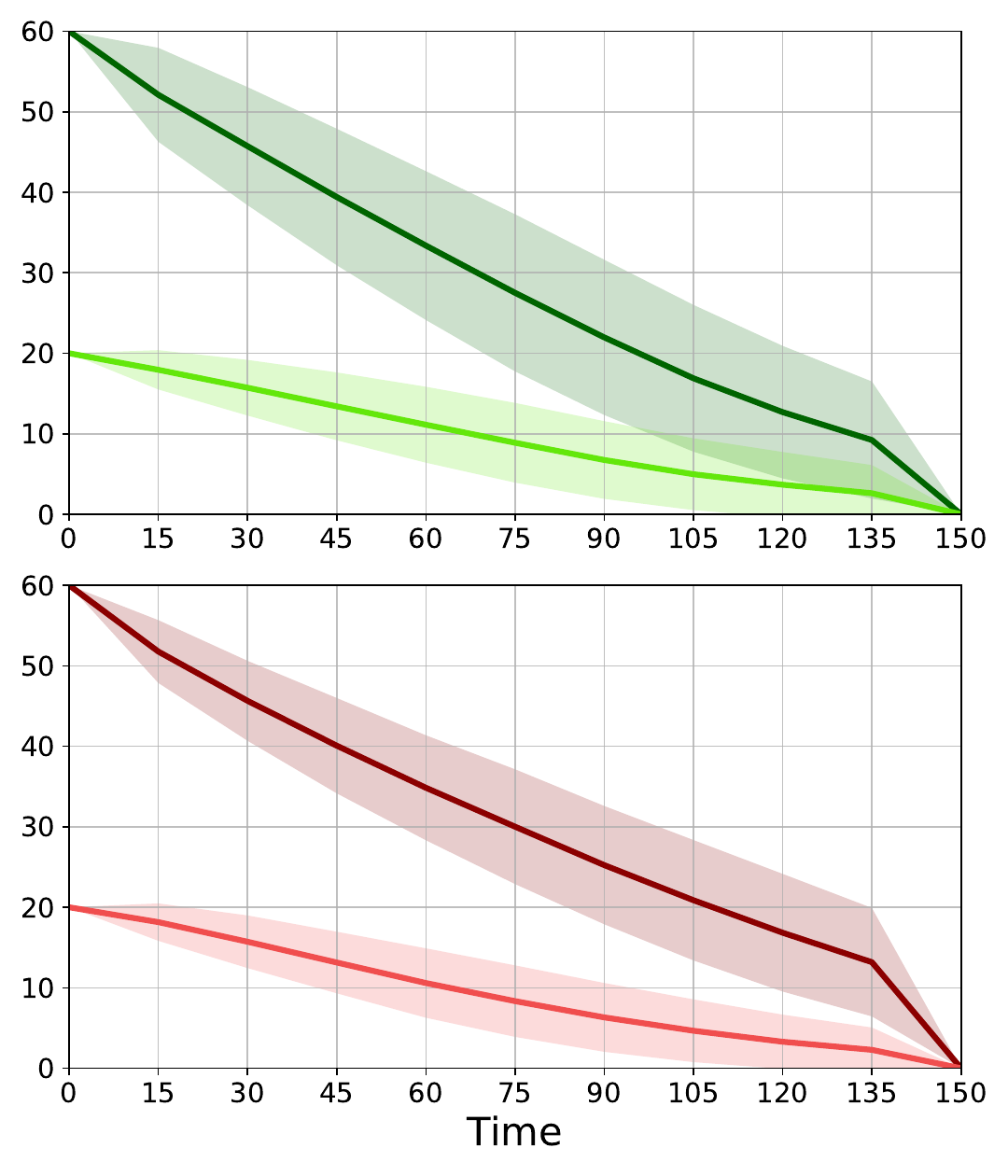}
		\caption{Noise \& Tactical Traders}
		\label{fig:vcurves_vlow}
	\end{subfigure}
	\hfill
	\begin{subfigure}[t]{0.3\textwidth}
		\includegraphics[width=\textwidth]{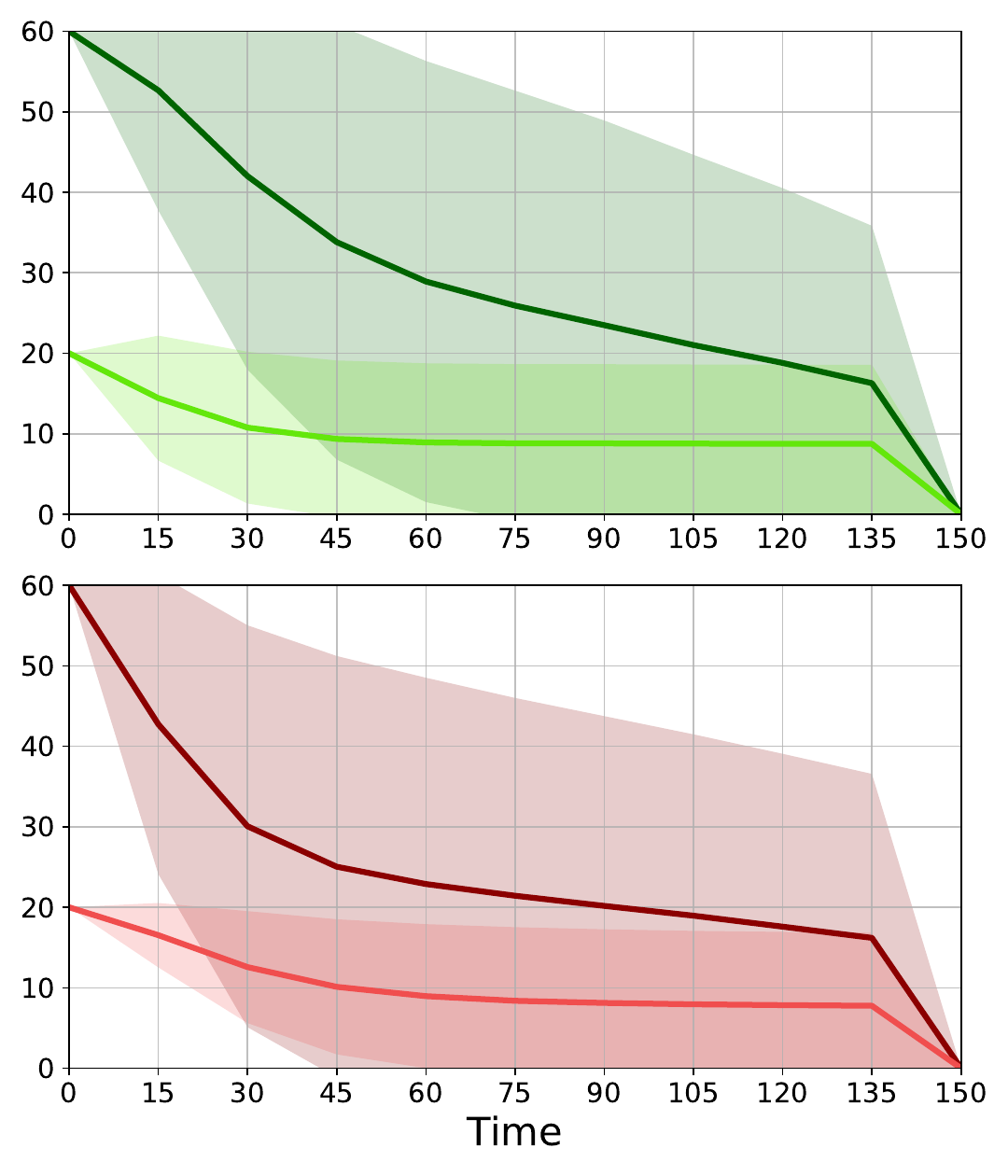}
	        \caption{Noise, Tactical, \& Strategic Traders}	
		\label{fig:vcurevs_strategic}
	\end{subfigure}
	
	\caption{Average inventory curves are displayed in bold, and the shaded areas around them indicate standard deviations.
    The green curves correspond to the LN algorithm, whereas the red curves correspond to the DR algorithm.  Deeper color tones correspond to larger initial positions. 
	}
	\label{fig:volume}
\end{figure}

The first experiment concerns the algorithms' inventories. Figure~\ref{fig:volume} shows the average inventory curves for the LN and DR algorithms in bold. The inventory is the number of lots $M(t)$ that the algorithm still has to sell at time $t,$ defined in Section~\ref{sec:trade_execution_problem}.
The green curves correspond to the LN algorithm, whereas the red curves correspond to the DR algorithm. The shaded areas indicate standard deviations. 
We observe that the LN algorithm trades 60 lots more patiently in the market with both noise and tactical traders than in the market with noise traders only. In contrast, the DR algorithm exhibits roughly the same trading speed in both markets. In the market with noise, tactical, and strategic traders, we observe significantly larger inventory standard deviations for both algorithms. This is because the algorithms liquidate faster or slower depending on whether the strategic trader is buying or selling. For the 60-lot case in the market with noise, tactical, and strategic traders, we observe that the LN algorithm holds back more inventory at the beginning of the execution period than the DR algorithm, indicating that the algorithms have different order placement strategies.

\begin{figure}[htbp]
	\centering
	\begin{subfigure}[t]{0.3\textwidth}
		\includegraphics[width=\textwidth]{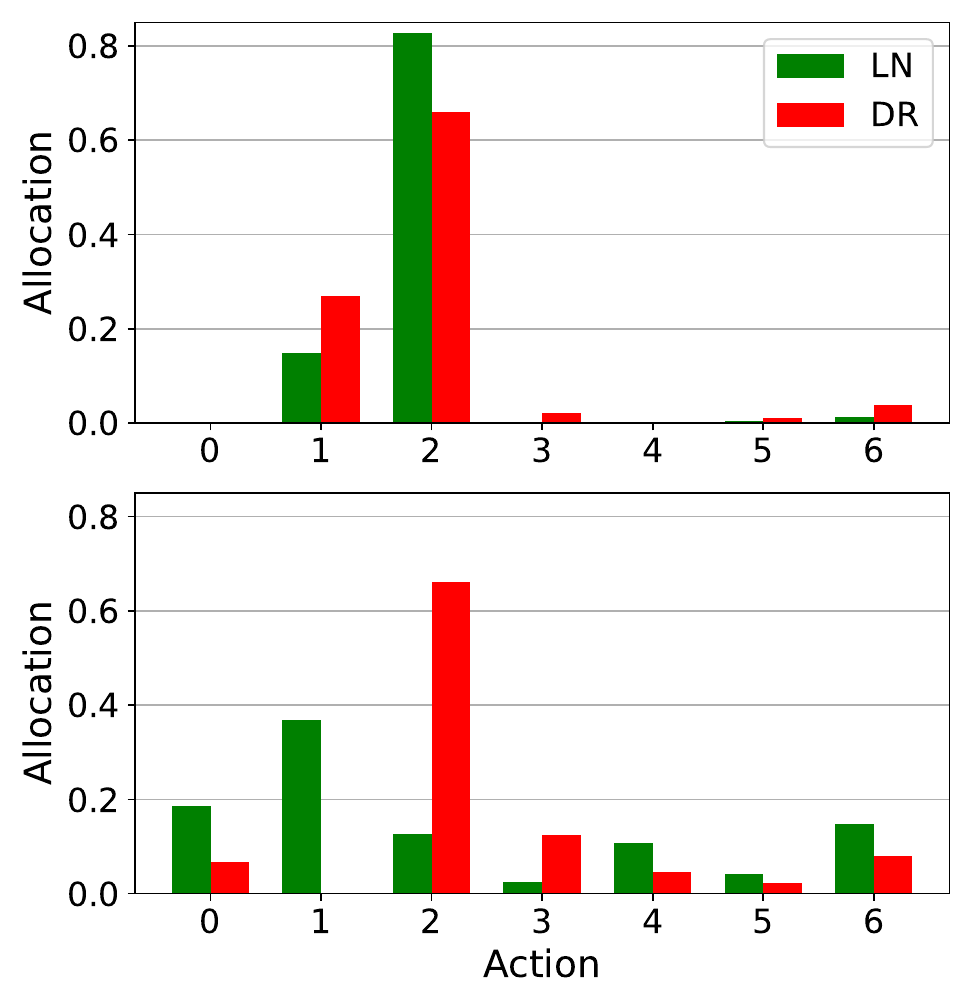}
		\caption{Noise Traders}
		\label{fig:actions_noise}
	\end{subfigure}\textbf{}
	\hfill
	\begin{subfigure}[t]{0.3\textwidth}
		\includegraphics[width=\textwidth]{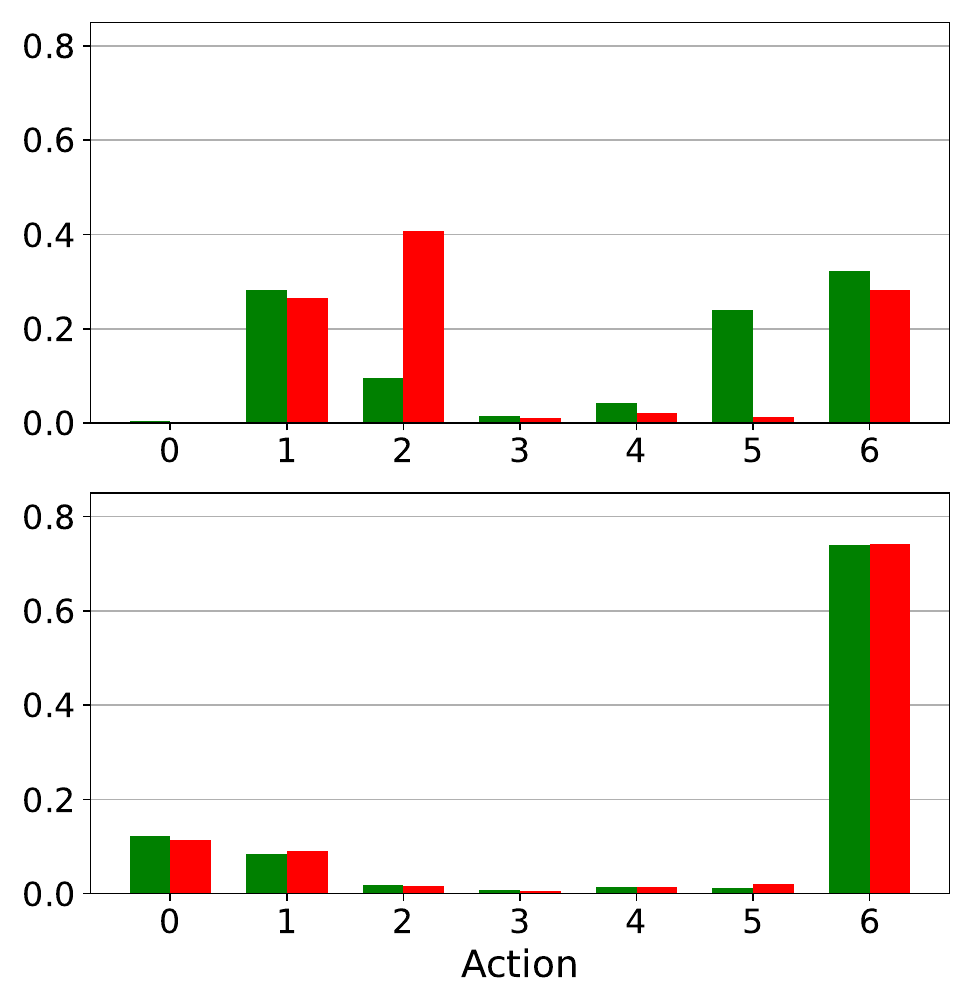}
		\caption{Noise \& Tactical Traders}
		\label{fig:actions_flow}
	\end{subfigure}
	\hfill
	\begin{subfigure}[t]{0.3\textwidth}
		\includegraphics[width=\textwidth]{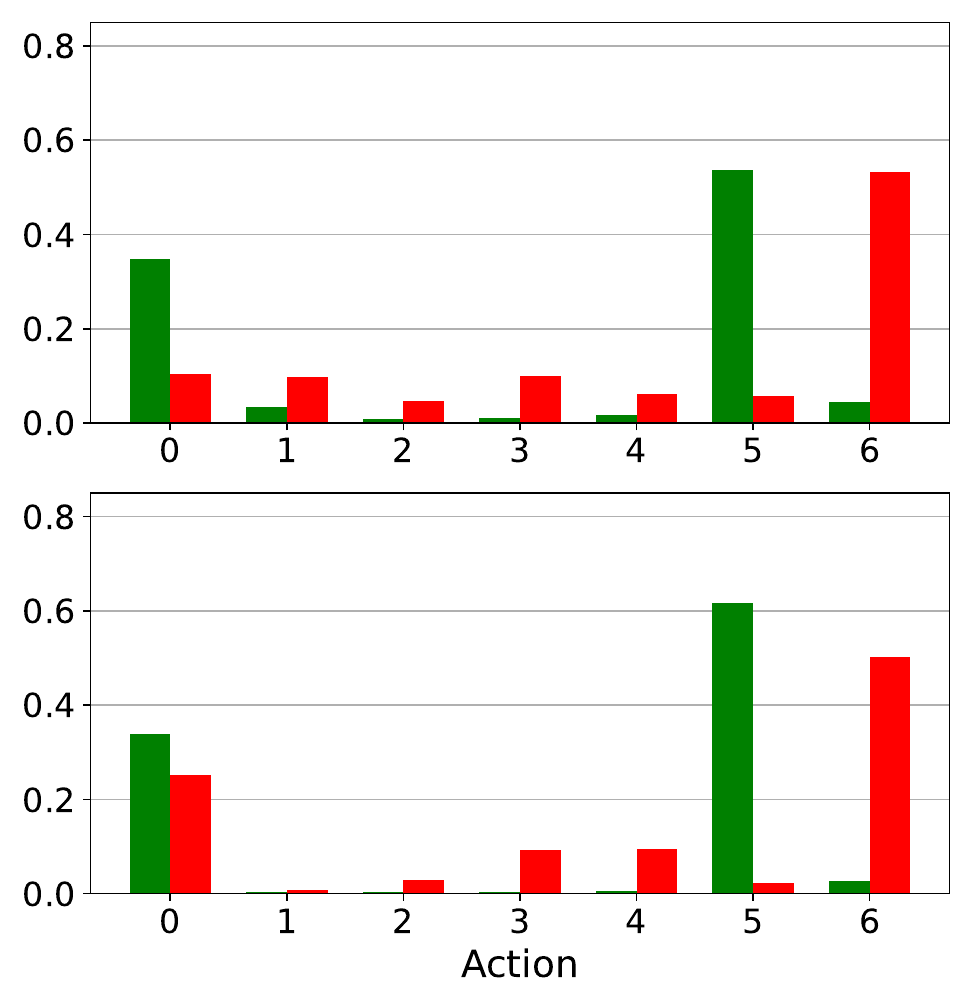}
	        \caption{Noise, Tactical, \& Strategic Traders}	
		\label{fig:actions_strategic}
	\end{subfigure} 
	\caption{Average actions at time $t=0$ for the LN and DR algorithms across all simulated markets. The green and red bars correspond to the average actions chosen by the LN and DR algorithms, respectively. The upper panel shows the actions for 20 lots, whereas the lower panel shows them for 60 lots.}
	\label{fig:actions}
\end{figure}

The second experiment concerns the algorithms' actions. 
Figure~\ref{fig:actions} shows the average actions at the initial time $t=0$ for the LN and DR algorithms, and for different initial positions. Figure~\ref{fig:actions_noise} demonstrates that both algorithms place limit orders aggressively, using actions $a_1$ and $a_2$ in the market with noise traders only. 
This is because limit orders placed at the best ask price have little market impact, given the absence of tactical traders. Interestingly, for 60 lots, they also send market orders using action $a_0.$ Figure~\ref{fig:actions_flow} demonstrates a more balanced passive placement in the market with noise and tactical traders, with greater use of actions $a_5$ and $a_6$ for both the LN and DR algorithms. This occurs because placing large limit orders near the best ask price triggers reactions from tactical traders. For 60 lots, both algorithms allocate many orders to action $a_6$ to avoid provoking such reactions. Figure~\ref{fig:actions_strategic} shows large allocations to both market and defensive limit orders, using actions $a_0, a_5$ and $a_6,$  in the market with noise, tactical, and strategic traders. This is because it is more profitable to send a market order when the strategic trader is selling, as the price is more likely to move down. When the strategic trader is buying, it is better to place a patient limit order far from the best ask using action $a_5$ or to withhold orders from the market using action $a_6$ to wait for the price to rise. 

\section{Conclusion} 
\label{sec:conclusion}

In this paper, we develop a trade execution algorithm based on a novel actor-critic policy gradient approach that uses the logistic-normal distribution to model random allocations. Its state space captures all relevant market features, including current and past states of the order book and queue positions. The action space allows the algorithm to place market orders and limit orders across multiple price levels, adapting to changing market conditions. The algorithm outperforms competitive benchmarks in realistic simulations that account for direct and indirect market impact. This is the first time the logistic-normal distribution has been combined with the actor-critic algorithm. Our simulations capture key stylized facts of limit order book dynamics. But our method can be used with any market simulation framework.


\bibliographystyle{plainnat}
\bibliography{rlte.bib}

\appendix 

\section{Implementation Details} 
\subsection{Parameters for market simulations}
\label{sec:parameters_market_sim}
We assume that the noise and tactical traders submit orders (market orders, limit orders and cancellations) up to level $D=30,$ and that all of their orders have the same shifted half-normal distribution with parameters $\traderstd^M_{\text{noise}} = \traderstd^{L,k}_{\text{noise}} = \traderstd^{C,k}_{\text{noise}} =2,$ and $\traderstd^M_{\text{tactical}} = \traderstd^{L,k}_{\text{tactical}} = \traderstd^{C,k}_{\text{tactical}} =2,$ for $k \in \{1, 2, \dots ,D \}.$ 
We cap order sizes at five standard deviations. We always start the simulations at $-\Delta t=-15s,$ from an equilibrium state and run them until time $T=150s.$ The equilibrium states are explained in more detail in Appendix \ref{subsec:stationary_shapes} below.  
The LN algorithm, DR algorithm, and the two heuristic benchmark algorithms trade during the time interval [0, T], updating their decisions at the times $n \Delta t$ for $n = 0, 1, \dots, N-1,$ with $\Delta t = 15s$ and $N = 10.$
The simulation parameters are summarized in Table \ref{table:simulation_parameters}.

\begin{table}[htbp] 
	\caption{Hyperparameters for market simulation.}
	\label{table:simulation_parameters}
	\begin{center}
		\begin{small}  	
			\begin{sc}
				\begin{tabular}{l | c }
					\toprule 
					start time for noise, tactical \& strategic traders $-\Delta t$ 
                    & $-15s$ \\ 
                    initial bid price  $p^{b}(-\Delta t)$ & 1000 \\
                    initial ask price  $p^{a}(-\Delta t)$ & 1001 \\
					$T$ & $150s$ \\ 
					$\Delta t$ & $ 15s $ \\ 
					    $N$ & 10 \\ 
                    
                    $\traderstd^M_{\text{\normalfont noise}} = \traderstd^{L,k}_{\text{\normalfont noise}} = \traderstd^{C,k}_{\text{\normalfont noise}} $  & 2  \\
                    $\traderstd^M_{\text{\normalfont tactical}} = \traderstd^{L,k}_{\text{\normalfont tactical}} = \traderstd^{C,k}_{\text{\normalfont tactical}} $  & 2 \\ 
					$\lambda^M$ & $0.1237$ \\ 
					$\lambda^{C,k}, \lambda^{L,k}$ & see table below \\								
					$d^M = d^{L,k} = d^{C,k} $ & 2 \\ 
					$\ordersizestrategic^M$ & 1 \\ 
					$\ordersizestrategic^L$ & 2 \\ 
					$\Delta t^L$ & $3s$ \\ 
					$\Delta t^M$ & $3s$ \\ 
					probability  Buy or sell strategic trader & $1/2$ \\ 	
					start time for RL and benchmark algorithms & $t=0$  \\ 
					small initial position & 20 \\
					large initial position & 60 \\ 
					\hline 
				\end{tabular}
			\end{sc}
		\end{small}
	\end{center}
	\vskip -0.1in
\end{table}

\paragraph{Market with noise traders}
The intensities of limit order arrivals $\lambda^{L,k}$ and cancellations $\lambda^{C,k}$ for the noise traders are given in Table \ref{table:intensities_noise}. The market order intensity is $\lambda^M=0.1237.$ Those intensities are taken from the paper \cite{abergel2013mathematical} with one modification. Instead of the cancellation intensities in that paper, we increase the cancellation intensities by a factor of $100.$ This leads to smaller queue sizes, which makes the simulation more efficient. 
At $- \Delta t$, we set the limit order book equal to its equilibrium shape. It is displayed in Figure \ref{fig:shape_noise}. 
\begin{table}[htbp]
	\caption{Intensities of limit order and cancellation arrivals for noise traders.}
	\label{table:intensities_noise}
	\begin{center}
		\begin{small}  	
			\begin{sc}
				\begin{tabular}{c c c c }
					\toprule 
					$k$ (ticks) & $\displaystyle \lambda^{L,k}$ & $10 \lambda^{C,k}$  \\
					\hline
					1 & 0.2842 & 0.8636 \\
					2 & 0.5255 & 0.4635 \\
					3 & 0.2971 &0.1487 \\
					4 & 0.2307 & 0.1096 \\
					5 & 0.0826 & 0.0402 \\
					6 & 0.0682 &  0.0341 \\
					7 & 0.0631 & 0.0311 \\
					8 & 0.0481 & 0.0237 \\
					9 & 0.0462 & 0.0233 \\
					10 & 0.0321 & 0.0178 \\
					11 & 0.0178 & 0.0127 \\
					12 & 0.0015 & 0.0012 \\ 
					13 & 0.0001 & 0.0001 \\ 
					14 & 0.0000 & 0.0000 \\ 
					\vdots & \vdots & \vdots \\ 
					30 & 0.0000 & 0.0000 \\ 
					\hline 
				\end{tabular}
			\end{sc}
		\end{small}
	\end{center}
	\vskip -0.1in
\end{table}

\paragraph{Market with noise \& tactical traders}
For the tactical traders, we choose a damping factor $c=0.65$ and constant imbalance reaction factors $d^{L,k} = d^{C,k} = d^M = 2$ for $k\in\{1,2, \dots D\}.$ 
To compensate for the presence of tactical traders, we reduce the intensities of the noise traders by 15\%. With this intensity reduction, the average total traded volume over the trading horizon is roughly the same for the market with noise traders only and the market with noise and tactical traders.
We start the simulation at $-\Delta t$ from its equilibrium state, displayed in Figure \ref{fig:shape_flow}.

\paragraph{Market with noise \& tactical  \& strategic traders}
We combine noise, tactical, and strategic traders for the third market. The parameters for noise and tactical traders remain the same as in the previous market simulation.
The strategic trader sends a limit order of size $\ordersizestrategic^L=2$ every $\Delta t^L=3$ seconds and a market order of size $\ordersizestrategic^M=1$ every $\Delta t^M=3$ seconds. Due to the presence of the strategic trader, this market does not have an equilibrium state. At time $-\Delta t,$ we start the order book from the same shape as in the market with noise and tactical traders. The trade direction of the strategic trader (buying or selling) is drawn randomly with probability $1/2$ at the time $-\Delta t.$ 

\subsection{Stationary order book shapes} 
\label{subsec:stationary_shapes}
\label{sec:intensities}

We always start the market simulation from an equilibrium state of the limit order book. The equilibrium states are described by the average volumes and the average spread: 
\newcommand{\tv}{\tilde{v}}
\newcommand{\ts}{\tilde{s}}
\begin{equation}
\tv = (\tv^b, \tv^a) 
= 
(\tv^{b,1}, \dots, \tv^{b,D}, 
\tv^{a,1}, \dots, \tv^{a,D})
\in \R_+^{2D}
,
\quad 
\ts \in \R_+. 
\label{average_shape} 
\end{equation}
Here, for $k\in\{1,\dots,D\}$ the quantity $\tv^{b,k}$ is the average volume $k-1$ ticks below the best bid price and $\tv^{a,k}$ is the average volume $k-1$ ticks above the best ask price. Since the average spread $\ts$ in our simulations is always close to one tick, we start all simulations with an initial bid price $p^{b}(-\Delta t)=1000$ and an initial ask price $p^{a}(-\Delta t)= 1001$. 

\begin{figure}[htbp]
	\centering
	
	\begin{subfigure}[t]{0.49\textwidth}
		\includegraphics[width=\textwidth]{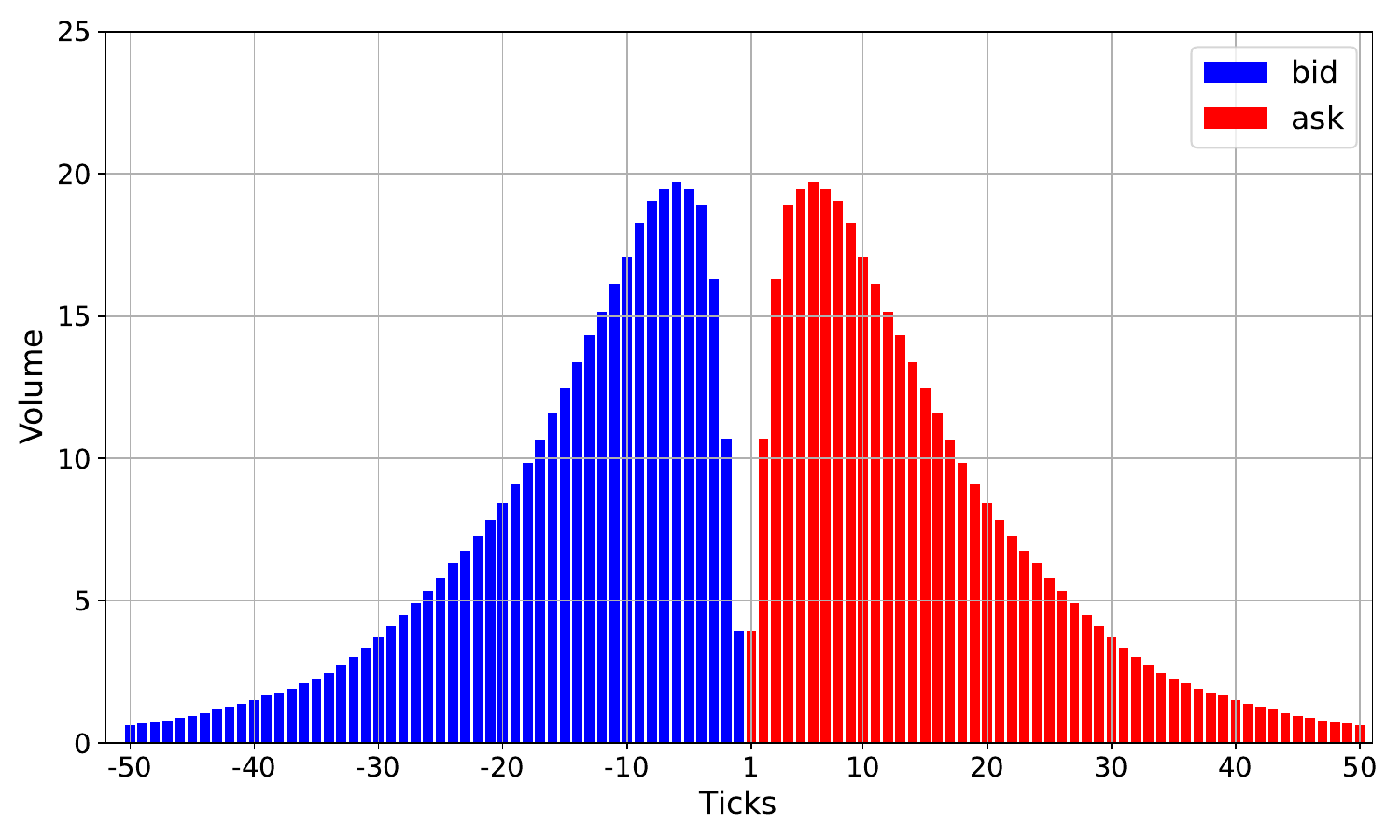}
		\caption{Average shape for market with noise traders}
		\label{fig:shape_noise}
	\end{subfigure}
	\hfill
	\begin{subfigure}[t]{0.49\textwidth}
		\includegraphics[width=\textwidth]{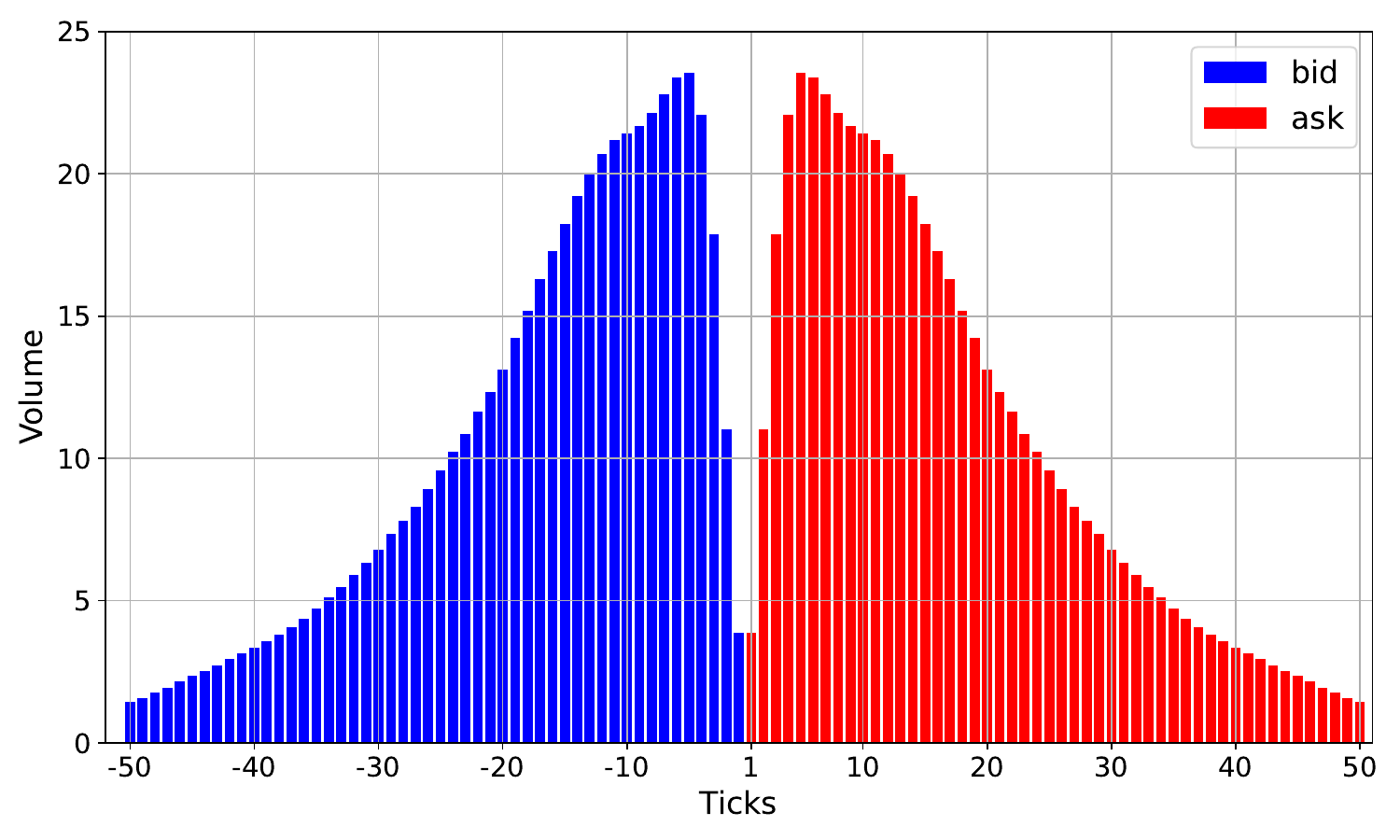}
		\caption{Average shape for market with noise and tactical traders}
		\label{fig:shape_flow}
	\end{subfigure}
	
	\caption{Average order book shapes for the market consisting of noise traders and both noise and tactical traders.}
	\label{fig:average_shape}
\end{figure}

We find such an equilibrium state by running the market simulation for a long time and then averaging. This equilibrium state is shown in Figure \ref{fig:shape_noise} for the first market consisting only of noise traders. For the second market, consisting of noise and a strategic trader, this equilibrium state is shown in Figure \ref{fig:shape_flow}. We observe that the shapes look similar but are not quite the same. In particular, the market with noise and tactical traders has more volume at the best bid and ask prices. The third market, consisting of noise traders, tactical traders, and a strategic trader, does not have an equilibrium state due to the presence of the strategic trader. 

\subsection{Traded volumes}
\label{sec:market_statistics}

In this section, we briefly document traded volumes and event statistics, which are computed over 10,000 trading simulations for all markets with a time horizon of 150 seconds. 
The statistics are documented in Table \ref{table:market_statistics}, which consists of five columns. The second column shows the average number of events. An event is a market order placement, a limit order insertion, or a cancellation. The third column shows the average traded volume. The fourth and fifth columns show the average buy and sell volumes, which sum to the total volume. There are many more events than traded volume, because limit order insertions and cancellations do not increase the traded volume. The total average traded volume in the first two markets is roughly the same. The presence of the strategic trader increases the average traded volume in the market with noise, tactical, and strategic traders. 

\begin{table}[htbp]
\caption{Average number of events (limit, market, and cancellation orders), the average number of trades, and the average total buy and sell volume for all market simulations.}
\label{table:market_statistics}
\begin{center}    
\begin{scriptsize}        
\begin{sc}\begin{tabular}{lcccc}
\toprule
market & \#events & total volume & buy volume & sell volume \\
\midrule
noise & 1162 & 95 & 48 & 47 \\
noise \& tactical & 1158  & 98 & 49 & 49 \\
noise \& tactical \& strategic & 1389  & 149 & 73 & 76 \\
\bottomrule
\end{tabular}
\end{sc}    
\end{scriptsize}\end{center}
\end{table}

\section{Implementation details for actor-critic reinforcement learning algorithms} 
\subsection{Logistic-normal algorithm}
\label{sec:hyper_parameters} 
\label{sec:hyper_parmaters_for_LN}

We implement an actor-critic reinforcement learning algorithm combined with the logistic-normal distribution for the trade execution problem. Our implementation is based on the implementation of the proximal policy optimization algorithm (PPO) by \cite{huang2022cleanrl}; see \url{https://github.com/vwxyzjn/cleanrl}.
We implement the following changes. First, we are not using clipping.
 Second, we implemented a custom logistic-normal action policy. Third, we implement a custom initialization of the mean of the underlying normal distribution and variance scaling as in \ref{covariance_scaling}.
 The algorithm's hyperparamters are summarized in Table~\ref{table:hype_params_sample_collection} and Table~\ref{table:hyper_params_neural_networks}.
\begin{table}[htbp] 
	\caption{Hyperparameters for sample collection and learning phase.}
	\label{table:hype_params_sample_collection}
	\vskip 0.15in
	\begin{center}
		\begin{small}  	
			\begin{sc}
				\begin{tabular}{l | c }
					\toprule 
					Number of parallel environments & $ 128 $ \\ 
					Steps per environment & $ 100 $ \\					
					Number of Trajectories $\tau$   & 1,280 \\
					Length of Trajectories & $10$ \\ 
					Training Iterations H & 400 \\
					Learning Rate $\eta$ & $0.0005$ \\ 						
					Linear Variance scaling & True \\ 
					Initial variance & 1 \\
					Final variance & 0.1 \\ 
					\hline 
				\end{tabular}
			\end{sc}
		\end{small}
	\end{center}
	\vskip -0.1in
\end{table}

\begin{table}[htbp] 
	\caption{Hyperparameters for neural networks.}
	\label{table:hyper_params_neural_networks}
	\vskip 0.15in
	\begin{center}
		\begin{small}  	
			\begin{sc}
				\begin{tabular}{l | c }
					\toprule 
					Simplex Dimension $K$ & 6 \\
					Number of hidden layers & $ 2 $ \\ 
					Nodes of in hidden layers & $128$ \\ 
					Nodes of output layer for value function network & $1$ \\ 
					Nodes of output layer for policy function network & $K$\\ 
					Activation functions for hidden layers & $ \tanh $ \\
					Activation functions for last layers  & None\\				
					\hline 
				\end{tabular}
			\end{sc}
		\end{small}
	\end{center}
	\vskip -0.1in
\end{table}

\paragraph{Neural network architecture}
For the value function $V_\vartheta$ and the policy $\pi_\theta,$ we use two separate neural networks with two hidden layers with $128$ nodes. The hidden layers have a
$\tanh$-activation function and the last layers have no activation function. The last layer of the policy network has $K$ nodes corresponding to the mean vector $\mu=(\mu^0, \dots, \mu^{K-1})$ of the normal distribution with density $\varphi_\theta.$ The last layer of the value function network has one node. We initialize the weights of both neural networks with an orthogonal initialization scheme; see e.g.\ \cite{saxe2013exact}. For the last layer of the policy  
network, we set the bias to $ b = (-1, -1, \dots, -1) \in \mathbb{R}^K $ and initialize the other weights using orthogonal initialization with a gain factor $10^{-5}.$ Therefore, the weights of the last layer of the policy network are close to zero, except for the bias term, and the output is effectively determined by the bias term. With our choice of the bias term, the action $a^K$ is more likely than the other actions at initialization, as explained in \eqref{bias}. For all other layers in the policy and value function neural networks, we initialize the bias terms to zero and use an orthogonal initialization scheme with a gain factor $0.01$ for all other weights.

\paragraph{Vectorized sample collection} We use a vectorized architecture to collect samples from the market simulation. We use $128$ parallel markets, using 128 CPUs, where each market collects 10 trajectories.  This results in a total number of $\tau=\text{1,280}$ trajectories. In each training iteration, we sample a batch of trajectories  
\begin{equation}
\mathcal{T}=\{(s_{n,k}, a_{n,k}, r(s_{n,k}, a_{n,k}, )), n\in\{0, \dots, N-1 \},\; k\in\{1,\dots,\tau\} \}.
\end{equation}

\paragraph{Learning phase}
We use the Adam optimizer \citep{kingma2017adam} with a learning rate (also referred to as step-size) of $\eta=0.0005.$ In all experiments, we make $H=400$ gradient steps. 
We choose the initial and final variance $\sigma_{\text{init}}=1.0$ and $\sigma_{\text{final}}=0.1,$ and reduce action variance following the schedule \eqref{covariance_scaling}. 

\paragraph{Computational resources}
The experiments were conducted on a workstation equipped with an AMD EPYC 7763 64-Core Processor with 128 threads, and a single NVIDIA GeForce RTX 4090 GPU with 24 GB VRAM. The system contains four NUMA nodes and 256 MiB of L3 cache, running with CUDA version 12.8 and driver 570.144. 

\paragraph{Training times} Simulating the order book is an expensive operation. We need to keep track of all orders on all price levels. Whenever new market, limit, or cancellation orders are received, they must be matched with the resting limit orders in the order book. To achieve this, we employ dictionaries to keep track of resting limit orders. Market orders require explicit iteration over price levels, whereas cancellations and limit orders can be processed without for-loops. Due to these computational complexities, training times for all our algorithms are relatively long. They are summarized in Table \ref{table:training_times} for the LN and DR algorithms. Training times are measured in hours and minutes. We notice that the DR algorithm trains faster than the LN algorithm in most cases. However, training duration also depends on the overall load on our computing machines, so these measurements should be interpreted only as rough indicators of computational cost. The DR algorithm was likely trained overnight, when system load was lower.
\begin{table}[htbp]
\caption{
Training times for the LN and the DR algorithms in hours and minutes (hh:mm). The results are displayed for the different market environments and initial positions. 
}
\label{table:training_times}
\begin{center}    
\begin{scriptsize}        
\begin{sc}\begin{tabular}{lcccc}
\toprule
Market &Lots& DR & LN \\
\midrule
Noise & 20 &  1:35 & 1:28\\
 & 60  & 1:13 & 1:29 \\
Noise \& Tactical & 20 & 1:36 & 1:28 \\
 & 60  & 1:19 & 1:43 \\
Noise \& Tactical & 20  & 1:32 & 1:52 \\
\& Strategic & 60 & 1:26 & 1:55  \\
\bottomrule
\end{tabular}
\end{sc}    
\end{scriptsize}\end{center}

\end{table}

\subsection{Dirichlet algorithm}
\label{sec:dirichlet_rl_implementation_details}

\paragraph{Dirichlet distribution}
Let $\alpha = (\alpha^0, \alpha^1, \alpha^2, \dots, \alpha^K)$ with $\alpha^k > 0,$ for $k=0,1,\dots,K.$ The Dirichlet distribution $\text{Dir}(\alpha)$ is defined on the probability simplex:
\[
\SI^{K} = \left\{ (a^0, a^1, \dots, a^K) \in \mathbb{R}^K : \, a^k \geq 0, \; \sum_{k=0}^K a^k = 1 \right\}
\]
Setting $\bar{\alpha}=\sum_{k=0}^{K} \alpha^k,$ the probability density function is defined by
\[
\pi(a^0, a^1, \dots, a^K) = \frac{1}{B(\alpha)} \prod_{k=0}^K (a^k)^{\alpha^k - 1}
\;
\text{ with }
\;
B(\alpha) = \frac{\prod_{k=0}^K \Gamma(\alpha^k)}{\Gamma\left( \bar{\alpha} \right)}.
\]
Let $a$ be a random variable with Dirichlet distribution, then the expected value for each component is given by 
\begin{align}
\label{eq:expected_value_dirichlet}
\E\big[a^k\big] &= \frac{\alpha^k}{\bar{\alpha}}, 
\quad 
\text{for } k=0,\dots,K.
\end{align}
The covariance for $i,j\in\{0,1,\dots,K\}$ is given by 
\begin{align}
\label{eq:covariance_of_dirichlet}
\mathrm{Cov}
\left[a^i, a^j\right] = 
\begin{cases}
   -\frac{\alpha^i \alpha^j}{{\bar{\alpha}}^2 ({\bar{\alpha}}  + 1)} \quad &i \neq j,  \\
   \frac{\alpha^i (\bar{\alpha}- \alpha^i)}{{\bar{\alpha}}^2 (\bar{\alpha} + 1)} \quad &i=j.
\end{cases} 
\end{align}

\paragraph{RL algorithm}
As for the logistic-normal distribution, we use the policy gradient formula \eqref{policy_gradient} to train the algorithm. The neural network architecture for the value function network is identical to that of the LN algorithm, which was defined in Section~\ref{sec:hyper_parmaters_for_LN}. The parameters of the density $\alpha_\theta = (\alpha^0_\theta, \alpha^1_\theta,\dots,\alpha^K_\theta)$ are the output of a neural network (policy network) with weights $\theta$. The network architecture is the same as for the LN algorithm, except for the output layer. To ensure that the outputs are positive, we use the soft-plus function, defined as $\mathrm{softplus}(x)=\log(1+\exp(x)),$ as an activation function in the output layer. We choose the soft-plus function over the exponential function to prevent the parameters from becoming excessively large, which could lead to a concentrated policy due to \eqref{eq:covariance_of_dirichlet}. As for the logistic-normal distribution, we initialize the weights of the final layer of the policy network with an orthogonal initialization scheme using a gain factor of $10^{-5},$ leading to small weights. We set the bias of the final layer to $b=\textrm{softplus}^{-1}(1,1, \dots, 1, 10)\in\R^{K+1},$ where the inverse is applied to each component. Thus, after applying the soft-plus activation function, the output is close to $\alpha_\theta=(1,1,\dots,1,10)\in\mathbb{R}^{K+1}.$ Then \eqref{eq:expected_value_dirichlet} implies 
\[
\mathbb{E}[a_\theta^K] \approx 10 \mathbb{E}[a_\theta^k] \quad \text{for } k= 0,\dots,K-1.
\]
Therefore, the action $a_\theta^K$ (not placing orders) is more likely than the other actions $a_\theta^0,\dots,a_\theta^{K-1}.$ Hence, the algorithm is more likely to observe full trajectories up to terminal time $T$ at the beginning of the training cycle, which helps to prevent it from getting stuck in local minima. When evaluating the algorithm, we use a deterministic version. More precisely, we use the expected value \eqref{eq:expected_value_dirichlet} with parameters $\alpha_\theta$ as actions, where $\theta$ are the final trained weights of the neural network.   

\subsection{Feature normalization} 
\label{sec:feature_normalization} 
The features described in Section \ref{subsec:observation_space} are normalized in our implementation.  
We aim to normalize all features to a common scale by transforming them to the interval $[-1,1]$ or $[0,1] $. We present the normalized counterpart of each feature in the following bullet points. Furthermore, we introduce an additional feature. 

\paragraph{Market states}
\begin{itemize}
	\item We normalize bid and ask prices around the initial prices. More specifically, we use $(p^a(t)-p^a(0))/10$ and $(p^b(t)-p^b(0))/10.$ In our simulations, the prices rarely move more than five ticks, which justifies our normalization. 	
	\item  
	We recall from Section \ref{subsec:observation_space} that the algorithm observes the first $K-1$ entries of the order book volumes defined in equation \eqref{volume}. Instead of the absolute, we consider the normalized volumes on the bid and ask sides $\bar{v}^b(t) \in \R^{K-1}_+ $ 
	and $\bar{v}^a(t) \in \R_+^{K-1}.$ We normalize the volumes by using the long-term average queue sizes $\tv,$ defined in Appendix \ref{subsec:stationary_shapes}, as follows:  
	\[
	\bar{v}^{b,k}(t)=v^{b,k}(t)/\tilde{v}^{b,k}
	\quad
	\text{and}
	\quad
	\bar{v}^{a,k}(t)=v^{a,k}(t)/\tilde{v}^{a,k}.
	\]
	The average queue sizes are estimated by running the simulation for a long time and computing averages as described in Appendix \ref{subsec:stationary_shapes}.  For the market consisting of noise traders only, we normalize $v(t)$ with the long-term average shape of that market, which is displayed in Figure \ref{fig:shape_noise}. For the market consisting of noise and tactical traders, we normalize $v(t)$ with the long-term average shape of that market, which is displayed in Figure \ref{fig:shape_flow}. There is no stable long-term average shape for the market consisting of noise, tactical, and strategic traders as the market drifts up or down. In this case, we normalize the volumes using the average shapes corresponding to the market with noise and tactical traders (Figure \ref{fig:shape_flow}).		
	\item We normalize the market order flow $\Delta^M(t)$ in $(t-\Delta t, t]$ by dividing it by the total volume of market buy and sell orders in $(t-\Delta t, t]$. Then, the normalized market order flow is contained in $[-1,1].$
	\item We normalize the limit order flow $\Delta^L(t)$ in $(t-\Delta t, t]$ by dividing it by the total volume of limit buy and sell orders in $(t-\Delta t, t]$. Then, the normalized limit order flow is contained in $[-1,1].$
    \item We normalize the cancellation order flow $\Delta^C(t)$ in $(t-\Delta t, t]$ by dividing it by the total volume of limit sell and buy order cancellations in $(t-\Delta t, t]$. Then, the normalized cancellation order flow is contained in $[-1,1].$
	\item We normalize the mid-price drift relative to the previous mid-price by considering $(p(t)-p(t-\Delta t))/p({t-\Delta t}).$ 	
	
\end{itemize}

\paragraph{Private states}

\begin{itemize}
	\item We normalize the current time relative to the final time by considering $t/T \in [0,1].$ 
	\item We normalize the inventory relative to the initial position by considering $\inventory(t)/M \in [0,1].$ 
	\item We normalize the number of active limit orders in the book relative to the inventory by considering $m(t)/M(t)\in[0,1].$
	\item If there are $m\leq M(t)$ active limit orders in the book, and $u\leq M(t)$ is the number of lots that are inactive, and the remaining orders have already been filled, then we encode the levels of all limit orders by
	\[
	\bar{l} = (l^1, \dots, l^m, l^{m+1}, \dots, l^{m+u}, l^{m+u+1}, \dots, l^M)/K.
	\]
	Here, for $k\in\{1,\dots m\},$ the quantity $l^k$ is the level of the active limit orders, ordered from the smallest to the highest price levels. For 
	$k\in\{m+1, \dots, m+u\},$ we set $l^k=K,$ where $K$ is the dimension of $\SI^K.$ For $k\in\{m+u+1, \dots, M\},$ we set $l^k=-K.$ 
	\item If there are $m\leq M(t)$ active limit orders in the book, and $u\leq M(t)$ is the number of lots that are inactive, and the remaining orders have already been filled, then we encode the queue positions of all limit orders by
	\[
	\bar{q} = (q^1, \dots, q^m, q^{m+1}, \dots, q^{m+u}, q^{m+u+1}, \dots, q^M)/50.
	\]
	Here, for $k\in\{1,\dots m\},$ the quantity $q^k$ is the queue position of the active limit orders, ordered from the smallest to the highest price levels and queue positions. For 
	$k\in\{m+1, \dots, m+u\},$ we set $q^k=50.$ For $k\in\{m+u+1, \dots, M\},$ we set $q^k=-50.$ 	
	In our simulations, the queue sizes are rarely larger than $50$ (see Figures \ref{fig:shape_noise} and \ref{fig:shape_flow}), which justifies the normalization of the queue positions by $50.$  
\end{itemize}

\paragraph{Additional feature}
We use one additional feature, representing the number of active orders per price level, given by 
	$\gamma(t) \in [0, 1]^{K-1},$ where $K$ is the dimension of the simplex containing the actions. For $k\in\{1, \dots, K-1\},$ the quantity $\gamma^k(t)$ is the fraction of the inventory $M(t)$ that is placed at $k-1$ ticks above the best ask and $\gamma^{K}(t)$ is the fraction of the inventory $M(t)$ that is either placed higher than $K-1$ ticks away from the best ask price or not placed in the book. We note that the feature $\gamma(t)$ is redundant since the algorithm already has information on the levels of its limit orders through the vector $\bar{l}(t).$ However, the inclusion of $\gamma(t)$ improves the convergence speed of the algorithm. Furthermore, we need to compute $\gamma(t)$ to implement the new order allocation of the algorithm after choosing action $a(t) $, so there are no added computational resources required to compute this feature. 

\section{Additional experiments}

\subsection{Generalized advantage function estimates}
\label{sec:gae_experiments}

In equation \eqref{advantage_estimate}, we use a version of the {\sl generalized advantage function estimate (GAE)}, which was introduced in \cite{schulman2015high}. The GAE depends on two parameters: The discount factor $\gamma^\text{GAE}$ and a parameter $\lambda^\text{GAE}\in(0,1).$ The parameters are called $\gamma$ and $\lambda$ respectively in \cite{schulman2015high}. For the trade execution problem, we have $\gaegamma=1.$ The parameter $\gaelambda$ controls the tradeoff between bias and variance of the advantage function estimate. A lower value of $\gaelambda$ leads to a low-variance/high-bias estimate. A high value for $\gaelambda$ leads to a high-variance/low-bias estimate. In \eqref{advantage_estimate}, we choose a high-variance/low-bias estimate by setting $\gaelambda=1.$ We control the variance by choosing a larger number of trajectories $\tau=1,280$ per gradient step. In this section, we evaluate whether it is possible to achieve similar performance by choosing a smaller number of trajectories $\tau$ while controlling variance with a smaller value for $\gaelambda.$ In \cite{schulman2015high}, the authors recommend a choice of $\gaelambda\in[0.9,0.99].$

We conduct two experiments with a smaller number of trajectories $\tau=320.$ 
We choose $\gaelambda=0.95$ to reduce variance and compare it with the baseline $\gaelambda=1.0$. The results are displayed in Table \ref{table:small_batch}. In both cases, the performance of the algorithm is worse than in the case where it is using $\tau=1,280$ trajectories; see Table \ref{table:results}. Furthermore, we observe that setting $\gaelambda=0.95$ does not significantly improve the algorithm's performance. An explanation for this might be that we are using only 10 time steps. In cases where the execution problem lasts over more time steps (like 100 or 1000 time steps), the variance of the advantage function estimate \eqref{advantage_estimate} will be larger, and choosing a smaller value for $\gaelambda=0.95$ might be beneficial.

\begin{table}[htbp]
\caption{Results when the algorithm is trained with $\tau=320$ trajectories per gradient step, using  $\gaelambda=1.0$ and $\gaelambda=0.95.$ The results for $\gaelambda=1.0$ are displayed in the upper panel of the table, while the results for $\gaelambda=0.95$ are displayed in the lower panel. The best expected values across both values for $\gaelambda$ are highlighted in bold. If two expected values are equal, we highlight both of them.}
\label{table:small_batch}
\begin{center}    \begin{scriptsize}        \begin{sc}\begin{tabular}{lccccccc}
\toprule
Market & \#Lots & $\mathbb{E}[\text{SL}]$ & $\sigma[\text{SL}]$ & $\mathbb{E}[\text{TWAP}]$ & $\sigma[\text{TWAP}]$ & $\mathbb{E}[\text{LN}]$ & $\sigma[\text{LN}]$ \\
\midrule  
\midrule 
$\gaelambda=1.0$ & & & & &   & &  \\
\midrule
Noise & 20 & 0.52 & 1.19 & -0.06 & 0.94  & \textbf{0.53} & 1.1 \\
 & 60 & -1.09 & 1.34 & -1.4 & 0.98  & \textbf{-0.78} & 0.89 \\
 \midrule 
Noise \& Tactical & 20 & 0.1 & 1.43 & 0.48 & 0.68  & \textbf{0.74} & 0.64 \\ 
 & 60 & -3.36 & 0.99 & -0.96 & 0.95 & -0.36 & 0.64 \\
 \midrule 
Noise \& Tactical & 20 & -1.64 & 2.95 & -0.36 & 3.03 & 1.1 & 2.14 \\
\& Strategic & 60 & -2.51 & 3.67 & -1.45 & 3.46 & \textbf{-0.01} & 2.39 \\
\midrule 
\midrule 
$\gaelambda=0.95$ & & & &  & & &    \\
\midrule
Noise & 20 & 0.52 & 1.19 & -0.06 & 0.94 & 0.6 & 0.82 \\
 & 60 & -1.09 & 1.34 & -1.4 & 0.98 & \textbf{-0.78} & 0.86 \\
 \midrule 
Noise \& Tactical & 20 & 0.1 & 1.43 & 0.48 & 0.68 & 0.72 & 0.72 \\
 & 60 & -3.36 & 0.99 & -0.96 & 0.95 & \textbf{-0.34} & 0.66 \\
 \midrule 
Noise \& Tactical & 20 & -1.64 & 2.95 & -0.36 & 3.03 & \textbf{1.12} & 2.13 \\
\& Strategic & 60 & -2.51 & 3.67 & -1.45 & 3.46  & -0.24 & 2.38 \\
\bottomrule
\end{tabular}
\end{sc}    
\end{scriptsize}
\end{center}
\end{table}

\subsection{Variance learning}
\label{section:std_learning}

In this section, rather than using the schedule \eqref{covariance_scaling}, we let the logistic-normal algorithm learn variances. As in Section \ref{sec:initialization_scaling_parameters}, let $\Sigma=(\Sigma^{ij})_{i,j=0,1,\dots,K-1}$ be the covariance matrix of the logistic-normal distribution. We set $\Sigma^{ij}=0$ for $i\neq j$ and encode the logarithms of the variances  
\begin{equation}
\label{eq:learnable_variances}
\log(\Sigma^{kk}) = \varsigma^{k} \in \R, \quad k=0,1,\dots,K-1.
\end{equation}
We then treat the vector $(\varsigma^0, \dots, \varsigma^{K-1})\in\R^K$ as a learnable but state independent parameter.  We initialized the log-variances to $\varsigma^k=0$ for $k=0,1,\dots,K-1$  at the beginning of the training cycle. In each step, the parameters $\varsigma^k,$ for $k=0,\dots,K-1,$ are then updated as part of the policy updating step using the loss function \eqref{logistic_normal_loss}. We could also consider state-dependent variances, but the authors in \cite{andrychowicz2021matters} find that this does not yield better results.

We conducted an additional test, using the same parameter settings as above for the market simulation and the logistic-normal algorithm. When evaluating the LN algorithm with scheduled variance scaling, we use a deterministic version by using $h(\mu_\theta)$ as the actions, as in Section~\ref{sec:rl_setup}. When evaluating the LN algorithm with learned action variances, we use the stochastic policy for evaluation by sampling from the logistic-normal distribution with mean $\mu_\theta$ and variance $(\exp(\varsigma^0), \exp(\varsigma^1), \dots,  \exp(\varsigma^{K-1}).$

\begin{figure}[htbp]
	\centering
	\begin{subfigure}[t]{0.3\textwidth}
		\includegraphics[width=\textwidth]{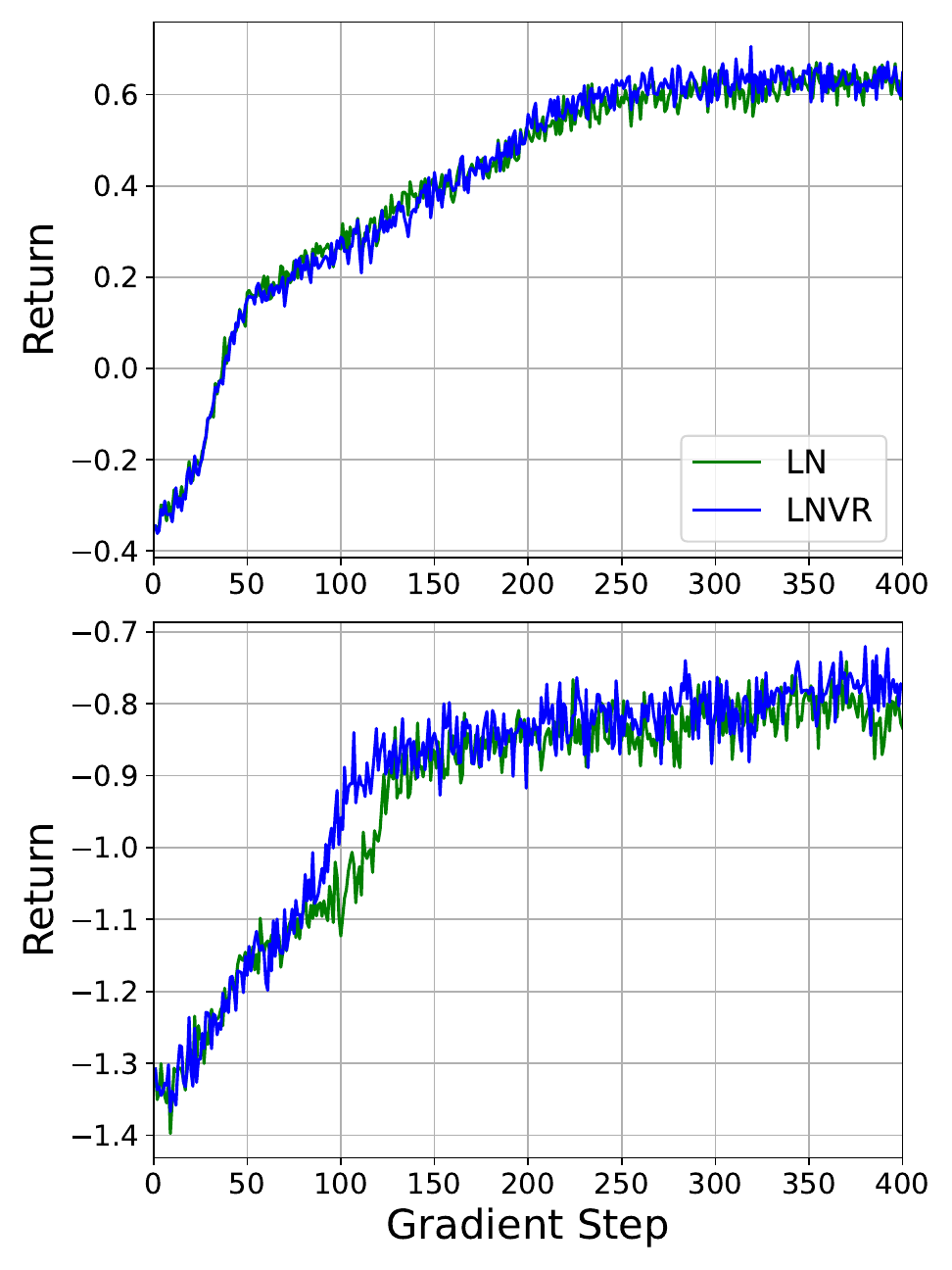}
		\caption{Noise Traders}
		\label{fig:convergenc_learn_var}
	\end{subfigure}
	\hfill
	\begin{subfigure}[t]{0.3\textwidth}
		\includegraphics[width=\textwidth]{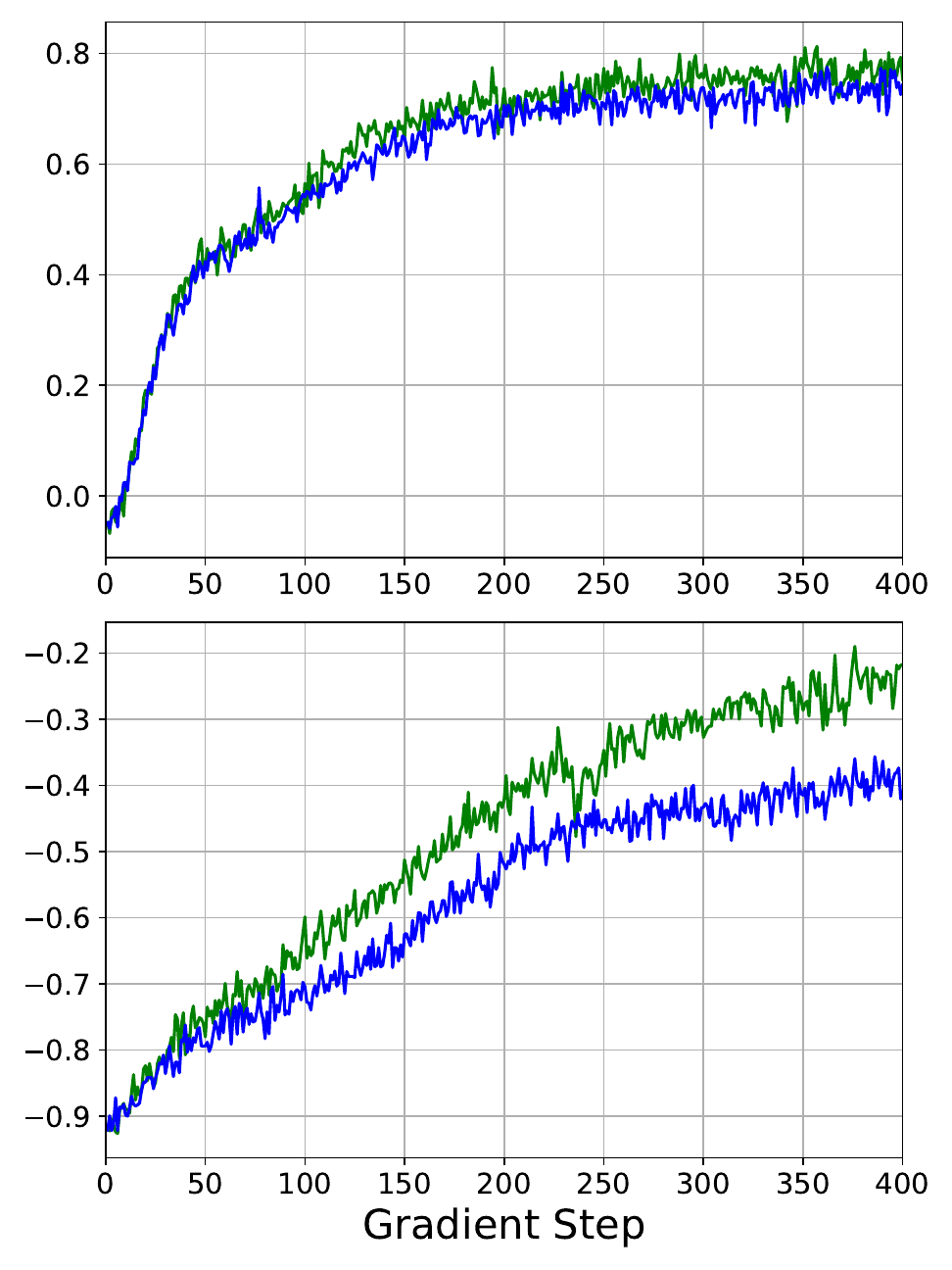}
		\caption{Noise \& Tactical Traders}
		\label{fig:tactical_convergence_learned_std}
	\end{subfigure}
	\hfill
	\begin{subfigure}[t]{0.3\textwidth}
		\includegraphics[width=\textwidth]{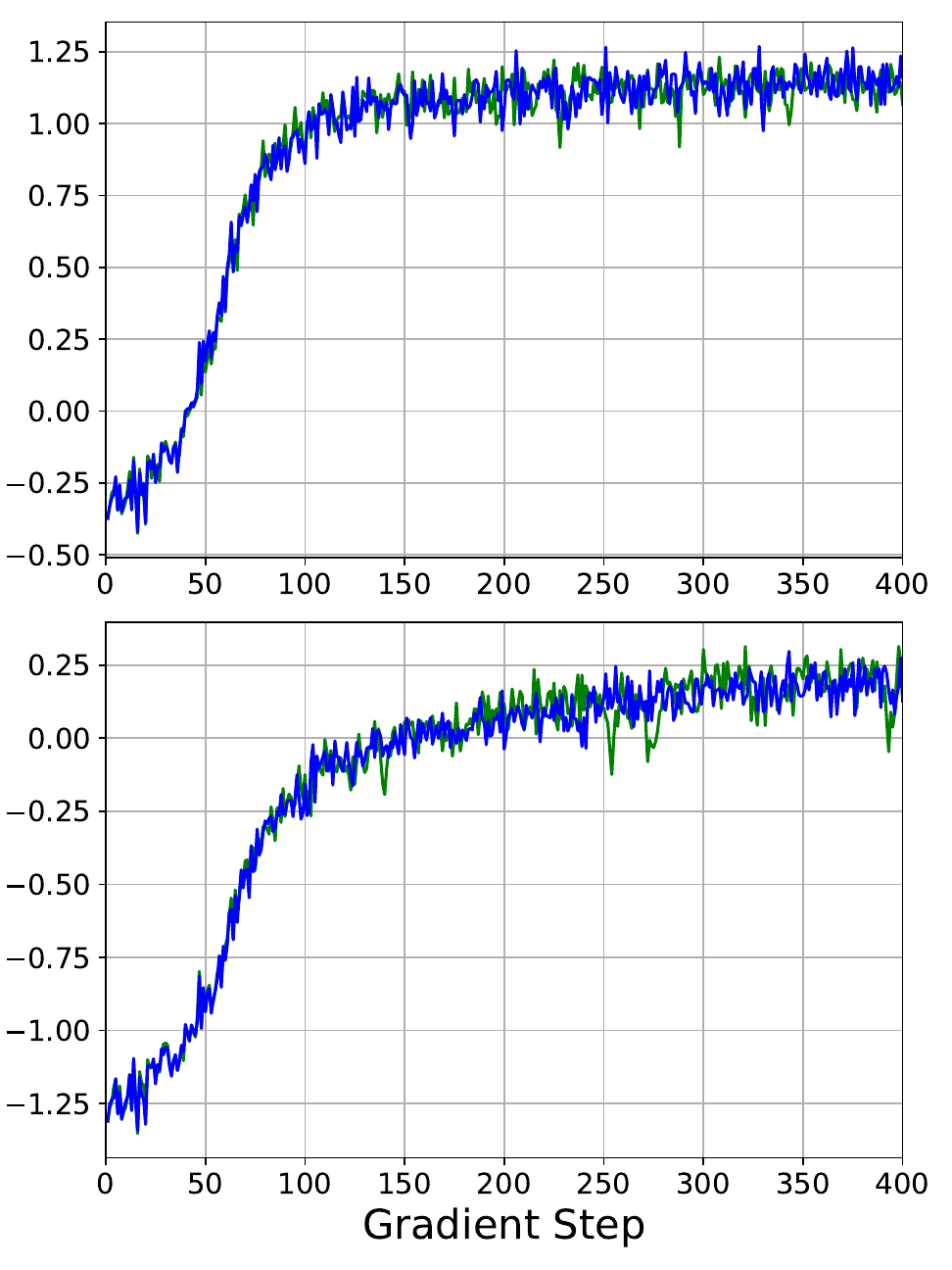}
	        \caption{Noise, Tactical, \& Strategic Traders}	
		\label{fig:strategic_convergence_learned_std}
	\end{subfigure}
	
	\caption{
		Average episode reward per batch during training for different environments and initial position sizes. The x-axis shows the number of gradient steps. The top row shows the rewards for 20 lots. The bottom row shows the rewards for 60 lots. The symbol LN denotes the logistic-normal algorithm. LNVR denotes the logistic-normal algorithm with learnable variance parameters.}
	\label{fig:return_convergence_learned_std}
\end{figure}

We find that the performance of the algorithm with manual variance scaling (LN) and learnable variances (LNVR) is similar across all cases except when the algorithm sells 60 lots in the market consisting of noise and tactical traders. In this case, we see better convergence of the average returns per batch in Figure \ref{fig:tactical_convergence_learned_std} and significantly better performance in Table \ref{table:results_learn_var} of the LN algorithm compared to the LNVR algorithm. Therefore, we prefer to manually reduce action variances at each gradient step, but in new applications both approaches should be tested. 
\begin{table}[htbp]
\caption{The table shows returns of the LN algorithm and all benchmark algorithms. Additionally, we include the LNVR algorithm, which also learns the action variances. The best expected returns per row are highlighted in bold.}  
\label{table:results_learn_var}
\begin{center}    \begin{scriptsize}        \begin{sc}\begin{tabular}{lccccccccc}
\toprule
Market & \#Lots & $\mathbb{E}[\text{SL}]$ & $\sigma[\text{SL}]$ & $\mathbb{E}[\text{TWAP}]$ & $\sigma[\text{TWAP}]$ & $\E[\text{LNVR}]$ & $\sigma[\text{LNVR}]$ & $\mathbb{E}[\text{LN}]$ & $\sigma[\text{LN}]$ \\
\midrule
Noise & 20 & 0.52 & 1.19 & -0.06 & 0.94 & \textbf{0.64} & 0.93 & 0.61 & 1.03 \\
 & 60 & -1.09 & 1.34 & -1.4 & 0.98 & -0.77 & 0.99 & \textbf{-0.72} & 0.9 \\
Noise \& Tactical & 20 & 0.1 & 1.43 & 0.48 & 0.68 & 0.74 & 0.7 & \textbf{0.81} & 0.64 \\
 & 60 & -3.36 & 0.99 & -0.96 & 0.95 & -0.41 & 0.69 & \textbf{-0.25} & 0.67 \\
Noise \& Tactical & 20 & -1.64 & 2.95 & -0.36 & 3.03 & \textbf{1.15} & 2.14 & 1.13 & 2.08 \\
\& Strategic & 60 & -2.51 & 3.67 & -1.45 & 3.46 & 0.14 & 2.04 & \textbf{0.23} & 2.15 \\
\bottomrule
\end{tabular}
        \end{sc} 
        \end{scriptsize}\end{center}\end{table}

\begin{figure}[htbp]
	\centering
	\begin{subfigure}[t]{0.3\textwidth}
		\includegraphics[width=\textwidth]{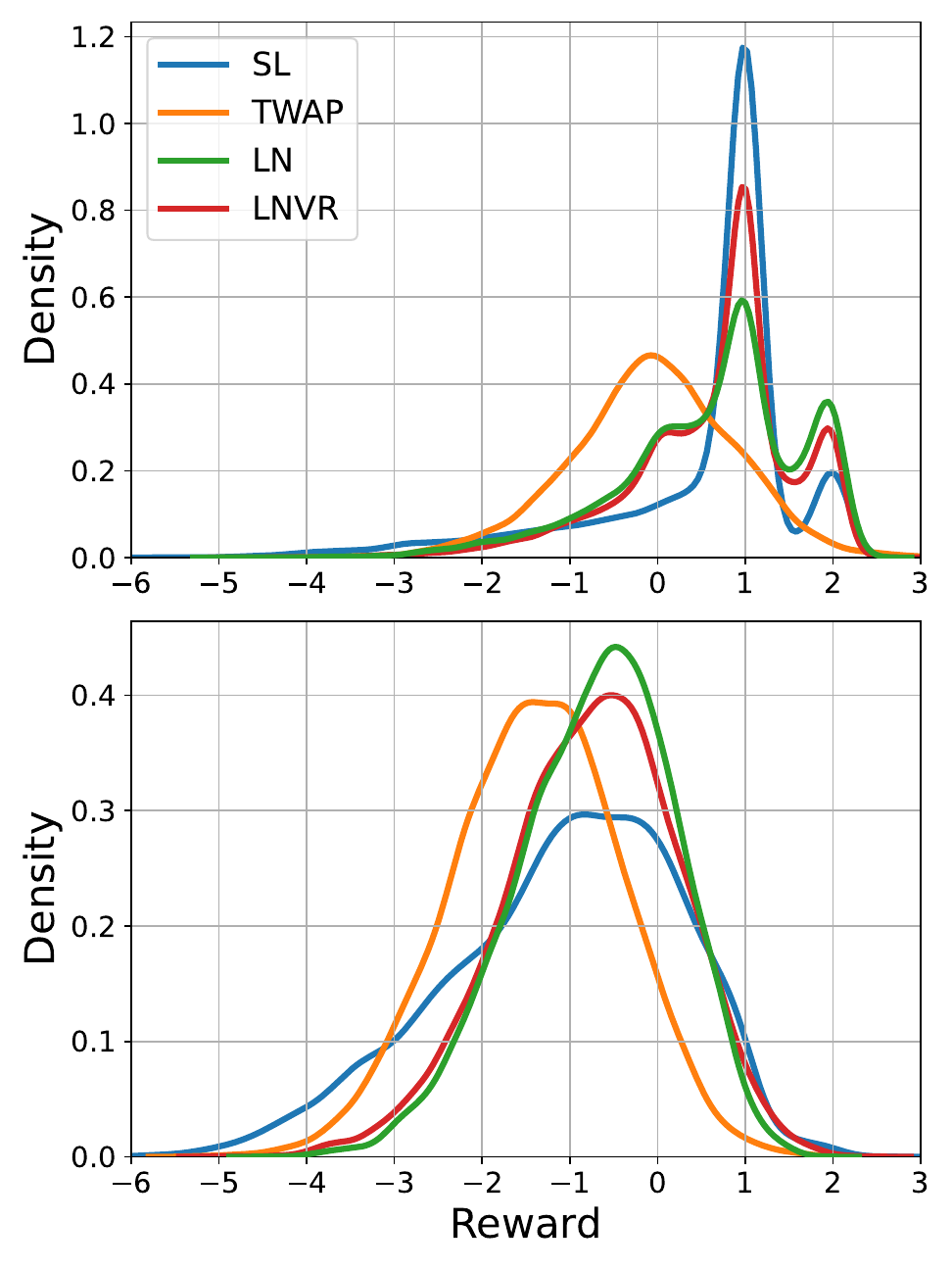}        
		\caption{Noise Traders}
		\label{fig:noise_results_learned_var}
	\end{subfigure}
	\hfill
	\begin{subfigure}[t]{0.3\textwidth}
		\includegraphics[width=\textwidth]{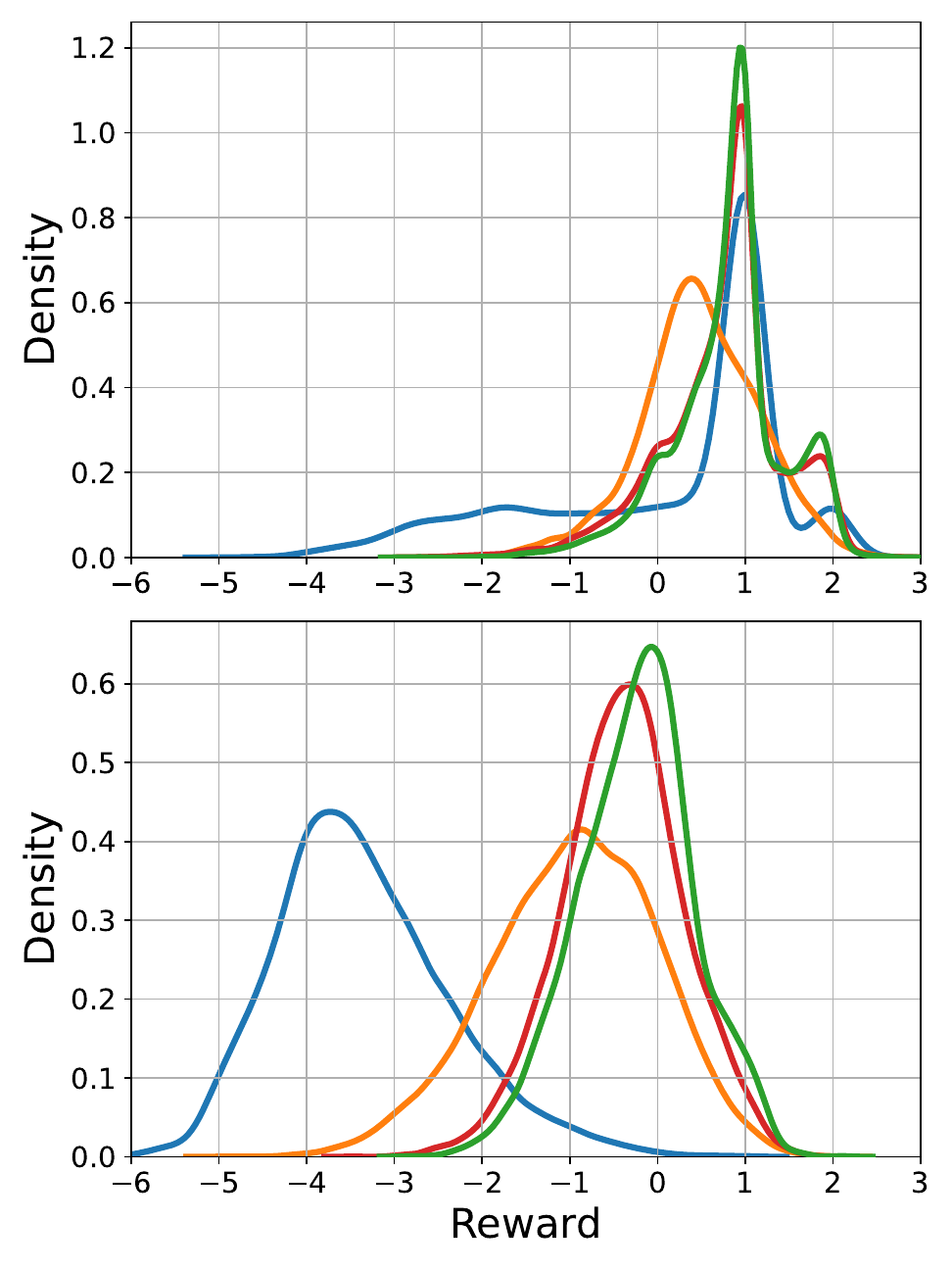}
		\caption{Noise \& Tactical Traders}
		\label{fig:tactical_results_learned_var}
	\end{subfigure}
	\hfill 
	\begin{subfigure}[t]{0.3\textwidth}
	\includegraphics[width=\textwidth]{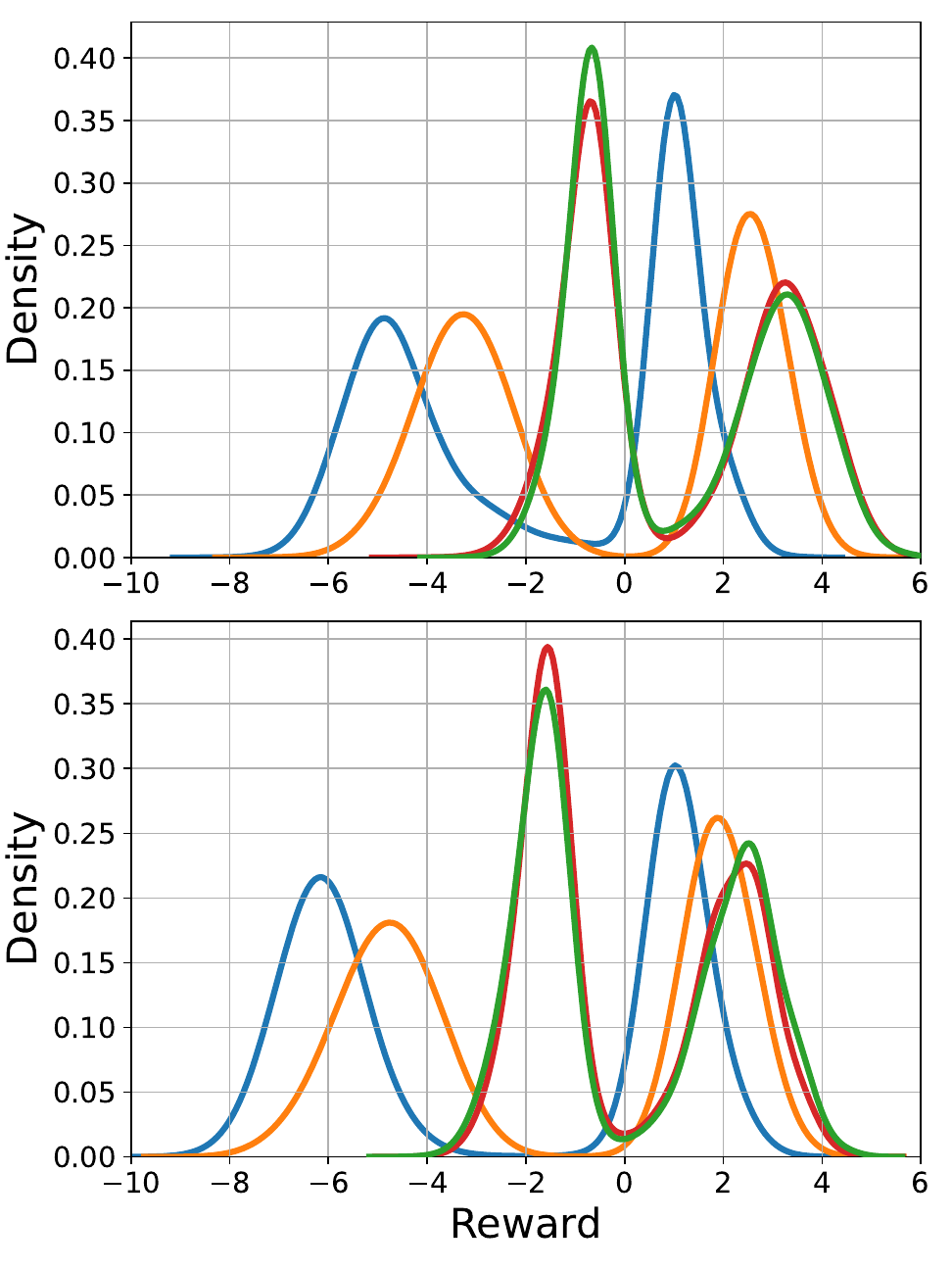}
	\caption{Noise, Tactical \& Strategic Traders}
	\label{fig:strategic_results_learned_var}			
	\end{subfigure}	
	\caption{Reward histograms for the heuristic benchmark strategies and RL algorithms. The upper panel shows the results for an initial position of 20 lots, whereas the lower panel shows those for an initial position of 60 lots. The symbol LN refers to the algorithm with manual variance scaling, and LNVR refers to the algorithm that learns action variances.
    }
	\label{fig:densities_learned_variance}
\end{figure}

\subsection{Long time horizon}
\label{sec:long_time_horizon_experiment}

In this section, we test a longer execution horizon of $T=300s.$ Since the algorithm has more time to sell its inventory, we adjust the initial position sizes to 40 lots and 120 lots. To maintain the same number of decision steps per execution, we set the algorithms' decision interval to $\Delta t = 30s.$  All other simulation parameters remain the same as before. 

\begin{figure}[htbp]
	\centering
	\begin{subfigure}[t]{0.3\textwidth}
		\includegraphics[width=\textwidth]{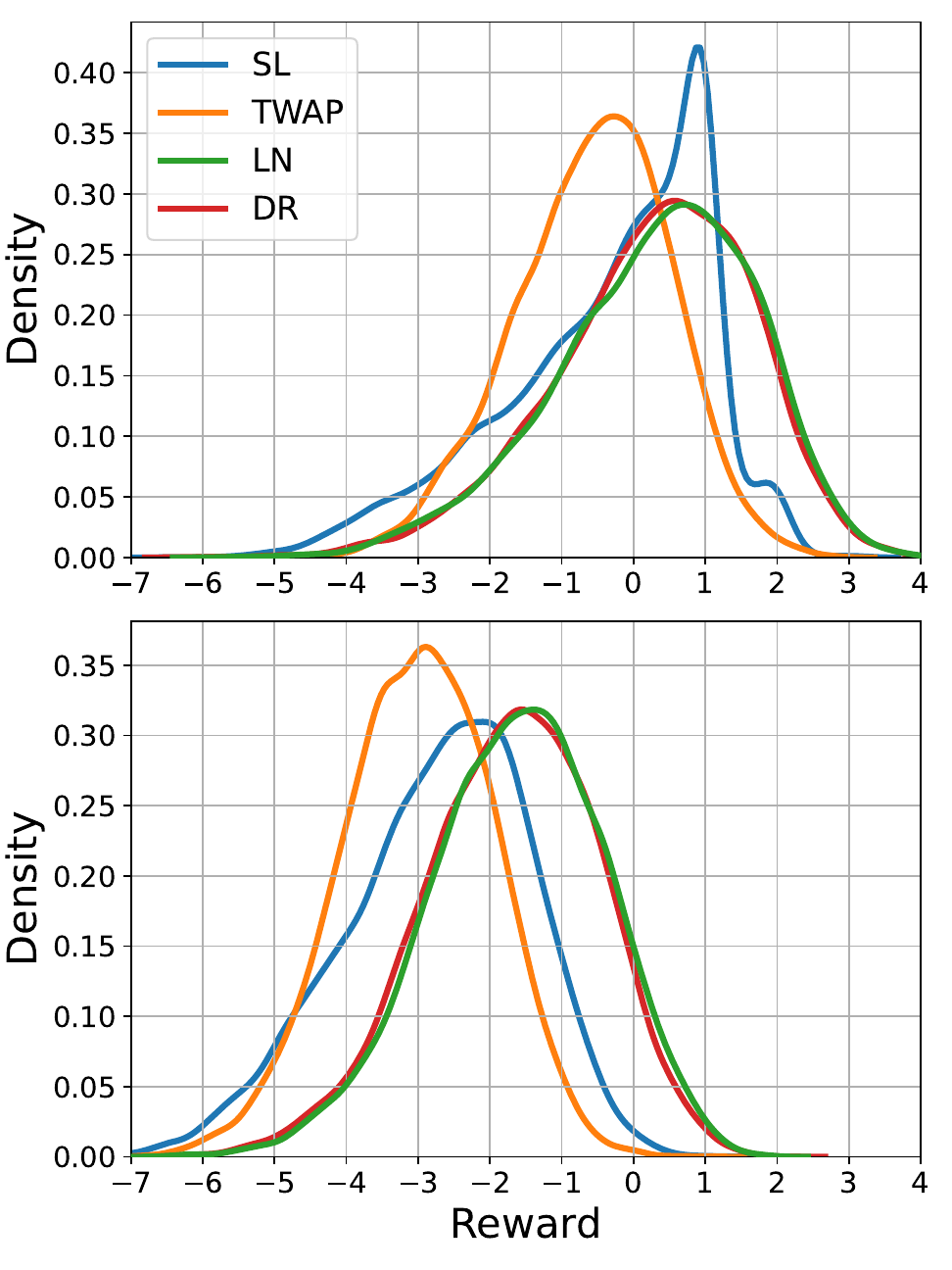}        
		\caption{Noise Traders}
		\label{fig:noise_results_long}
	\end{subfigure}
	\hfill
	\begin{subfigure}[t]{0.3\textwidth}
		\includegraphics[width=\textwidth]{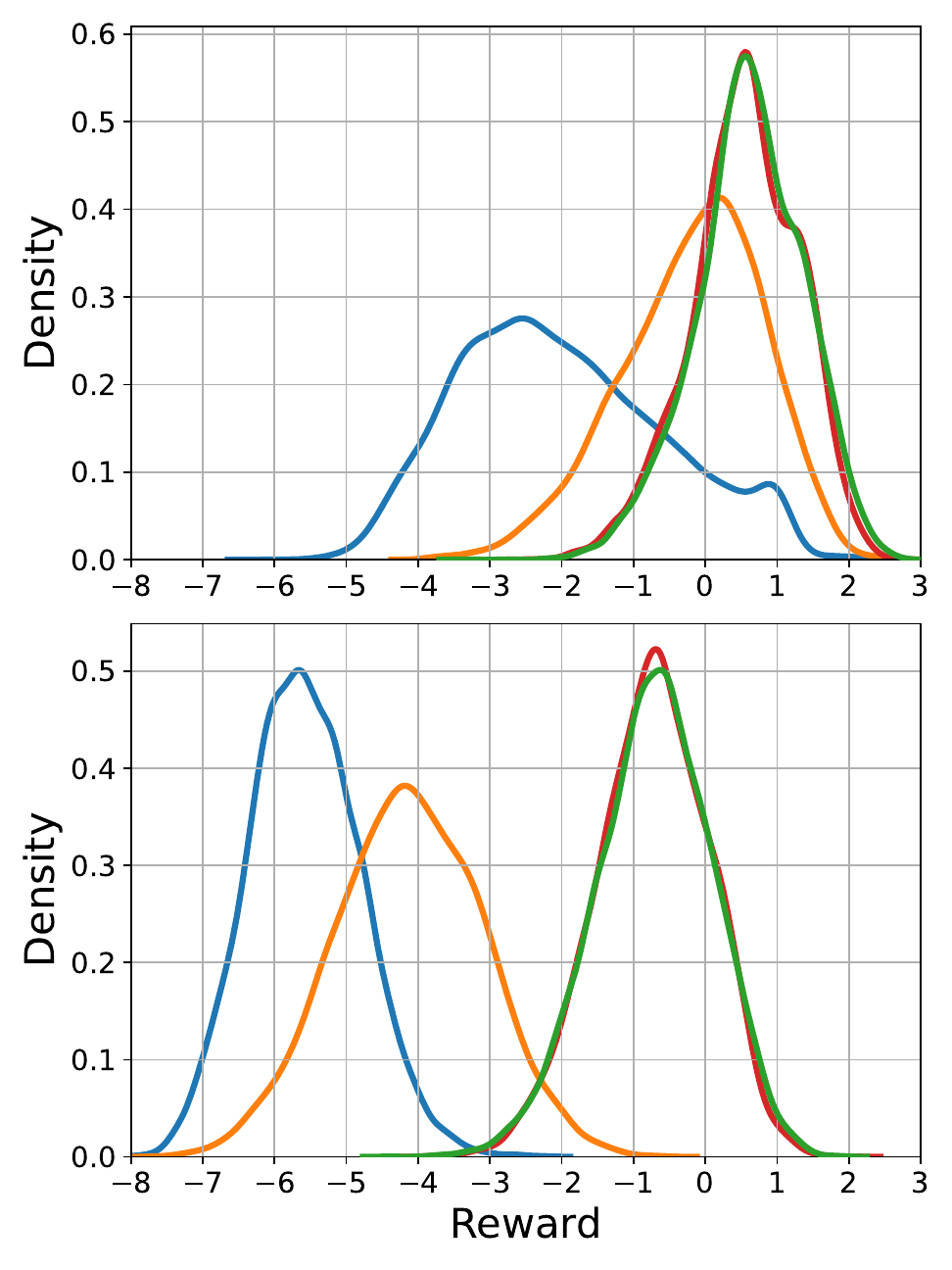}
		\caption{Noise \& Tactical Traders}
		\label{fig:tactical_results_long}
	\end{subfigure}
	\hfill 
	\begin{subfigure}[t]{0.3\textwidth}
	\includegraphics[width=\textwidth]{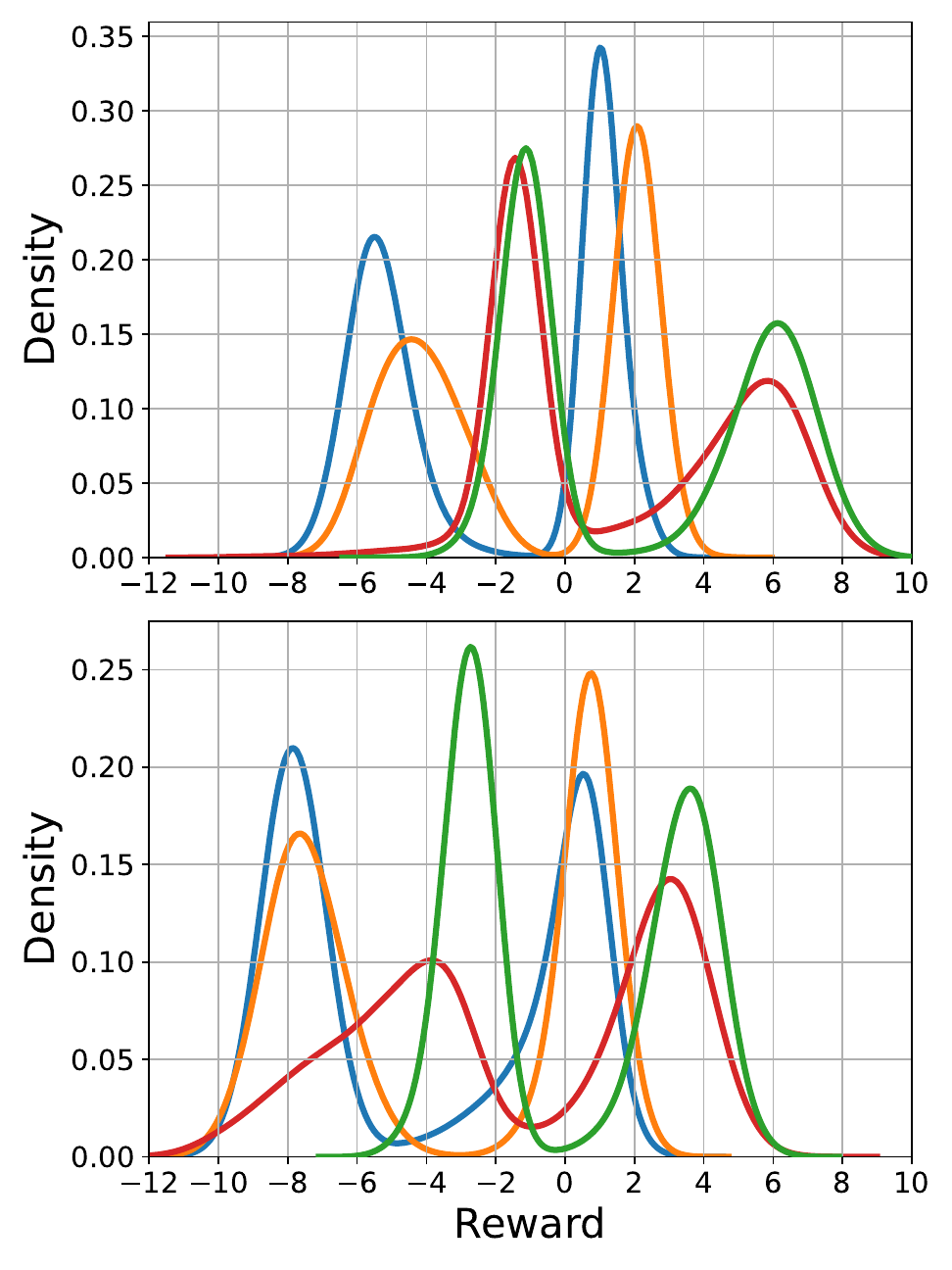}
	\caption{Noise, Tactical \& Strategic Traders}
	\label{fig:strategic_results_long}			
	\end{subfigure}	
	\caption{Reward histograms for all heuristic benchmark strategies and the logistic-normal and Dirichlet algorithms, for an initial position of 40 and 120 lots, and an execution horizon of 300 seconds. The upper panels show the results for 40 lots. The lower panels show the results for 120 lots.}
	\label{fig:densities_long_horizon}
\end{figure}

\begin{figure}[htbp]
	\centering
	\begin{subfigure}[t]{0.3\textwidth}
		\includegraphics[width=\textwidth]{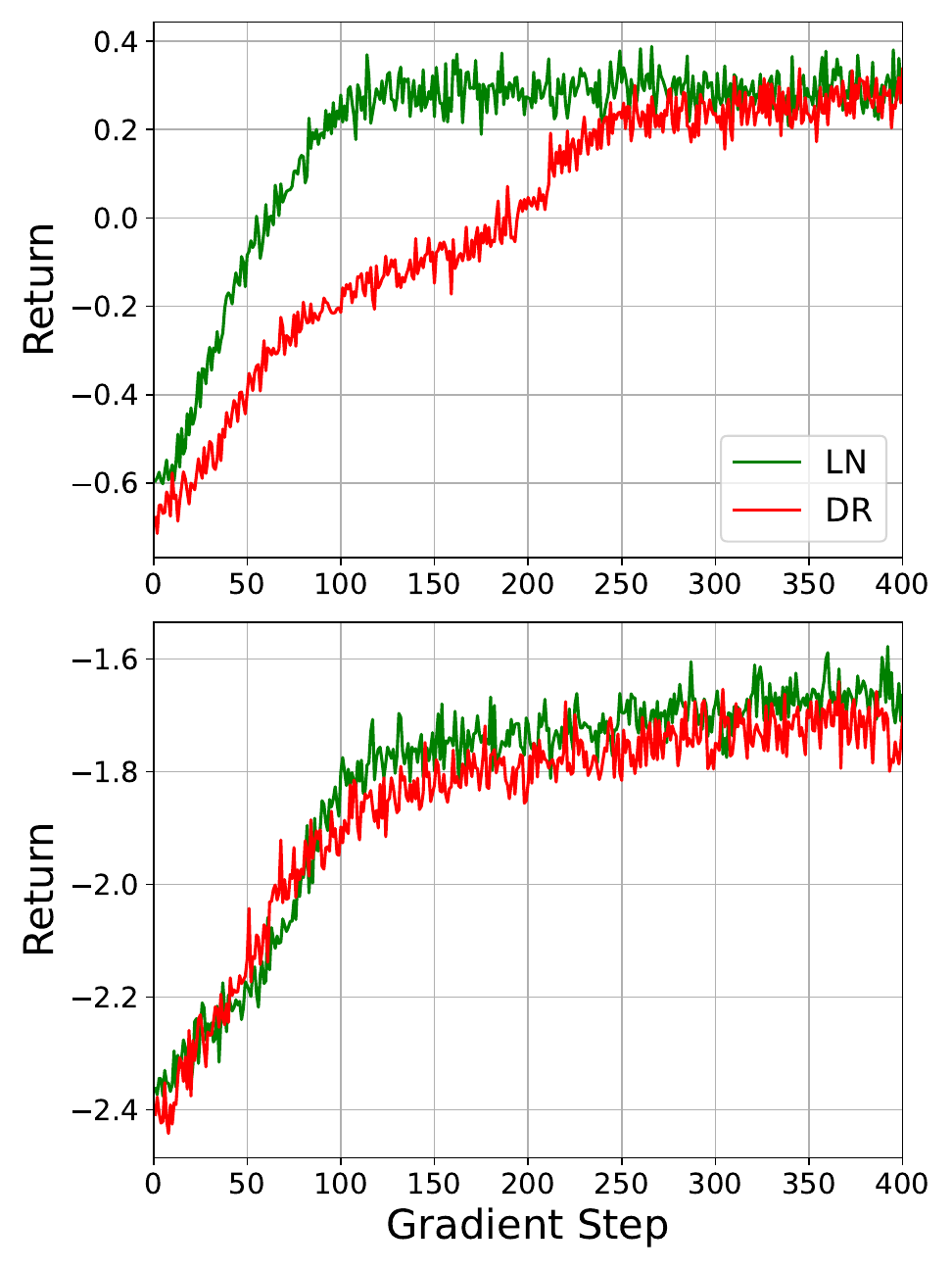}
		\caption{Noise Traders}
		\label{fig:1}
	\end{subfigure}
	\hfill
	\begin{subfigure}[t]{0.3\textwidth}
		\includegraphics[width=\textwidth]{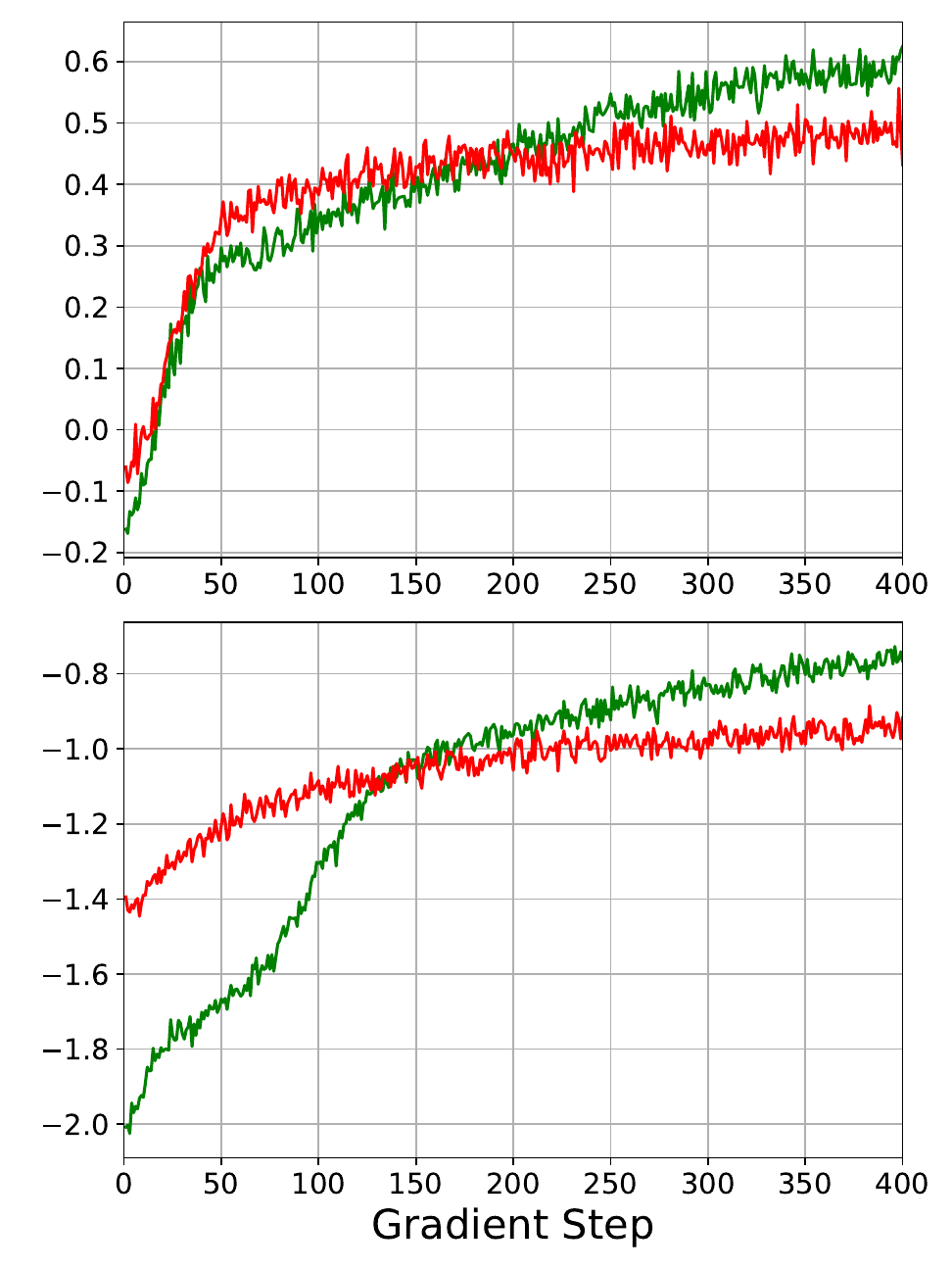}
		\caption{Noise \& Tactical Traders}
		\label{fig:2}
	\end{subfigure}
	\hfill 
	\begin{subfigure}[t]{0.3\textwidth}
	\includegraphics[width=\textwidth]{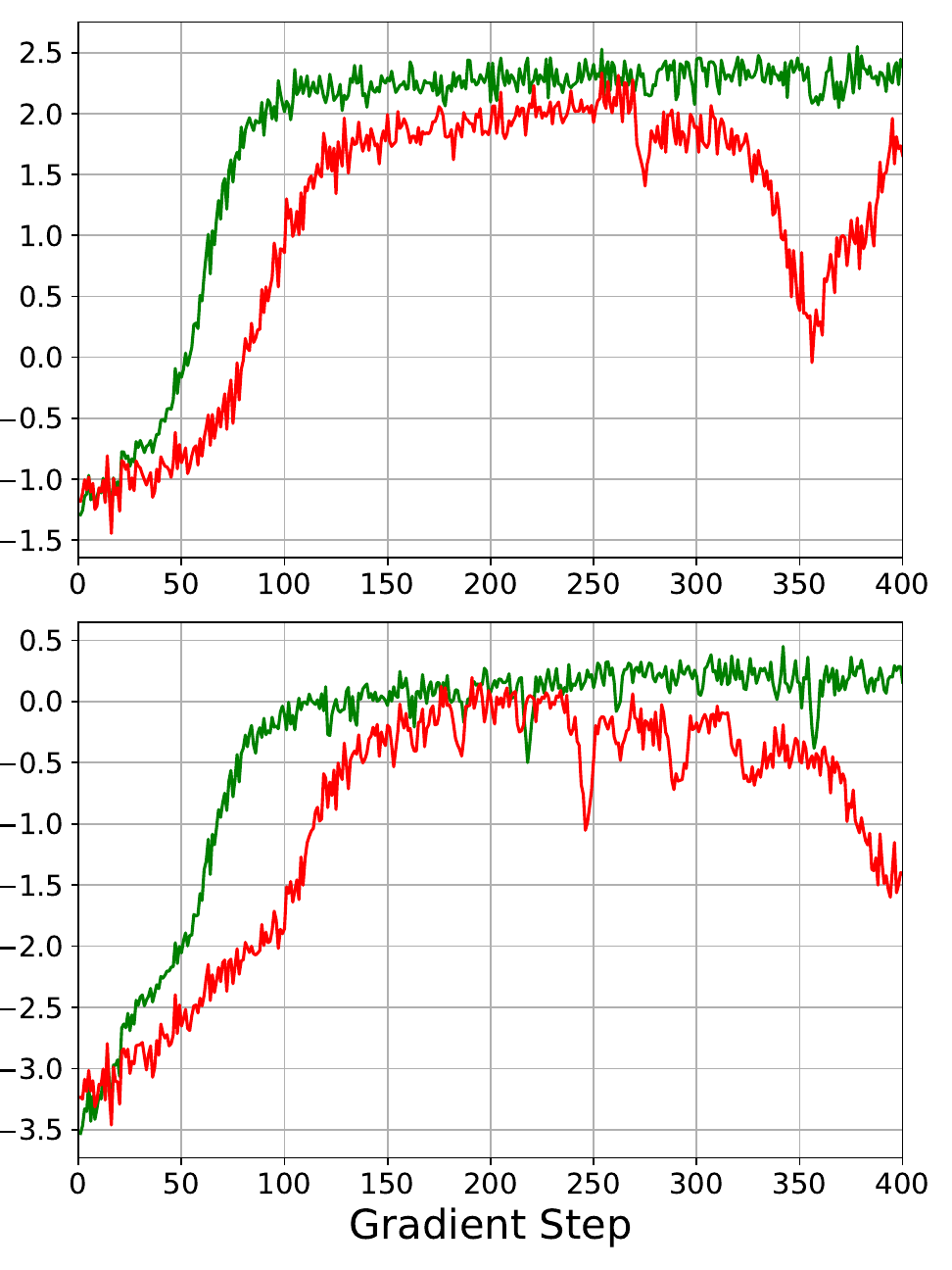}
	\caption{Noise, Tactical \& Strategic Traders}
	\label{fig:3_long}			
	\end{subfigure}	
	\caption{Average episode returns per batch for an execution horizon of 300 seconds and initial positions of 40 and 120 lots. The x-axis shows the number of gradient steps, and the y-axis shows the average returns per batch. The upper panels correspond to an initial position of 40 lots, and the lower panel to 120 lots. LN denotes the logistic-normal algorithm, and DR the Dirichlet algorithm.}
    \label{fig:convergence_long_horizon}
\end{figure}

The LN algorithm outperforms all other algorithms, as shown in Table \ref{table:long_horizon}. Figure \ref{fig:densities_long_horizon} shows the reward histograms for all algorithms. The reward convergence plots in Figure~\ref{fig:convergence_long_horizon} also demonstrate a more stable and faster convergence of the LN algorithm compared to the DR algorithm. In the market with strategic traders (Figure~\ref{fig:3_long}), we observe that the DR algorithm's rewards become unstable toward the end of the training cycle. This may be mitigated by clipping, as in PPO \cite{schulman2017proximal}, but we leave this to future research. 
 
\begin{table}[htbp]
\caption{Returns of all heuristic benchmark algorithms and the logistic-normal and Dirichlet algorithm, for an initial position of 40 and 120 lots, and an execution horizon of 300 seconds. The best expected values per row are highlighted in bold.}
\label{table:long_horizon}
\begin{center}    
\begin{scriptsize}        
\begin{sc}
\begin{tabular}{lccccccccc}
\toprule
Market & \#Lots & $\mathbb{E}[\text{SL}]$ & $\sigma[\text{SL}]$ & $\mathbb{E}[\text{TWAP}]$ & $\sigma[\text{TWAP}]$ & $\mathbb{E}[\text{DR}]$ & $\sigma[\text{DR}]$ & $\mathbb{E}[\text{LN}]$ & $\sigma[\text{LN}]$ \\
\midrule
Noise & 20 & -0.35 & 1.43 & -0.58 & 1.1 & 0.27 & 1.38 & \textbf{0.3} & 1.4 \\
 & 60 & -2.75 & 1.29 & -3.02 & 1.06 & -1.72 & 1.22 & \textbf{-1.64} & 1.21 \\
Noise \& Tactical & 20 & -2.0 & 1.43 & -0.19 & 1.0 & 0.55 & 0.75 & \textbf{0.61} & 0.75 \\
 & 60 & -5.57 & 0.77 & -4.13 & 1.03 & -0.74 & 0.78 & \textbf{-0.72} & 0.81 \\
Noise \& Tactical & 20 & -2.11 & 3.29 & -1.09 & 3.28 & 1.62 & 3.61 & \textbf{2.32} & 3.69 \\
\& Strategic & 60 & -3.96 & 4.01 & -3.42 & 4.18 & -1.35 & 4.33 & \textbf{0.26} & 3.21 \\
\bottomrule
\end{tabular}
\end{sc}    
\end{scriptsize}\end{center}\end{table}

\subsection{Feature importance}
\label{sec:feature_importance}

Finally, we conduct experiments to evaluate feature importance. We tested the LN algorithm across all markets and for all initial positions, while omitting certain features. More precisely, we train the algorithm in three cases. The first case uses all features (LN). The second case uses all features except those that reflect order price levels and queue positions, defined in the fourth item of the private states in Section \ref{subsec:observation_space}, and is referred to as LN-O. The second case uses all features, except the order book volume features, defined in the second item of the market states in Section \ref{subsec:observation_space}, and is denoted by LN-V. The third case uses all features, except the market, limit, and cancellation order flow imbalance, and the mid-price price drift feature, defined in the last four items of the market states in Section \ref{subsec:observation_space}, and is denoted by LN-D. 

\begin{figure}[htbp]
	\centering
	\begin{subfigure}[t]{0.3\textwidth}
		\includegraphics[width=\textwidth]{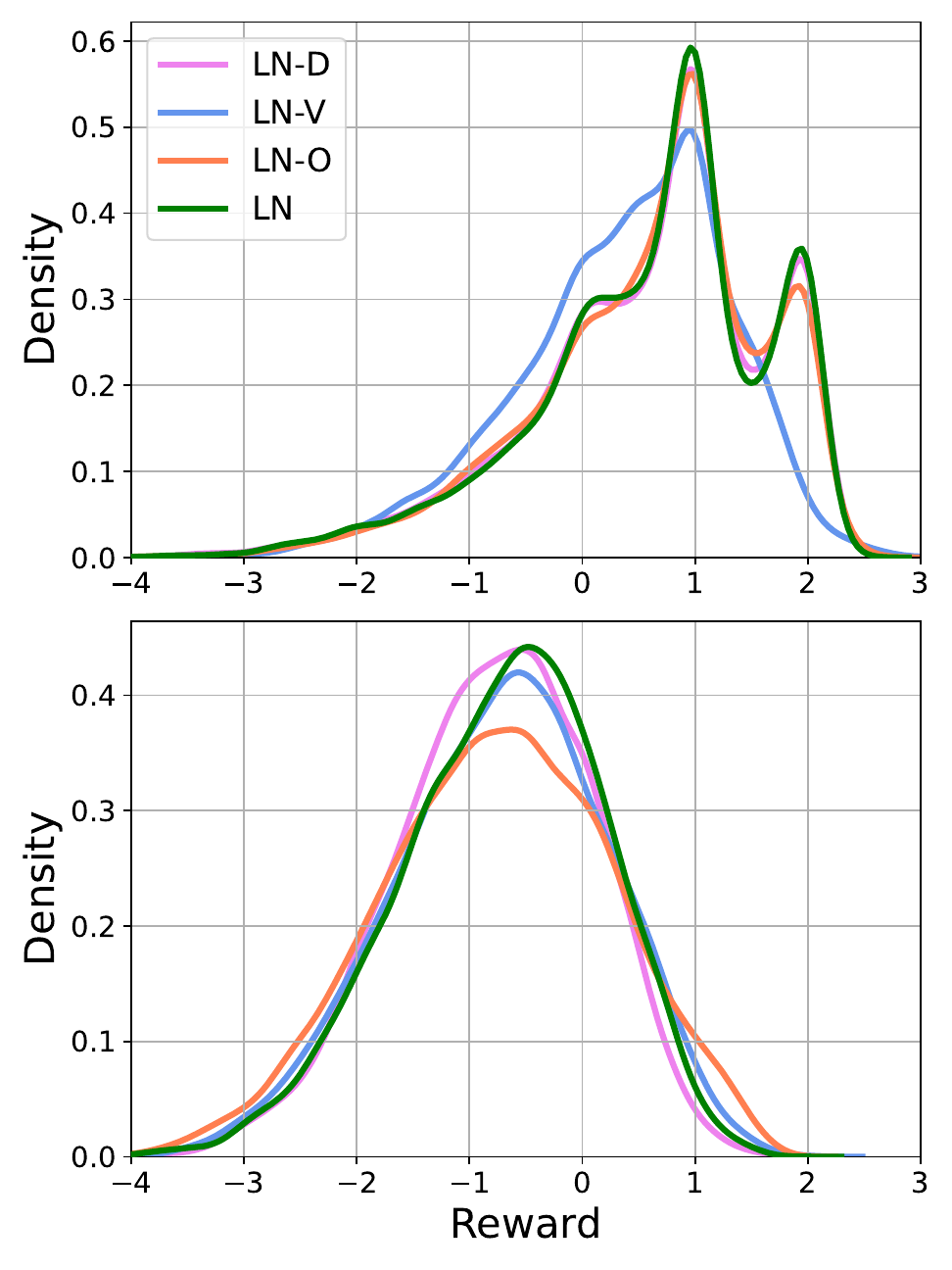}        
		\caption{Noise Traders}
		\label{fig:noise_results_feature}
	\end{subfigure}
	\hfill
	\begin{subfigure}[t]{0.3\textwidth}
		\includegraphics[width=\textwidth]{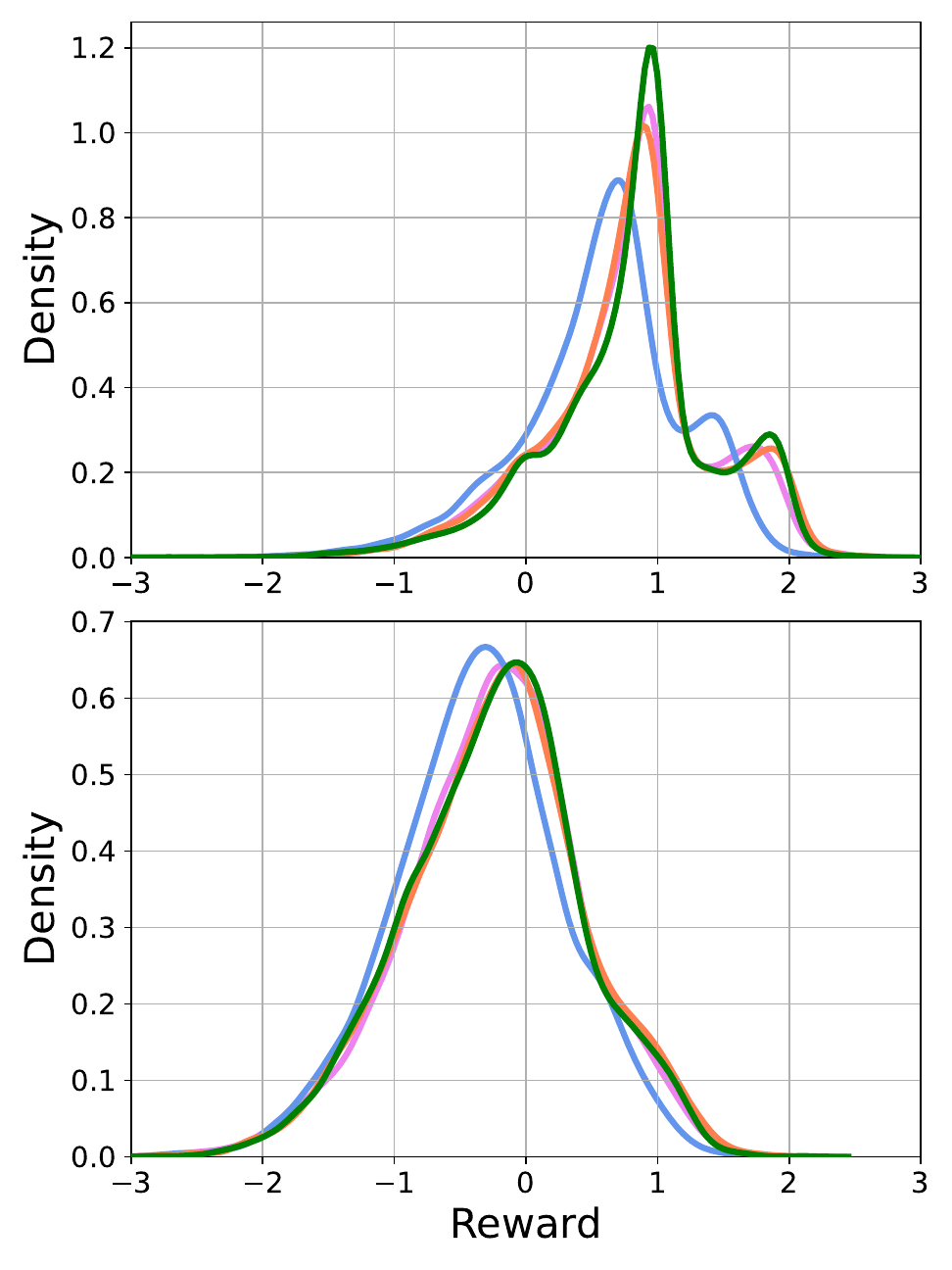}
		\caption{Noise \& Tactical Traders}
		\label{fig:tactical_results_feature}
	\end{subfigure}
	\hfill 
	\begin{subfigure}[t]{0.3\textwidth}
	\includegraphics[width=\textwidth]{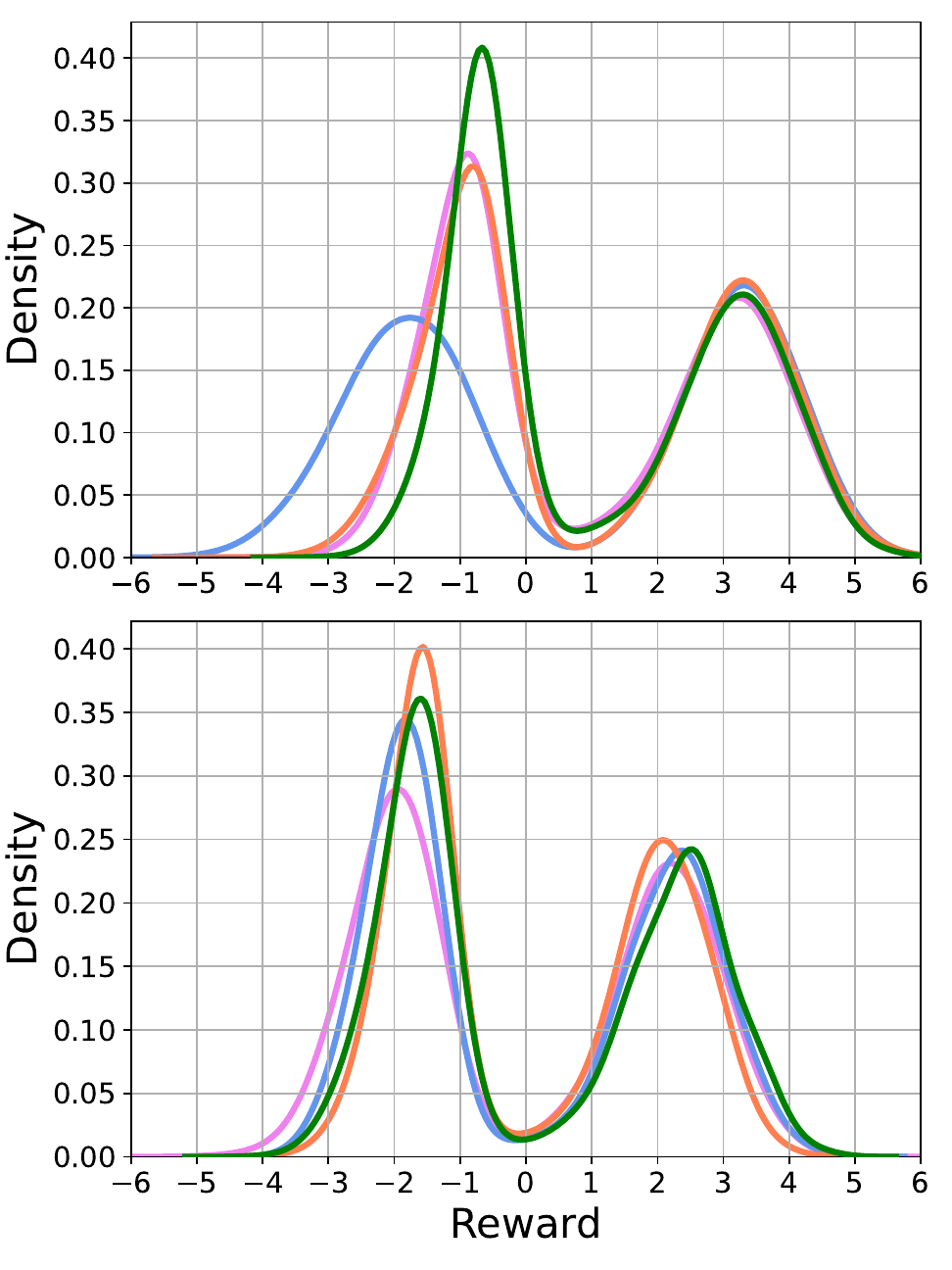}
	\caption{Noise, Tactical \& Strategic Traders}
	\label{fig:strategic_results_feature}			
	\end{subfigure}	
	\caption{Reward densities of the different RL algorithms. LN corresponds to the logistic–normal algorithm that uses all features. LN-O corresponds to the algorithm without level and queue position features. LN-V corresponds to the algorithm without volume features. LN-D corresponds to the algorithm without the order flow imbalance and mid-price drift features. The upper panels correspond to an initial position of 20 lots. The lower panel corresponds to an initial position of 60 lots. }
	\label{fig:densities_feature_importance}
\end{figure}

The results are reported in Table~\ref{table:feature_importance} and Figure~\ref{fig:densities_feature_importance}. The most important features are the order book volumes. Removing them leads to the worst performance across all markets and initial positions, except for an initial position of 60 lots in the market with noise and tactical traders, and for an initial position of 60 lots in the market with noise, tactical, and strategic traders. This is expected, as volume imbalance is a strong predictor of short-term price movements. Interestingly, the volume features remain important even in the market with strategic traders, where one might expect order-flow or mid-price drift information to dominate. This suggests that order book volumes also help detect the presence and direction of strategic traders. The order-flow and mid-price drift features are the second-most-important set of inputs, particularly in markets with strategic traders, where dropping them leads to substantially worse performance. They are less influential in the market with only noise traders, and in the market with noise and tactical traders, where using all features offers only marginal improvement. The least important features are those related to the levels and queue positions of the algorithm’s own orders. In the market with noise and tactical traders, when 60 lots are traded, omitting these features even yields slightly better performance, indicating that volume information dominates in this environment. However, in the market with strategic traders, removing these features significantly degrades performance. In this market, queue position information becomes important for execution timing. If the strategic trader is selling, the market is more likely to drift downward, making it advantageous for a seller to be near the front of the queue and secure a fill before prices fall. Conversely, if the strategic trader is buying and the market is likely to drift upward, being further back in the queue is preferable, as the order can still be canceled to capture a better selling price if the queue size at the best ask price decreases and an upward price move becomes more likely.

\begin{table}[htbp]
\begin{center}    
\caption{Rewards for different RL algorithms, omitting certain features. LN corresponds to the logistic–normal algorithm that uses all features. LN-O denotes the algorithm without level and queue-position features. LN-V corresponds to the algorithm without volume features. LN-D corresponds to the algorithm without market, limit, and cancellation order imbalance features, and the mid-price drift feature. The best expected values per row are highlighted in bold.}
\label{table:feature_importance}
\begin{scriptsize}        
\begin{sc}\begin{tabular}{lccccccccc}
\toprule
Market & \#Lots & $\mathbb{E}[\text{LN-D}]$ & $\sigma[\text{LN-D}]$ & $\mathbb{E}[\text{LN-V}]$ & $\sigma[\text{LN-V}]$ & $\mathbb{E}[\text{LN-O}]$ & $\sigma[\text{LN-O}]$ & $\mathbb{E}[\text{LN}]$ & $\sigma[\text{LN}]$ \\
\midrule
Noise & 20 & \textbf{0.61} & 1.04 & 0.38 & 0.94 & \textbf{0.61} & 1.01 & \textbf{0.61} & 1.03 \\
 & 60 & -0.79 & 0.86 & -0.73 & 0.95 & -0.77 & 1.03 & \textbf{-0.72} & 0.90 \\
Noise \& Tactical & 20 & 0.76 & 0.64 & 0.58 & 0.62 & 0.78 & 0.64 & \textbf{0.81} & 0.64 \\
 & 60 & -0.25 & 0.67 & -0.36 & 0.64 & \textbf{-0.23} & 0.68 & -0.25 & 0.67 \\
Noise \& Tactical & 20 & 0.99 & 2.2 & 0.69 & 2.73 & 1.09 & 2.28 & \textbf{1.13} & 2.08 \\
\& Strategic & 60 & 0.00 & 2.22 & 0.11 & 2.17 & 0.10 & 1.96 & \textbf{0.23} & 2.15 \\
\bottomrule
\end{tabular}
\end{sc}    
\end{scriptsize}
\end{center}
\end{table}

\end{document}